\documentclass[11pt]{article}
\usepackage{cite}
\pdfoutput=1
\usepackage{amsmath,amsfonts,amssymb,hyperref}
\usepackage[small,bf,hang]{caption}
\usepackage{slashed}
\input epsf.sty
\usepackage{epsfig}
\usepackage[titletoc,toc]{appendix}

\def\hybrid{
        \topmargin -20pt
        \oddsidemargin 0pt
        \headheight 0pt \headsep 0pt
        \textwidth 6.55in 
        \textheight 9.5in 
        \marginparwidth .875in
        \parskip 5pt plus 1pt \jot = 1.5ex}

\hybrid
 \csname
@addtoreset\endcsname{equation}{section}

\newcommand{\sectiono}[1]{\section{#1}\setcounter{equation}{0}}

\def\moth{\mathsurround=0pt}
\newdimen\zo \zo=0pt

\def\tick{\leaders\hrule height 0.5ex depth 0pt \hskip 0.5pt}
\def\upboxfill{$\moth \setbox\zo\hbox{\tick}%
  \hskip 3pt\hbox to 0pt{$\tick$\hss}\hrulefill \hbox to 7.5pt{$\tick$\hss}$}

\def\dtick{\leaders\hrule height .34pt depth 0.5ex \hskip 0.5pt}
\def\downboxfill{$\moth \setbox\zo\hbox{\dtick}%
  \hskip 2pt\hbox to 0pt{$\dtick$\hss}\hrulefill \hbox to 2pt{$\dtick$\hss}$}

\def\bec{\begin{center}}
\def\ec{\end{center}}

\def\be{\begin{equation}}
\def\ee{\end{equation}}
\def\bea{\begin{eqnarray}}
\def\eea{\end{eqnarray}}
\def\ba{\begin{array}}
\def\ea{\end{array}}

\begin{document}

\begin{titlepage}

\rightline{\tt BRX-TH-6332} 
\rightline{\tt MIT-CTP-4897} 
\begin{center}
\vskip 1.5cm

{\Large \bf {Minimal-area metrics on the Swiss cross and punctured torus
}}

\vskip 0.5cm

  \vskip 1.0cm
 {\large {Matthew Headrick}}
 
{\em  \hskip -.1truecm Martin Fisher School of Physics \\
Brandeis University\\
Waltham MA 02143, USA\\
\tt headrick@brandeis.edu}

 \vskip 0.5cm

 {\large {and}}

  \vskip 0.5cm
 {\large {Barton Zwiebach}}

{\em  \hskip -.1truecm 
Center for Theoretical Physics \\
Massachusetts Institute of Technology\\
Cambridge MA 02139, USA\\
\tt zwiebach@mit.edu \vskip 5pt }

\vskip 1.0cm
{\bf Abstract}

\end{center}


\noindent
\begin{narrower}

\baselineskip15pt

The closed string theory
minimal-area problem asks for the conformal 
metric of least area on a Riemann surface with the condition that 
all non-contractible closed curves 
have length greater
than or equal to an arbitrary constant that can be set 
to one.  Through every point in such a metric there is a geodesic that saturates the length condition, and the saturating geodesics in a given homotopy class form a band. The extremal  metric is unknown when bands of geodesics cross, as it happens for surfaces of non-zero genus. We use recently proposed convex programs to numerically find the minimal-area metric on the square torus with a square boundary, for various sizes of the boundary.  For large enough boundary the problem is equivalent to the ``Swiss cross'' challenge posed by Strebel. We find that the metric is positively curved in the two-band region and flat in the single-band regions. 
For small boundary the metric develops a third band of geodesics wrapping around it, and has both 
regions of positive and of negative curvature. This 
surface can be completed to provide the minimal-area metric on 
a once-punctured torus, representing a closed-string tadpole diagram.

\end{narrower}

\end{titlepage}

\baselineskip11pt

\tableofcontents


\baselineskip15pt

\section{Introduction and summary}

A class of minimal-area metrics on Riemann 
surfaces is the key ingredient in the definition of
a field theory of closed strings~\cite{Zwiebach:1992ie}.  
In this paper
we continue the investigation we 
began in \cite{headrick-zwiebach},
where we used the tools of convex optimization~\cite{boyd} 
to formulate programs whose solutions
are these minimal-area metrics.   Related ideas from
convex optimization have also proven useful in other
contexts~\cite{Freedman:2016zud,Headrick:2017ucz}. 
The goal of our work in \cite{headrick-zwiebach} was to address
the minimal-area problem~\cite{Zwiebach:1990ni,Zwiebach:1990nh}
that asks for the conformal metric of least area on a Riemann surface
under the condition that all noncontractible closed curves
be longer than or equal to a fixed length $\ell_s$ that can
be chosen arbitrarily.  The length $\ell_s$ is the
{\em systole} of the surface,  the length of the shortest 
non-contractible closed curve.  
All curves of length $\ell_s$ are geodesics and are called
systolic geodesics.   
 In general the surfaces have marked points 
and curves cannot be moved across those points.
The minimal-area metrics are known for spheres with
arbitrary number of marked points~\cite{Saadi:1989tb,Kugo:1989aa}.
They are also known in {\em some} regions of every moduli space of Riemann
surfaces with genus greater than or equal to one and some number of marked points.
  
The extremal metric is expected
to have systolic geodesics 
that cover the full surface.  Near the marked points the systolic geodesics are circles on
a flat semi-infinite cylinder, with the marked point at infinite distance. 
Each point on that cylinder belongs to just one systolic
geodesic.   On the rest of the surface there can exist
regions where points belong to just one 
systolic geodesic, and  regions where points belong to
several systolic geodesic.  
In general the surface is covered by 
multiple bands of intersecting and non-intersecting systolic geodesics.
It has been shown that the conformal metric in a region covered by a 
{\em single} band of systolic geodesics is flat~\cite{Ranganathan:1991qd,Wolf:1992bk}.
Generically, however,
 the metric is {\em not known} as soon as systolic bands intersect.
This will happen for surfaces over finite subsets of each
moduli space ${\cal M}_{g,n}$ of Riemann surfaces of
genus $g\geq 1$ and $n$ marked points.\footnote{The exception is ${\cal M}_{1,0}$,
the moduli space of genus one surfaces with no punctures, where all
minimal area metrics are known.  For the square torus $\tau=i$ 
each point on the surface
 lies on two systolic geodesics.  For the torus $\tau= e^{i\pi/3}$
each point lies on three systolic geodesics.}
All known minimal area metrics are locally flat, have curvature
singularities, and arise as the norm of Jenkins-Strebel
quadratic differentials~\cite{strebel}. 

A variant of this minimal area 
problem, as well as  a particular case of Gromov's isosystolic
problems~\cite{gromov},  asks for the {\em Riemannian} metric 
of minimal area on a two-dimensional 
surface of fixed topology and fixed systole. 
Here one minimizes over the conformal structure.  
 Calabi~\cite{calabi} proved that for this problem 
there are no regions covered by a single band of systolic
geodesics.  
Moreover, in regions covered by exactly two bands of systolic geodesics, 
the metric is flat and the geodesics in these bands are 
orthogonal to each other.   Calabi described the foliations using calibrations, 
and this approach was elaborated upon and extended by Bryant~\cite{bryant}. 
The proofs, however,
do not apply in the conformal case.  
In fact,  extremal conformal metrics have 
regions covered by single bands of geodesics and the conformal
metric in regions covered by two bands of geodesics is not
necessarily flat, as we will see.    For more information on
systolic geometry see~\cite{m_katz}.

\begin{figure}[!ht]
\leavevmode
\begin{center}
\epsfysize=6.5cm
\epsfbox{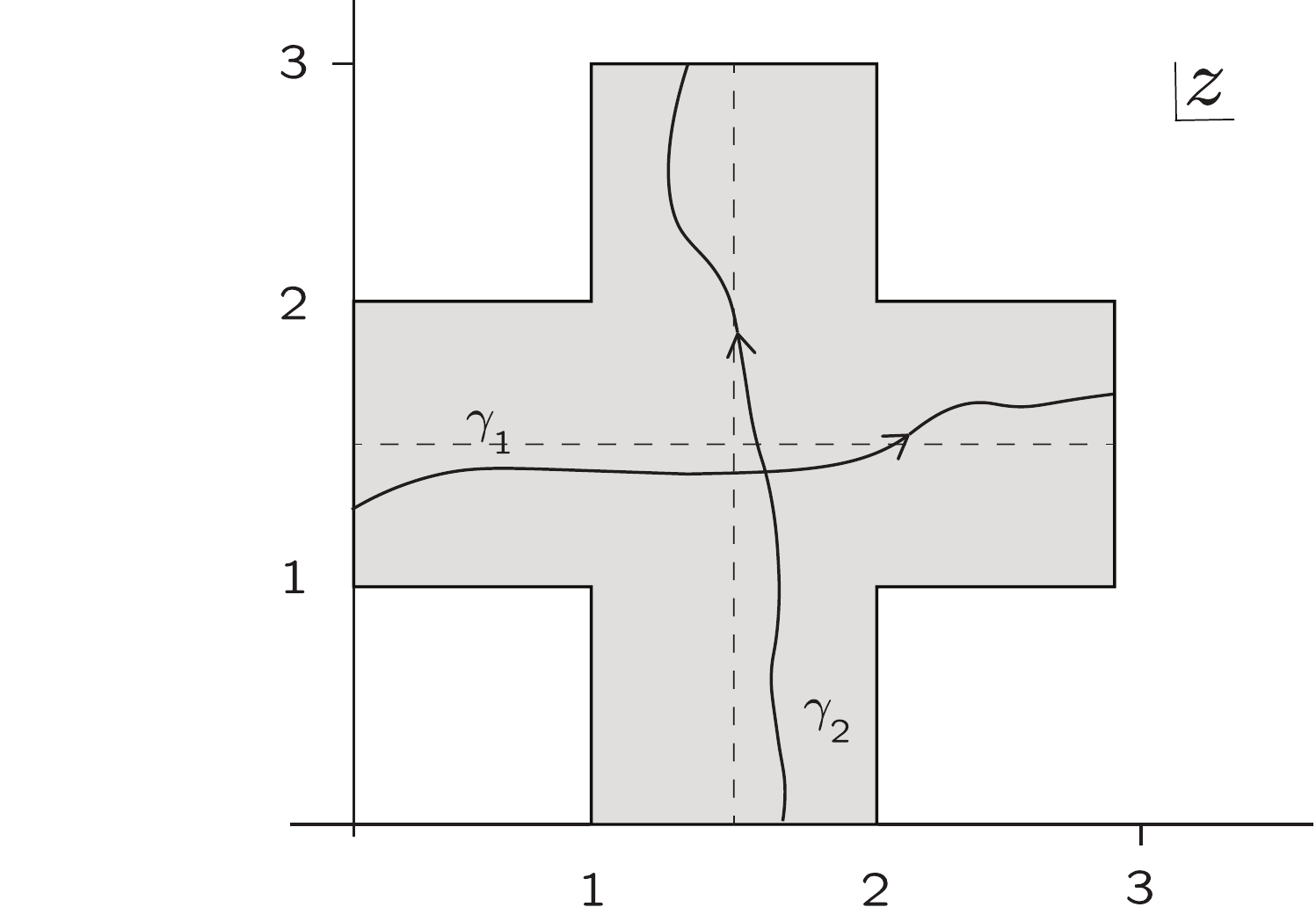}
\end{center}
\caption{\small  The region of the complex plane defining
a Swiss cross.   In Strebel's challenge the metric must have least
area while all curves
from the left edge to the right edge (like $\gamma_1$) and
all curves from the bottom edge to the top edge (like $\gamma_2$)
are longer than or equal to $\ell_s =3$. }
\label{ffjn1}
\end{figure}

For this paper we wanted to identify 
the simplest possible surface for which we have
intersecting bands of systolic geodesics and the extremal metric
is not known.  
A good candidate is provided by {\em Strebel's challenge:} Find
the extremal metric on a Swiss cross with two length conditions~\cite{strebel-pc}.   A Swiss cross, defined as the region of the
complex $z$ plane shown in Figure~\ref{ffjn1}, is a planar figure
that can be build bringing together five unit squares 
or by arranging two
identical rectangles perpendicular to each other.  
A Swiss cross has four edges: 
a left edge ($x=0, y\in [1,2]$), a right edge  ($x=3, y\in [1,2]$) 
a bottom edge ($x\in [1,2], y=0$) and a top edge ($x\in [1,2], y=3$).
We constrain the lengths of all curves beginning on the left
edge and ending on the right edge, like $\gamma_1$.  
Similarly, we constrain the lengths of all curves beginning on the bottom
edge and ending on the top edge, like $\gamma_2$.    We thus have constraints
on curves that necessarily intersect.  
For the Swiss cross shown in the figure, whose arms have parameter length 3,
a convenient
choice of systole is $\ell_s =3$.

Perhaps surprisingly, the
extremal metric for the Swiss cross was unknown (some attempts to find it were made in \cite{marnerides} and \cite{vigre}). 
Not only that, one was also lacking a qualitative understanding 
of the properties of the extremal metric. 
 It was not known if it is locally flat (except at 
 special points) or, if curved, whether the 
curvature is positive or negative, constant or non-constant.    Moreover, while one must have two
bands of saturating geodesics, it was not known if each band
would cover the full cross or if the bands intersect 
over some proper subregion of the cross.  We will provide answers to all these questions.

Recalling that a conformal metric $\rho(x, y)$ is defined 
by the length formula $ds = \rho |dz|$, with $z = x+ iy$,
we note that 
the flat constant metric $\rho=1$ is admissible because it satisfies
all length conditions.  Since it has area 5 we know
that the Swiss cross minimal area $A$ 
satisfies $A <  5$.   On the other hand, it was shown in \cite{marnerides}
that $A > 9/2$, so we have
\be
4.5 <    A  <   5 \,. 
\ee

\begin{figure}[!ht]
\leavevmode
\begin{center}
\epsfysize=6.5cm
\epsfbox{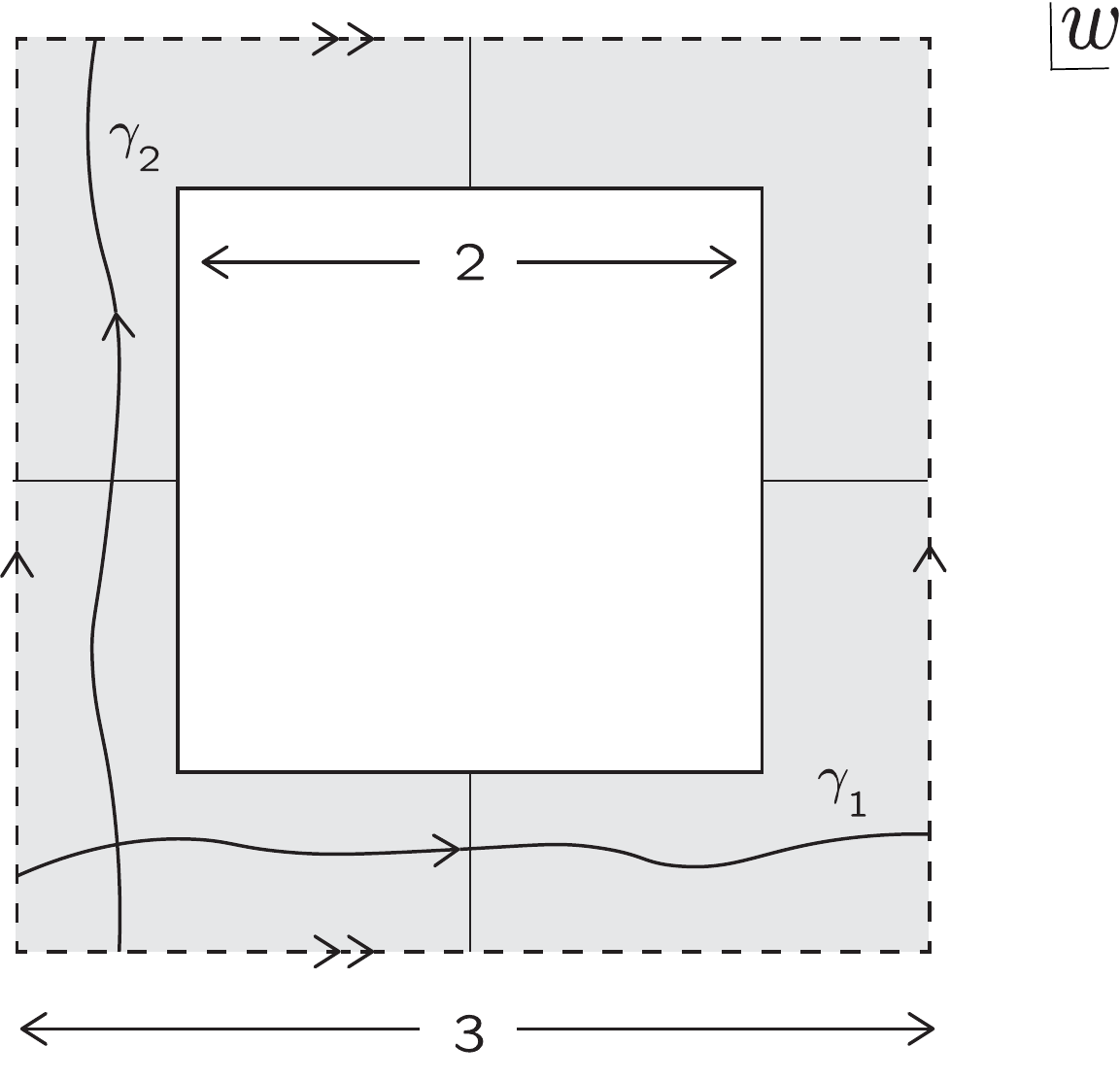}
\end{center}
\caption{\small  The square torus with one boundary obtained
by identification of edges on the Swiss cross of Figure~\ref{ffjn1}.
We constrain the curves homologous to $\gamma_1$ and those
homologous to $\gamma_2$ to be longer than or equal to $\ell_s =3$. }
\label{ffjn2}
\end{figure}

The Swiss cross is an excellent problem to try
the new programs proposed in~\cite{headrick-zwiebach}. 
As formulated above, however, it is not  
a closed string theory minimal-area problem.  We will show
that the solution of the Swiss cross problem actually provides
the minimal-area metric on a torus with a boundary.
The torus is obtained by identifying the left and right edges
and the top and bottom edges of the Swiss cross.   The resulting
surface (after cutting along the dotted lines 
in Figure~\ref{ffjn1} and rearranging the parts)  is shown in Figure~\ref{ffjn2}.  We have
a square torus with an aligned square boundary.  In this
torus, the minimal-area problem constrains the curves homologous
to $\gamma_1$ and the curves homologous to $\gamma_2$ to be longer than or equal to $\ell_s=3$.  

A torus with a boundary belongs to a moduli space 
relevant to {\em open-closed} string theory.
In fact, by making the arms of the Swiss cross of arbitrary length (while keeping
their width equal to one) we can introduce a modulus to the Swiss cross.  That modulus is also a modulus for the torus with a boundary, 
as it controls the ratio of the size of the boundary to the size of the torus.  The metrics we will find would be relevant to
some vacuum graphs in open-closed string field theory.\footnote{For such
surfaces the generalized minimal area problem~\cite{zwiebach-open-closed} 
would also impose
length conditions on non-contractible open curves that begin and
end on the boundary.  We will not study this problem here.} 
Equipped with a modulus, chosen as the size
of the square boundary,  we can also study the moduli
space of tori for the case of small boundary,  where we require
 that curves homotopic to the boundary are longer 
 than or equal to $\ell_s$.  In the limit when the size of the boundary
 goes to zero we find the metric on a once-punctured square torus.  This
 metric is relevant to closed string theory---it belongs to the one-loop
 once-punctured string vertex ${\cal V}_{1,1}$ of the closed  
 string field theory
---and was previously unknown.  
 A discussion of this vertex can be found in section~2.1 of~\cite{headrick-zwiebach}.

\begin{figure}[!ht]
\leavevmode
\begin{center}
\epsfysize=7.9cm
\epsfbox{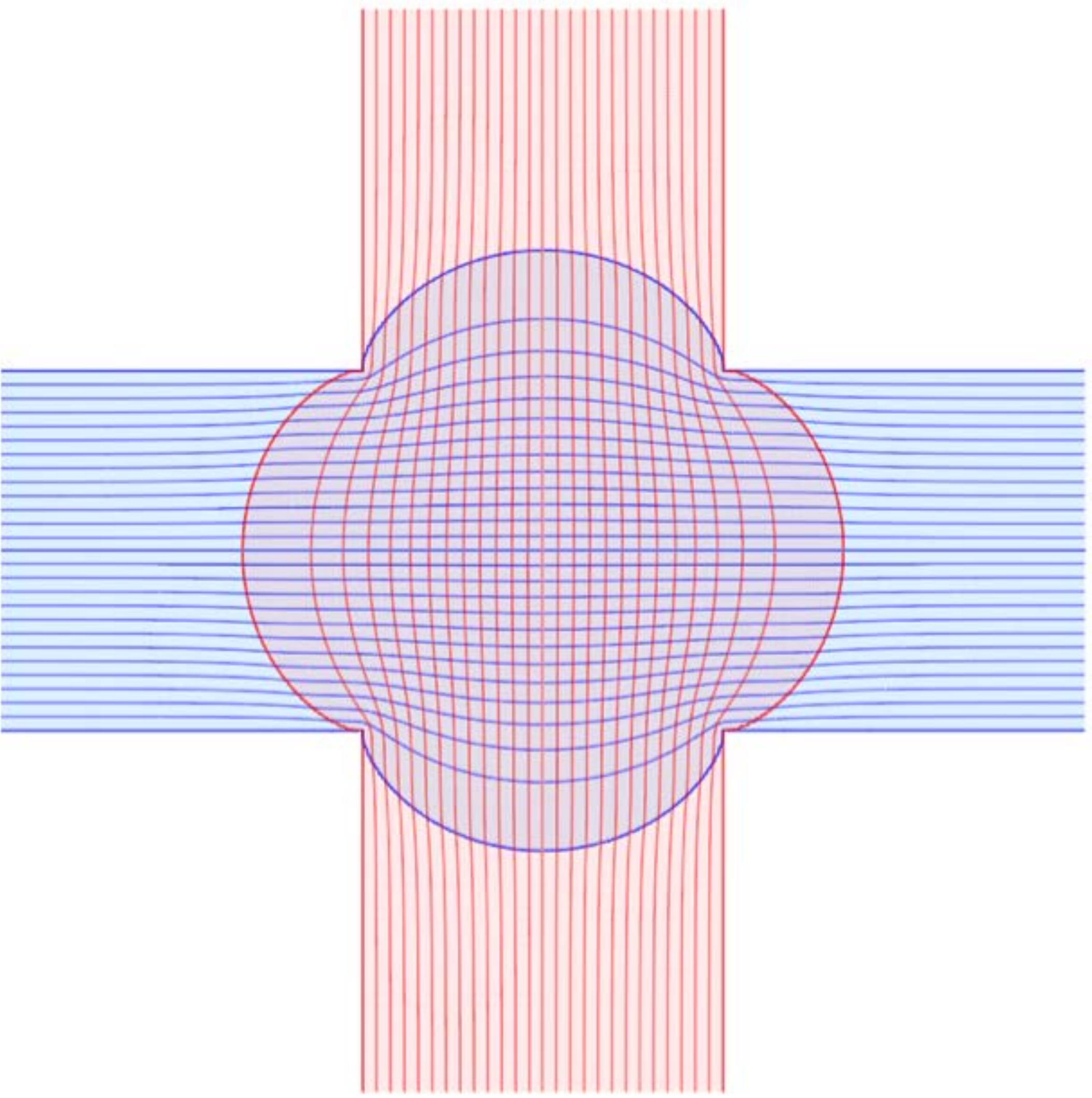} \hskip10pt
\epsfysize=7.9cm
\epsfbox{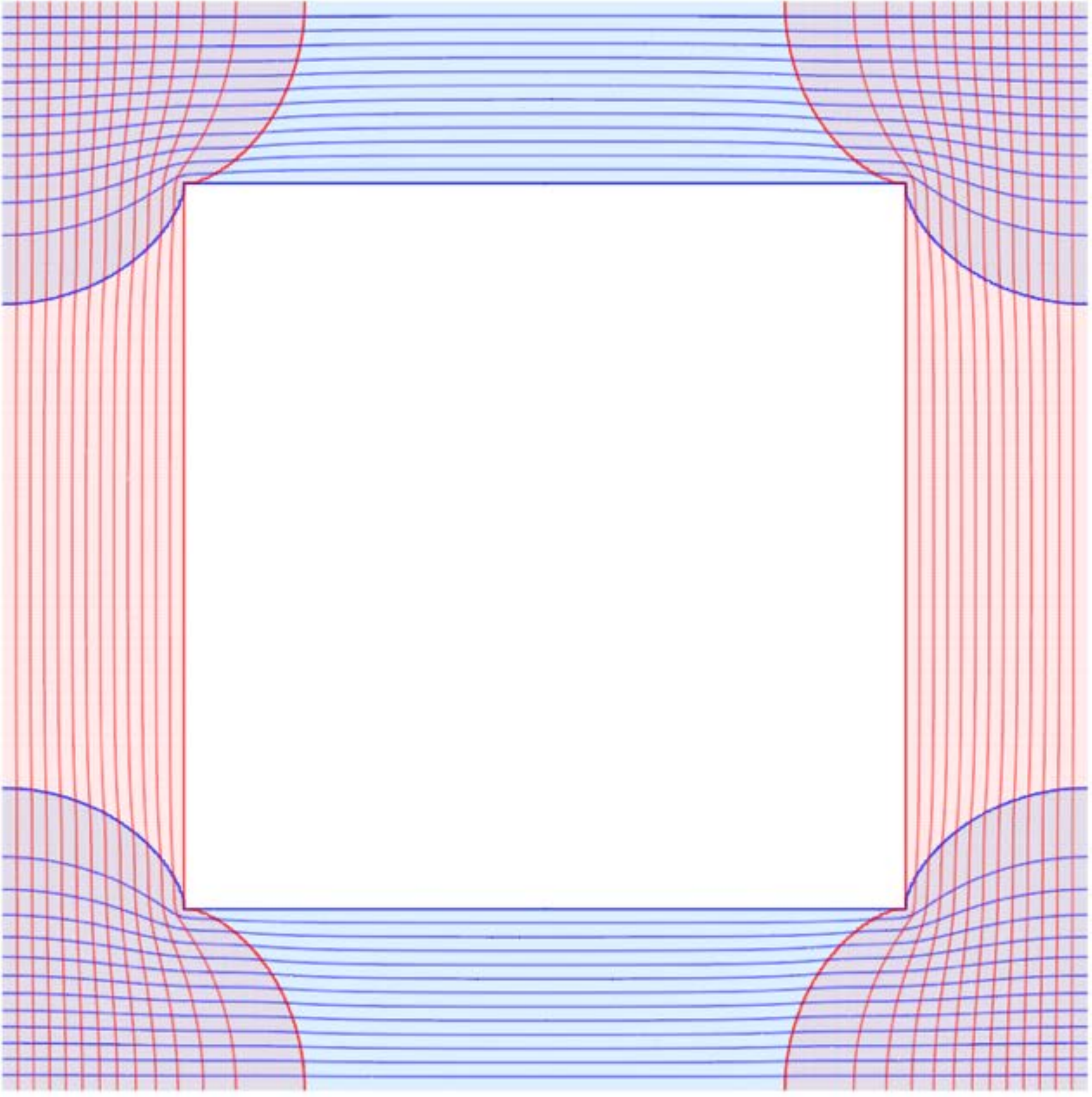}
\end{center}
\caption{\small  Left: The systolic geodesics for the extremal metric
on the $\ell_s =3$ Swiss cross.  
The blue geodesics running from left to right
are all of length 3 and so are the red geodesics running from bottom to top.  There is a central region where two geodesics go through each
point and regions near the edges where there is just one geodesic
through each point.  Right:  The Swiss cross on the left reconfigured 
as a torus with a boundary. The boundary of the region with a single
geodesic is in fact the boundary of the torus.}
\label{ffjn3}
\end{figure}

Our work begins by applying the two programs in~\cite{headrick-zwiebach} to find the extremal metric and
the value of the minimum area for the Swiss cross/torus-with-a-boundary problem.
As explained in detail in~\cite{headrick-zwiebach}, while the closed string minimal
area problem constrains homotopy classes of curves, the programs are most simply formulated to constrain {\em homology} classes of curves.  For the present problem, however,
the distinction has no effect;  the two homotopy classes of curves we wish to constrain belong to two homology classes, and constraints in these homologies do not impose additional length conditions on curves.  
Using the max flow-min cut
theorem  the length conditions on non-contractible
curves are implemented by 
a set of calibrations (closed one-forms $u$ with
fixed periods and satisfying $|u| \leq 1$). This turns nonlocal length constraints on the metric into local constraints, making the problem tractable for numerical analysis.
This first program, called the primal, minimizes the area 
while allowing the calibrations that constrain the metric to vary. Introducing Lagrange multipliers for the constraints and eliminating the original variables, we obtained a dual program. In this program
we {\em maximize} a functional defined in terms of a set of functions on the surface with prescribed discontinuities across curves that represent the homologies that are constrained.  By strong duality the minimum of the primal and the maximum
of the dual must agree.   Both programs are implemented by defining a lattice on the Swiss cross and writing a {\em Mathematica} program for
each case, using the functions FindMinimum and FindMaximum to obtain the extremum.   Our results indicate that the minimum area $A$ is 
\be
\label{extremal-area-value+ls=3}
A= 4.675\hskip1pt 145\,,  
\ee 
with an error of $\pm1$ in the last digit. Both the extremal metric and the systolic geodesics are determined
 by the programs.  In Figure \ref{ffjn3} we show the systolic geodesics on the $\ell_s=3$ Swiss cross and 
its torus version, respectively.

Here are a few features of the extremal metric, in terms of the
Swiss-cross presentation:
\begin{enumerate}

\item The saturating geodesics form bands;  a horizontal band (blue) and a vertical band (red).  The bands intersect over a central 
region. The boundaries of these bands are piecewise geodesics
going through the corner points of the central region.

\item  The surface has four regions covered by just one
band of systolic geodesics.  The metric is flat in these regions
and  the geodesics reach the edges orthogonally.

\item The metric has positive non-constant
Gaussian curvature throughout
the two-band central region.  The integral of Gaussian
curvature over this
region is equal to $2\pi$, the same as that of a round hemisphere. 
  
\item The boundary of the central region is the locus
of negative line-curvature. On each arm of the Swiss cross
this curvature integrates to $-\pi$, for a total of $-4\pi$. 

\item The metric 
diverges at each of the four corner points of the central region.  
This divergence straightens out the boundary, 
making its extrinsic curvature at those points zero.

\end{enumerate}

Our results suggest it may be challenging to find
an explicit closed-form expression for the extremal metric.
Nevertheless, the metric has a number of remarkable
properties.
Both the bulk curvature and the line curvature integrate
to simple quantities.  Moreover, the value of the extremal
area satisfies a simple relation.  
The optimum of the dual program includes parameters
$\nu^1$ and $\nu^2$, defined as the values
 of the discontinuity of the 
functions associated to the two constrained homologies.  
Because of symmetry, in the Swiss cross the two parameters
are the same, both equal to some value $\nu$.
As demonstrated in \cite{headrick-zwiebach}, section 7.3,  
 the value of $\nu$ at the optimum
 is the height of the band of systolic geodesics measured
 on a region where it is the only
 systolic band.  In the Swiss cross this height is the  
  length (on the extremal metric) of any of the edges. 
  The dual program predicts that the minimum area $A$
  is $2\nu \ell_s$, as if the surface was built from two
  rectangles, one for each band, each of systole length and
   height $\nu$!  Somehow the complicated metric arranges
   matters for this to hold.      The positive curvature on the central
   region is reminiscent of Pu's result~\cite{pu} for the constant
   positive curvature minimal-area metric on $\mathbb{RP}_2$.

\begin{figure}[!ht]
\begin{center}
\epsfysize=14.5cm
\epsfbox{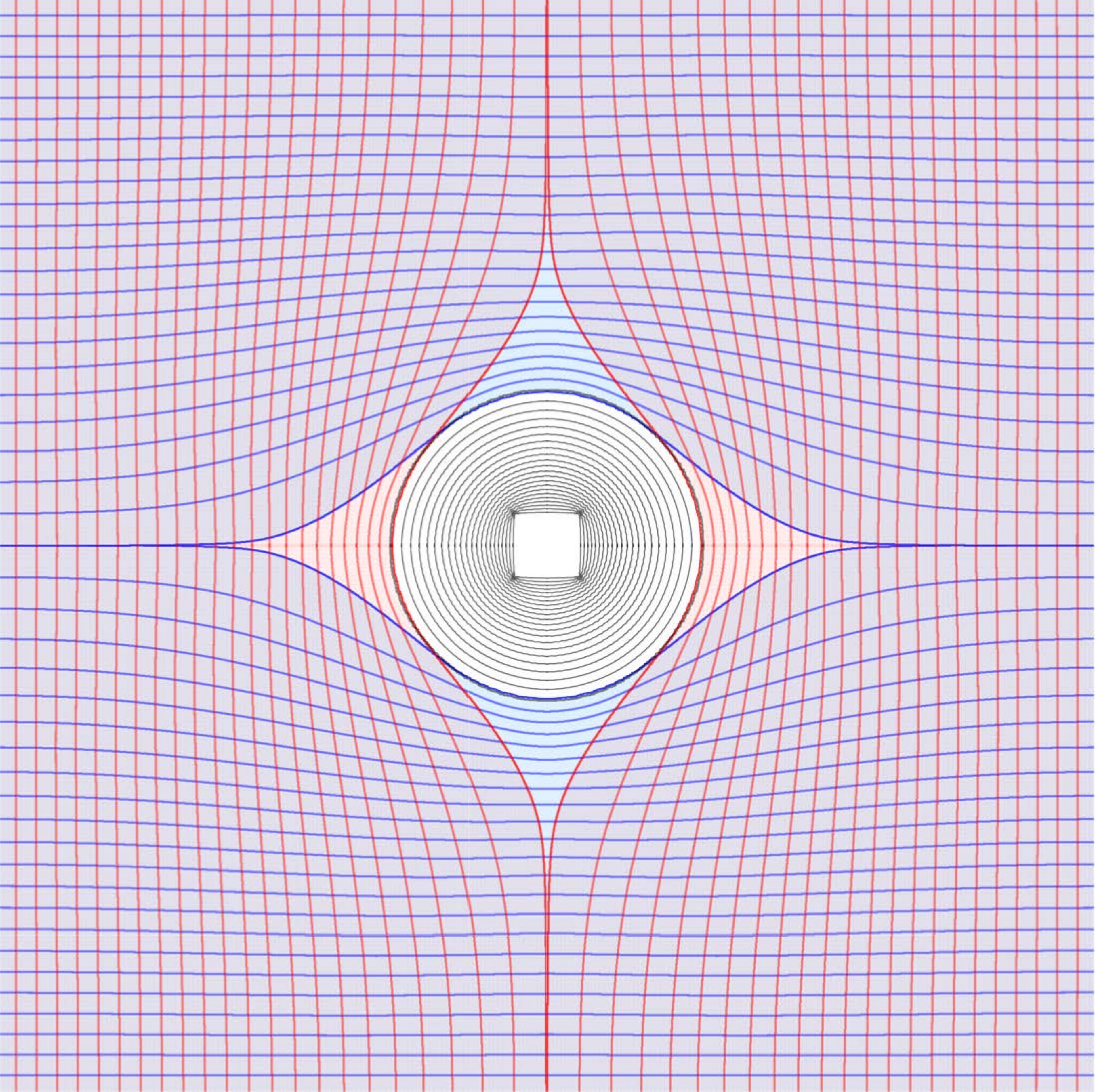}
\end{center}
\caption{\small Systolic geodesics on the torus with a hole removed. The hole is a square with side length $1/16$ that of the torus. There are geodesics in three different homotopy classes: 1-geodesics (blue), which run horizontally; 2-geodesics (red), which run vertically; and 3-geodesics (black) which wrap the hole.
}
\label{fig:punc-h1/16-fig}
\end{figure}

We explored the moduli space of the square torus with a boundary. To do so it is convenient to
describe the  torus as a square with an edge of size 
1 and a square boundary with an edge of size $h$.  
The systole is then chosen to be equal to $1$ for all 
values of $h$.  In Figure \ref{ffjn3}, for example, the edge of the boundary
has size 2, while the edge of the square has size 3.  With an overall scaling we conclude that
$h= 2/3$ for this torus.   The full moduli space of the square
torus with a boundary is 
$0 < h < 1$.  As $h\to 1$ the boundary is as large as possible, and this corresponds in string theory to long-time open string propagation. 
For all $h > 2/3$ the minimal area metric is qualitatively
similar to the $h=2/3$ metric whose properties were enumerated
above.  As we consider $h < 2/3$ no qualitative
changes occur until $h = h^{(1)} \simeq 0.4201$.  At this point
the arms of the Swiss cross have become short enough that the
central region hits itself (see Figure~\ref{fig:just-touch}).  
For $h < h^{(1)}$ we find a surprise: in the region with two systolic bands, the extremal metric displays both positive curvature and 
{\em negative} curvature.  

As $h \to 0$ the boundary turns into a puncture, or marked point,
and we have a once-punctured square torus. 
This is the simplest surface in closed string theory for which the minimal area metric is unknown.
We study this torus by considering the torus
with a boundary for small values of~$h$.  In closed string theory
\emph{all} non-contractible closed curves must be longer than or equal
to~$\ell_s$.  This includes the closed curves that go around the boundary.  In the previous discussion of the surfaces we only
constrained the curves in two homologies.  
But even then,  the curves
going around the boundary are in fact longer than or equal
to $1$ as long as $h \geq h^{(2)} \simeq 0.24469$. For those
values of $h$, therefore, the extremal metric solves
 the minimal area problem that also
constrains curves homotopic to the boundary.
For $h < h^{(2)}$, however, the minimal area metric with two constrained homologies fails to satisfy the length condition for curves
homotopic to the boundary.   This condition must now be
imposed explicitly.  The closed 
curves going around
the boundary, however, are homologically trivial.  Indeed, any such  simple curve $\gamma$  divides the surface into two regions, one that includes the boundary and the rest $R$;  the curve $\gamma$  is the boundary of $R$, thus trivial in homology.   As explained in~\cite{headrick-zwiebach}
to impose constraints on non-contractible curves that are trivial
in homology, we need to pass to a suitable cover 
of the surface in which the curves in question
are no longer trivial.   For the square torus a four-fold cover is particularly convenient, as it allows us to preserve some of the symmetries that simplify the problem.  We can then write both
primal and dual programs on the covering surface and 
obtain minimal area metrics for the original surface.  

As soon as $h < h^{(2)}$ the minimal area metric displays three systolic bands:  the first two along the original two homology classes, and 
the third wrapping around the puncture.  In Figure~\ref{fig:punc-h1/16-fig} we show the systolic geodesics for 
a torus with boundary size $h= 1/16$.   Here are the main features of the metric:

 \begin{enumerate}

\item There are three bands of saturating geodesics:  a horizontal blue band, a vertical red band, and a black band surrounding the boundary.  The black band does not intersect any other band.
The blue and red 
vertical bands intersect each other in a simply
connected region, whose boundary is made of piecewise geodesics.

\item  The black band represents a flat, finite-height cylinder of 
circumference~$1$.  One of the boundaries of the cylinder is
the boundary of the surface. The other boundary is
the {\em last geodesic} in this homotopy class.  In the limit $h\to 0$, this last curve defines the local coordinate at the 
resulting puncture. 
This is the information required for the calculation  
of off-shell string amplitudes or the evaluation of the string field theory
master action.  

\item  The black band is surrounded by 
four triangle-shaped flat regions covered
by single bands of geodesics (red and blue, alternating as
we go around the black band).   We encounter no curvature along
the last black geodesic,  although the lines of curvature 
on the boundary of the two-band region, meet it tangentially
at four points. 

\item Over the two-band region the metric exhibits
both positive and negative Gaussian curvature.  The latter
appears near the midpoints of the edges of the figure. 
 The integral of Gaussian
curvature over the two band 
region is equal to $2\pi$, the same as that of a round hemisphere. 
  
\item The boundary of the two-band region has eight segments,
four red and four blue, each carrying an integrated line
curvature of $-\pi/2$.  Thus we get $-4\pi$ from line
curvature.  There are no point curvature 
singularities\footnote{
This is in contrast to the 
metrics arising from Jenkins-Strebel quadratic differentials, since those metrics are locally flat and do not have line curvature.}  and therefore
the total integrated curvature is $-2\pi$, consistent with an Euler number of $-1$.  

\item The metric is finite and non-vanishing everywhere on the surface.  This confirms the absence of point curvature.  One can also confirm that all curves in other homotopy classes are longer than or equal to $1$, making this a full solution of the closed string theory minimal area problem.

\end{enumerate}

The methods of the present paper can also be applied to the study
of conformal metrics of non-positive curvature on Riemann surfaces.  These metrics are also of interest for string field theory
and exhibit surprising features revealed by the numerical study:  not all of the Riemann surface is necessarily covered by saturating geodesics~\cite{headrick-zwiebach3}\footnote{We have learned of work by Katz and Sabourau~\cite{k-s},
dealing with systolic geometry of surfaces with Riemannian metrics of non-positive curvature. Regions without systolic geodesics play an important role
in this work.}.

This paper is organized as follows.   In section~\ref{sec:homology} 
we review the results of \cite{headrick-zwiebach} that are needed for most of this paper.  The review is brief and for more details the reader should go back to the original reference.  In 
section~\ref{sec:torusboundarysc} we discuss the minimal area problem on the Swiss cross with its length conditions on open
curves 
and the minimal area problem on the associated torus with 
constraints on two homology classes of closed curves. We show
that the solutions of these two problems are the same. 
In  section~\ref{sec:programs}  we discuss the primal and dual programs
on the Swiss cross, explaining in detail the role of symmetry in simplifying the problem and reducing it to a fundamental domain 
one-fourth of the size of the surface.  We explore the use of other
conformal frames to study the extremal metric, in particular 
a frame in which the fundamental domain is a pentagon and the
central corner on the cross is straightened out. 
Our results for the tori with constraints on two homology classes
are presented in section~\ref{sec:results}.  We give the extremal 
area, metric, and curvature of the solution for $h=1/2$.  We also discuss the moduli space of these surfaces, as determined by the
size $h$ of the boundary when the systole is assumed to be $1$.
The important case of the once-punctured torus is studied in
section~\ref{once-p-torusvm}.  We show in detail the construction of the relevant covering spaces, which are also tori 
with boundaries, and exhibit the corresponding flows (dual to the calibrations).  We build the requisite doubly periodic calibrations  in terms of the Weierstrass zeta function.  We include in this section the results for the extremal metrics.  We conclude with an appendix where we give details on the numerical discretization
and discuss the convergence of our results.

\section{Review: Convex programs for the minimal-area problem}
\label{sec:homology}

In this section we review and summarize the results
of our earlier paper~\cite{headrick-zwiebach}  that are relevant to our 
present discussion.  In that paper we considered a
minimal-area problem with constraints on homotopy classes of curves
and replaced it by one with constraints on homology classes, showing
how the former could be solved by 
an extension of the latter.  The minimal-area problems were formulated
as convex programs. 

Let $M$ be a Riemann surface equipped with
a fiducial conformal metric  
$g^0_{\mu\nu}$.  The general conformal metric 
on $M$ takes the form  
$g_{\mu\nu} = \Omega \, g^0_{\mu\nu}$, with $\Omega$
a positive function on $M$.
We also define the fiducial volume form $\omega_0  = d^2 x \sqrt{g_0}$  
and the volume form $\omega = d^2x \sqrt{g}$.
The area of $M$ is $\int_M \omega = \int_M \omega_0\Omega \,. $
Parameterizing a closed curve $\gamma$ using 
$x^\mu: [0,1]\to M$,  we have
$\text{length} (\gamma) = \int_\gamma  \sqrt{\Omega}  |\dot x |_{{}_0}$.

 Letting $\Gamma$ denote the set of all non-contractible closed 
 curves on $M$ and letting $\ell_s$ be the systole (the length of the shortest
 non-contractible closed curve), the closed string field theory minimal area program (MAP) is
\begin{equation}
\label{csft-MAP}
\begin{split}
 \text{\bf Closed string field theory MAP: }\ \ \  & \text{Minimize } \,  
\int_M\omega_0 \,\Omega \ \  
\ \ \text{over }\Omega\ge0 \text{ (function)}\\
& \hbox{subject to} \ \ \ \ 
\, \ell_s-\int_\gamma\sqrt{\Omega} \, |\dot x |_{{}_0} \le0\,, \ \ 
  \forall\,   \gamma \in \Gamma \,. \ \ 
  \end{split} 
\end{equation}

The homology version of this problem considers 
non-trivial homology 1-cycles  $C_\alpha \in H_1 (M, \mathbb{Z})$, 
with $\alpha \in J$
an index labeling the cycles.  We aim to minimize the area of $M$ subject to the constraint that all representatives of each cycle $C_\alpha$ have length at least $\ell_\alpha$. 
In analogy to the homotopy MAP above, we write  
\begin{equation}
\label{firstprogram}
\begin{split}
 \text{\bf Homology MAP: }\ \ \  & \text{Minimize } \,  
\int_M\omega_0 \,\Omega \ \   
\ \ \text{over }\Omega\ge0 \text{ (function)}\\
& \hbox{subject to} \ \ \ \ 
\, \ell_\alpha-\int_m\sqrt{\Omega} \, |\dot x |_{{}_0} \le0\,, \ \ 
\forall \,  m\in C_\alpha\,, \  \forall\,   \alpha \in J \,.
  \end{split} 
\end{equation}
In both of the above programs the length condition is applied to an infinite set of curves. 
 
 In the first reformulation, we replaced the
 length constraints---conditions for the metric that
 are nonlocal on the surface---by
 local constraints that define, for each $C_\alpha$, a calibration $u^\alpha$ with 
 period $\ell_\alpha$ in $C_\alpha$.  
A  1-calibration  
on a manifold is a closed one-form $u$ obeying $|u|\le1$. 
The new program, called the {\em primal} 
(since there will be a dual program below), 
is also convex and
takes the form 
\begin{equation}\label{secondprogram}
\begin{split}
\text{\bf Primal MAP v1:} \ \   
\ \ &\text{Minimize }\ \, \int_M \omega_0 \,\Omega \quad 
\text{ over 
$\Omega$ (function), $u^\alpha$ (one-forms)}  \ \  \\
&   \hbox{subject to}  \hskip25pt |u^\alpha|_0^2-\Omega\le0\,,\\
&  \hskip99pt
  du^\alpha=0\, , \  \\
& \hskip66pt  \ell_\alpha-\int_{m_\alpha}\hskip-8pt u^\alpha=0\,, 
\ \forall \alpha \in J  \,.  \phantom{{x}_A}  \  \\
\end{split}  
\end{equation}
 There are no conditions on the calibrations at the boundary of $M$, if it exists. 
Any {\em feasible} $u^\alpha,\Omega$ provides a rigorous upper bound on the value of the minimum.

We can  solve the period constraints on the calibrations $u^\alpha$ 
using a basis of real closed one-forms $\omega^i$ with periods 
$\omega^i_\alpha  \equiv  \int_{C_\alpha}\omega^i$.
We then write $u^\alpha$ as as a linear combination
of the basis one-forms plus a trivial one-form $d\phi^\alpha$
where $\phi^\alpha$ is a function on $M$:
\begin{equation}
\label{calibration-ansatz}
u^\alpha = \sum_i  c_i^\alpha \omega^i+d\phi^\alpha\,.  
\end{equation}
The constants $c_i^\alpha$  are constrained by the period
conditions $\int_{C_\alpha} u^\alpha = \ell_\alpha$:
\begin{equation}\label{cconstraint}
\sum_i c_i^\alpha\omega^i_\alpha =\ell_\alpha\,. \end{equation}
Having replaced the calibrations 
by the constants $c_i^\alpha$ and the functions $\phi^\alpha$,
where the zero mode of $\phi^\alpha$ drops out, and
the program reads:
\begin{equation}\label{thirdprogram}
\begin{split}
\text{\bf Primal MAP v2:} \  \ 
\ &\text{Minimize }\, \int_M \omega_0\,\Omega\, \quad  \ 
\text{over $c^\alpha_i$ (constants), $\Omega$, $\phi^\alpha$ (functions)} \\
& \, \hbox{subject to} \ \ \ 
\Bigl|\sum_ic_i^\alpha \omega^i+d\phi^\alpha\Bigr|_0^2-\Omega\le0\,,\\
& \hskip99pt   \ell_\alpha - \sum_ic^\alpha_i\omega^i_\alpha=0\,, \quad\forall \alpha \in J\,.
\end{split}  
\end{equation}

In~\cite{headrick-zwiebach} we gave several derivations of a \emph{maximization} convex program that is equivalent to the above primal program.
One of the derivations simply applies Lagrangian 
duality to the primal, hence we term it the \emph{dual} minimal-area program. The variables are, for each homology class $C_\alpha$, a  function $\varphi^\alpha$ on the surface and a constant $\nu^\alpha$. 
Choosing a set $m_\alpha$ of representatives in $C_\alpha$ the program
takes the form
\begin{equation}\label{thirddual}
\begin{split} 
&\hbox{\bf Dual MAP:} \\
 &\quad\text{Maximize}\ \  
  \, 2\sum_\alpha \nu^\alpha \ell_\alpha-\int_{M'}
\omega_0\Bigl(\sum_\alpha\left|d\varphi^\alpha\right|_0\Bigr)^2\   \ \     \hbox{over $\nu^\alpha$ (constants), $\varphi^\alpha$ (functions)}
\\
&\quad \hbox{subject to} \ \ \ \ 
 \Delta\varphi^\alpha|_{m_\alpha}=-\nu^\alpha\,, \\
 &  \hskip70pt  \varphi^\alpha|_{\partial M}=\ 0\, ,  \quad \forall \alpha\in J \,. 
\end{split} 
\end{equation}
As stated above, the function $\varphi^\alpha$ 
has discontinuity $-\nu^\alpha$ across $m_\alpha$. 
Moreover, $M'= M\setminus \cup_{\alpha\in J}  m_\alpha$ is the manifold $M$ with the chosen representative curves $m_\alpha$ removed.
 This is just a way to tell us that in
 calculating the objective with our 
 discontinuous $\varphi^\alpha$ there is no delta-function 
 contribution in $d\varphi^\alpha$ on $m_\alpha$.  
 In trying to maximize the objective the first term tries to make 
$\nu^\alpha$ large; however, a non-zero jump forces $\varphi^\alpha$ to have a non-zero gradient somewhere, making the second term, which is always negative,  larger.  The former is linear while the latter is quadratic, so we expect there to exist a maximum.
Any trial values for $\nu^\alpha,\varphi^\alpha$ give a rigorous lower bound on the maximum. Using both the primal~\eqref{thirdprogram} and the dual~\eqref{thirddual}, we can thus bound the solution both above and below. Strong duality holds for this problem: the maximum in the dual coincides with the minimum in the primal; this is the extremal area $A$. 
 Note also that we don't need to be careful about the orientation of $m_\alpha$ or the sign of the jump, since the objective is invariant under $\varphi^\alpha\to-\varphi^\alpha$.

A special case occurs when, for some $\alpha$, a subset $m'_\alpha$ of the boundary $\partial M$ is a representative of the class $C_\alpha$. 
In this case the representative $m_\alpha$ where $\varphi^\alpha$ jumps
can be chosen to be $m'_\alpha$.  
Since $\varphi^\alpha$ is supposed to vanish at the boundary, placing
the discontinuity  at $m'_\alpha$ effectively sets the value of $\varphi^\alpha$ 
on $m'_\alpha$  equal to $\nu^\alpha$.  
The value of $\varphi^\alpha$ on the rest
of the $\partial M$ remains zero:
\begin{equation}\label{boundaryrep}
\left.\varphi^\alpha\right|_{\partial M\setminus m'_\alpha}=0\,,\qquad
\left.\varphi^\alpha\right|_{m'_\alpha}=\nu^\alpha\,.
\end{equation}

A representative of $C_\alpha$ that has length $\ell_\alpha$ saturates the length condition and must
be locally length-minimizing, else there exist representatives that violate the length condition. We call  a $C_\alpha$ representative of length $\ell_\alpha$ an  $\alpha$-\emph{geodesic}.  The location of the $\alpha$-geodesics can be readily extracted from any solution to either the primal or the dual program. 
An $\alpha$-geodesic in the metric $\Omega g^0$
is calibrated  by $u^\alpha$:
 the constraint $|u^\alpha|_0^2\le\Omega$ is saturated on that curve and furthermore the vector $\hat u^\alpha$  associated to the form $u^\alpha$ is tangent to the curve.  
Thus the $\alpha$-geodesics are the closed integral curves of $u^\alpha$ on which $|u^\alpha|^2_0=\Omega$ everywhere. 
In a solution to the dual program, wherever $d\varphi^\alpha\neq0$, the curves of constant $\varphi^\alpha$ are $\alpha$-geodesics. This implies that $u^\alpha$ is related to $d\varphi^\alpha$ (where $d\varphi^\alpha\neq0$) by
\begin{equation}\label{primalfromdual}
u^\alpha = -\frac{*d\varphi^\alpha}{|d\varphi^\alpha|}
\end{equation}
(the sign is by convention).

One can also show that
the Weyl factor $\Omega$ of the metric is
\begin{equation}\label{solution22}
\Omega = \Bigl(\sum_\beta|d\varphi^\beta|_0\Bigr)^2\,,
\end{equation}
and since we write $\Omega = \rho^2$, we have $\rho = \sum_\beta|d\varphi^\beta|_0$. 
Moreover
\begin{equation}
\label{area-heights}
\sum_\alpha\nu^\alpha \ell_\alpha = A\,.
\end{equation}
In other words, the area for the extremal metric equals the total area of a set of 
 {\em flat rectangles} of height $\nu^\alpha$ and length $\ell_\alpha$.  Equation \eqref{solution22} is also a powerful constraint on the solution and can be rewritten
 as a sum rule
\begin{equation}\label{solution2} 
\sum_\alpha|d\varphi^\alpha| = 1\,, \quad \text{at every point on $M$}\,. 
\end{equation}

\begin{figure}[!ht]
\leavevmode
\begin{center}
\epsfysize=6.0cm
\epsfbox{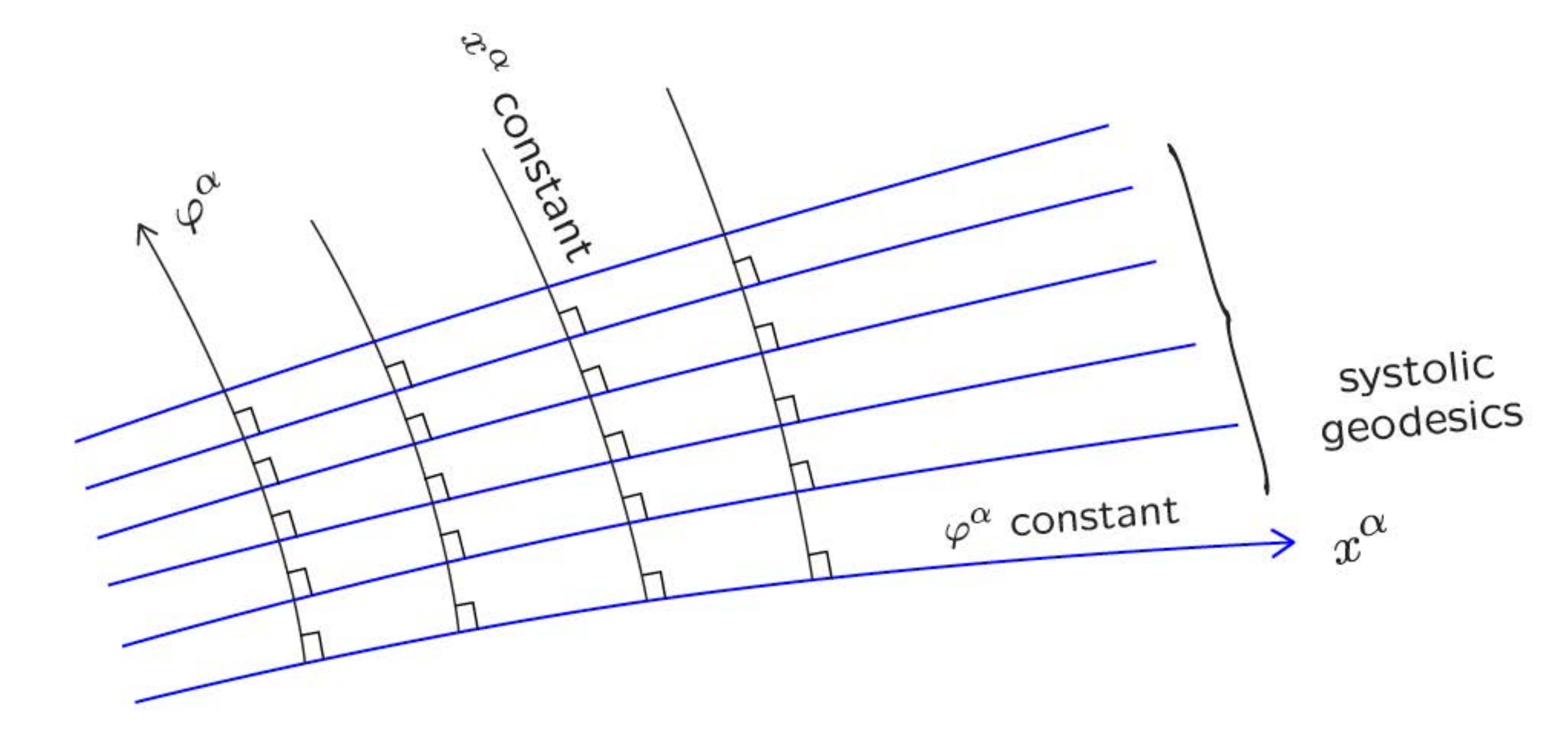}
\end{center}
\caption{\small  Normal coordinates associated with a collection
of homotopic systolic geodesics.  The coordinate $x^\alpha \in [0, \ell_\alpha]$ 
is a length parameter along the geodesics, which are 
curves of constant $\varphi^\alpha$.}
\label{ffsg1}
\end{figure}

As explained in~\cite{headrick-zwiebach} we can construct  Gaussian normal coordinates using a single band 
of systolic geodesics in the class $C_\alpha$.
We use a function $x^\alpha$ 
that parameterizes the $\alpha$-geodesics
by length as the first coordinate.
The calibration $u^\alpha$ can be identified locally
as $ u^\alpha   =  d x^\alpha$ 
Since the period of $u^\alpha$ is~$\ell_\alpha$, $x^\alpha\in[ 0, \ell_\alpha]$. 
The second coordinate is $\varphi^\alpha$
which, as required, is constant along the systolic geodesics (Figure~\ref{ffsg1}).  
 The cotangent  metric $g^{-1}$  is constrained
 by the conditions $| d x^\alpha| =1$ and 
$\langle d x^\alpha , d\varphi^\alpha \rangle = 0$:
\begin{equation}
\label{Gaussian}
g^{-1} \ = \ \begin{pmatrix} 1 & 0 \\ 0 &\  h_\alpha^2 (x,\varphi)  \end{pmatrix}\, , \quad   |d\varphi^\alpha| =  |h_\alpha|  \quad \to \quad ds^2 = (dx^\alpha)^2+\frac1{h_\alpha^2} (d\varphi^\alpha)^2\,.
\end{equation}  
This is the metric in Gaussian normal coordinates. As a consequence of the sum rule (\ref{solution2}),  which now reads $\sum_{\alpha:\, d\varphi^\alpha\neq0} | h_\alpha|  = 1$,  in a region where there is a single
set of systolic geodesics $|h|=1$ and the metric must be flat.   If there is more than one set of systolic geodesics then the metric can have curvature.
It is a simple consequence of the Gauss-Bonnet theorem that over a band
of closed systolic geodesics the integral of the Gaussian curvature must give zero.  This follows because the band is topologically
an annulus of zero Euler number and the boundaries of the annulus are geodesics. 

On a band $R_\alpha$ of geodesics in the class $C_\alpha$ we
can use the metric (\ref{Gaussian}) to define the height $b^\alpha(x)$ 
of the band as a function of the position $x^\alpha$ along the band:
\be
b^\alpha (x)  \equiv \int_0^{\nu^\alpha}  {d\varphi^\alpha \over |h_\alpha (x , \varphi) |}\,. 
\ee
Assume now there is a value $x_0^\alpha$ of $x^\alpha$ for 
which the integral is done over a region
where $R_\alpha$
is the {\em only} systolic band. Then $|h_\alpha | = 1$ along this segment~and 
\be
\label{identify-nu-height}
b_\alpha (x_0^\alpha ) =  \nu^\alpha \,.
\ee
This gives simple characterization of $\nu^\alpha$: it is the height
of the band in a region of the surface where it is the only band.

\section{The torus with a boundary and the associated Swiss cross}
\label{sec:torusboundarysc}

In here we show how a Swiss cross defines uniquely a torus with
a boundary, and vice versa.  We then formulate a minimal area problem
on the Swiss cross and a minimal area problem on the torus with
a boundary.  We show that the extremal metric is the same for both
problems.

\subsection{Constructing the surfaces}
\label{sec:constrthesurf}

\begin{figure}[!ht]
\leavevmode
\begin{center}
\epsfysize=11.5cm
\epsfbox{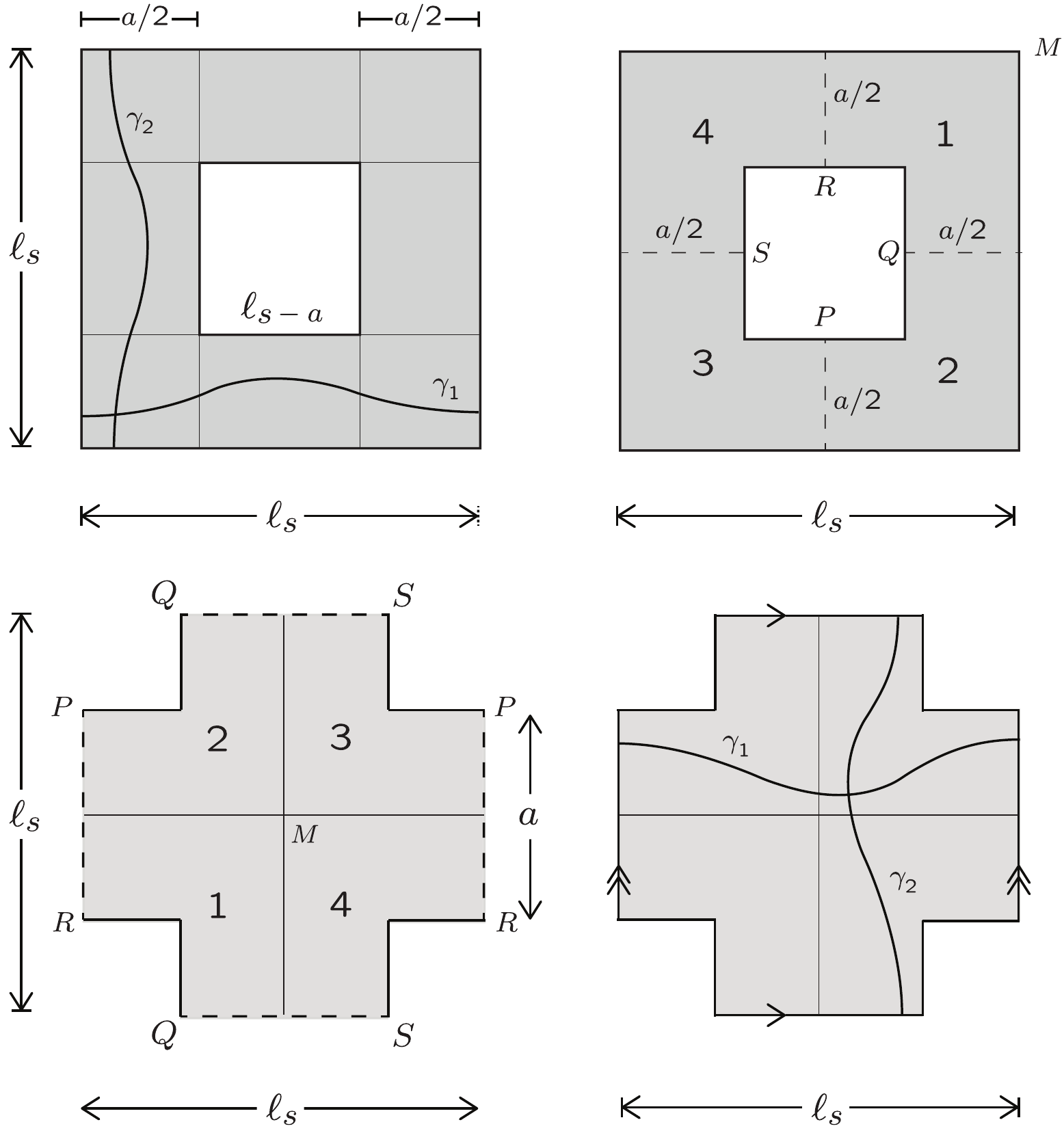}
\end{center}
\caption{\small  Upper-left:  A square torus with a square boundary.  The identified edges have length $\ell_s$ and the edge of the boundary has size $\ell_s -a$.
Upper right:  The same torus with four regions 1,2,3, and 4 defined.  
Lower left:  The four regions rearranged  into the shape of a
 Swiss cross.  The dashed
edges must be identified to form the torus. 
Lower right: The Swiss cross with the identifications of edges  indicated by arrows.  Two curves $\gamma_1$ and 
$\gamma_2$ representing two homology classes in the torus are shown
in the upper left and lower right pictures.}
\label{ff1}
\end{figure}

Consider first a square torus with a boundary.  The square
torus, a torus with modular parameter $\tau = i$, is defined by a square 
region in the complex plane with opposite 
edges identified.  We will take the length
of the edges to be $\ell_s >0$, a fixed quantity.   
The torus also has a boundary, 
obtained by removing a square region with side 
length $\ell_s-a$, where $0 < a < \ell_s$.   The torus square region
and the boundary square are aligned and have a common center.
The result is shown in the upper left of Figure \ref{ff1}.    We also show
in this figure two closed curves, $\gamma_1$ and $\gamma_2$ that 
are representatives of the homology classes $C_1$ and $C_2$,
respectively.

With  $\ell_s$  fixed one expects  different values of $a$ to represent conformally inequivalent tori with boundary.   
When $a \to \ell_s$ we have a square torus with 
a boundary that is becoming so small that is turning into a puncture.
When $a \to 0$  the boundary is becoming as big as possible-- in string field theory this would correspond to an open string diagram with very
long open-string propagators. 

We now show an alternative presentation of the surface.
For this purpose we cut the surface into four regions denoted as
$1,2,3,$ and $4$, as shown in the upper right of Figure \ref{ff1}.  The four
regions are glued back as indicated on the lower left, and
we obtain a ``Swiss cross'' presentation 
where horizontal and vertical dashed edges
must be identified to form a torus.   The Swiss-cross presentation
is shown again in the
bottom right, with identifications indicated by similar arrows, and 
with the closed curves $\gamma_1$ and $\gamma_2$ displayed.
We call this a {\em toroidal Swiss cross} and it is exactly the same
Riemann surface as the torus with a boundary, only the presentation is different. 

We are interested in the metric of minimal area on this surface with
the condition that all curves in the homology classes $C_1$ and $C_2$,
with representatives $\gamma_1$ and $\gamma_2$, respectively,  
have length greater than or equal to $\ell_s$.  With this choice of 
systole we see that the constant metric $\rho=1$ is an admissible
metric.   We will call this the {\em torus problem.}  One can discuss this
problem in the original presentation or 
in the presentation as a ``toroidal Swiss cross'',
as the surfaces are the same.  The minimal
area metric should be invariant under all the discrete transformations
that leave the torus invariant.  This follows from 
strict convexity of the area functional:  if the 
extremal metric and its transformed version are different, the 
average metric would have lower area while satisfying all the constraints.
The general version of this argument was given in~\cite{headrick-zwiebach}, 
at the end of section~3.1.

We also want to introduce the {\em conventional} Swiss cross, or just
Swiss cross.  This is a planar surface, with shape as shown at
bottom right of Figure~\ref{ff1}.  In the Swiss cross there 
are no identifications and the figure is 
conformally equivalent to the unit disk, with the twelve turning points
of the cross resulting in special points on the boundary of the disk.  We will call
``vertical edges''  and ``horizontal edges'' the segments that were
identified in the toroidal Swiss cross. 

It is clear from the construction in Figure \ref{ff1} that given
a square torus with a centered square boundary we can construct the associated
Swiss cross.  Moreover, given a Swiss cross, by identification of the two vertical edges and identification of the two horizontal
edges we obtain a square torus with a centered square boundary.

\subsection{The minimal-area problems on the Swiss cross
and the torus}

We have already stated the {\em torus problem}, namely, the
minimal-area problem on the
torus with a boundary, or equivalently, on the toroidal Swiss cross.
Since it has no identifications, on a  Swiss cross
there are actually two minimal-area problems that could be
posed.  Each uses a different set of curves to be constrained:
\begin{enumerate}
\item[1.]   Any curve beginning on the left vertical edge and ending on the right
vertical edge, and  any curve beginning on the bottom horizontal
 edge and ending on the top horizontal edge.
  \item[2.]  The curves above with the additional condition that
 beginning and ending points would be identified 
in the torus version of the Swiss cross.
\end{enumerate}
The set of curves (2) is contained in the set of curves (1).
The curves in (2) are the curves that would be closed if we made
our Swiss cross toroidal.  

We define the {\em Swiss cross problem} as the minimal
area problem that uses (1):  the constraints apply to all curves from
left to right and all curves from bottom to top.

\begin{figure}[!ht]
\leavevmode
\begin{center}
\epsfysize=5.5cm
\epsfbox{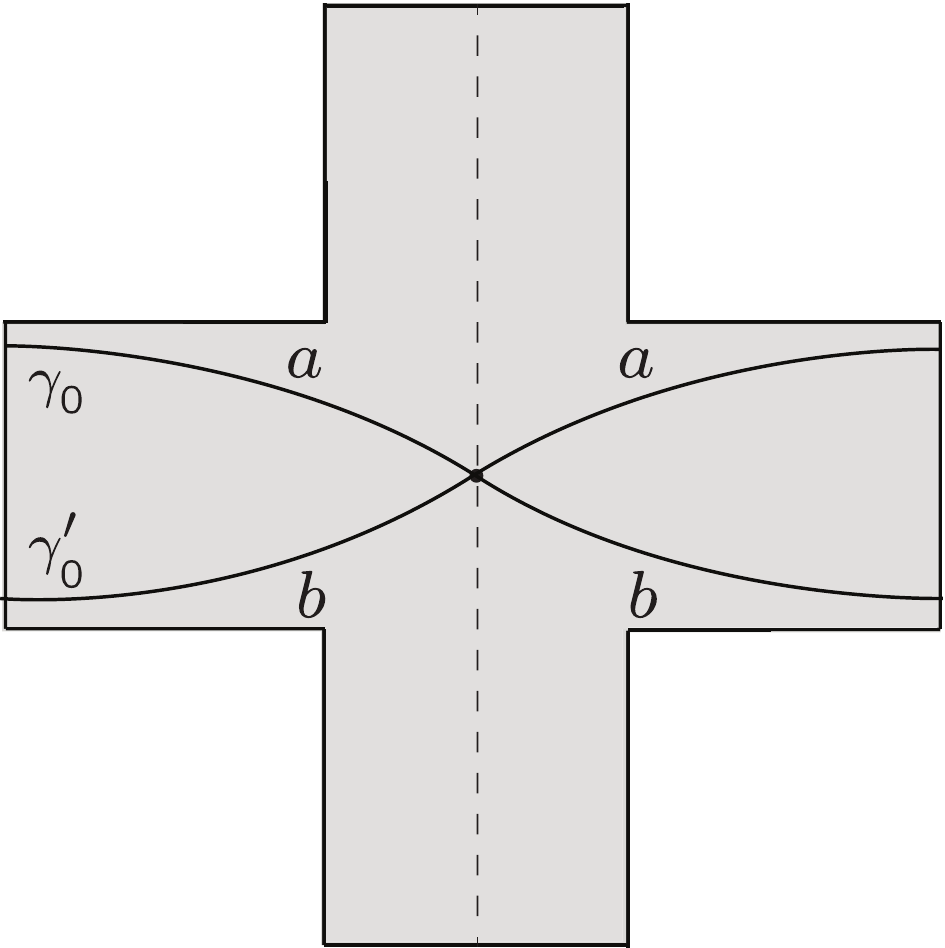} 
\end{center}
\caption{\small  The curves $\gamma_0$ and $\gamma'_0$ are
related by a reflection about the vertical axis of the Swiss cross.}
\label{ff12cx}
\end{figure}

We now claim that:  {\em The extremal metric for (2) is extremal for (1).}  Namely, constraining the would-be closed curves is enough to make sure all curves are good. 
To see this, work by contradiction.  Suppose we have an extremal metric for (2) in which there is a curve $\gamma_0$ in (1) that is too short.  
Let $\gamma_0$ be a curve joining the vertical edges, as shown in 
Figure~\ref{ff12cx}. Reflection
about the vertical axis going through the center of the cross is an isometry of the extremal metric.  
Therefore, this reflection gives us another curve
$\gamma'_0$ of the same length as $\gamma_0$ and intersecting
$\gamma_0$ on the  axis.  This cuts $\gamma_0$ into two
pieces of lengths $a$ and $b$ with $a+b <  \ell_s$, because $\gamma_0$ is too short.   The length
of the pieces of $\gamma'_0$ are also $a$ and $b$.  We can then
use a piece of each curve to  
 form two curves of type (2), one of length $2a$ and one of length
$2b$.  Whether $a<b$ or $a>b$, or $a=b$  at least one of the two
curves is shorter than $\ell_s$.
This is a contradiction, showing that
the extremal metric for (2) satisfies all the constraints (1) and 
is thus extremal for (1). \hfill$\square$
 
Consider now a square torus with a boundary and its 
associated Swiss cross.  We will show below that the extremal
metric for the Swiss cross problem provides an extremal metric
for the torus problem. 
 
\medskip
\noindent
{\em Claim:  
 The extremal metric for the Swiss cross problem
 is also extremal for the torus problem.}  
 
 \noindent
{\em Proof}:  Consider the Swiss cross with its extremal metric
and copy the metric onto the torus with a boundary.  Let the area
of (either) surface with this metric be $A_0$. 
We  first show that the resulting 
metric on the torus is admissible
for the torus problem.  
\begin{figure}[!ht]
\leavevmode
\begin{center}
\epsfysize=8.5cm
\epsfbox{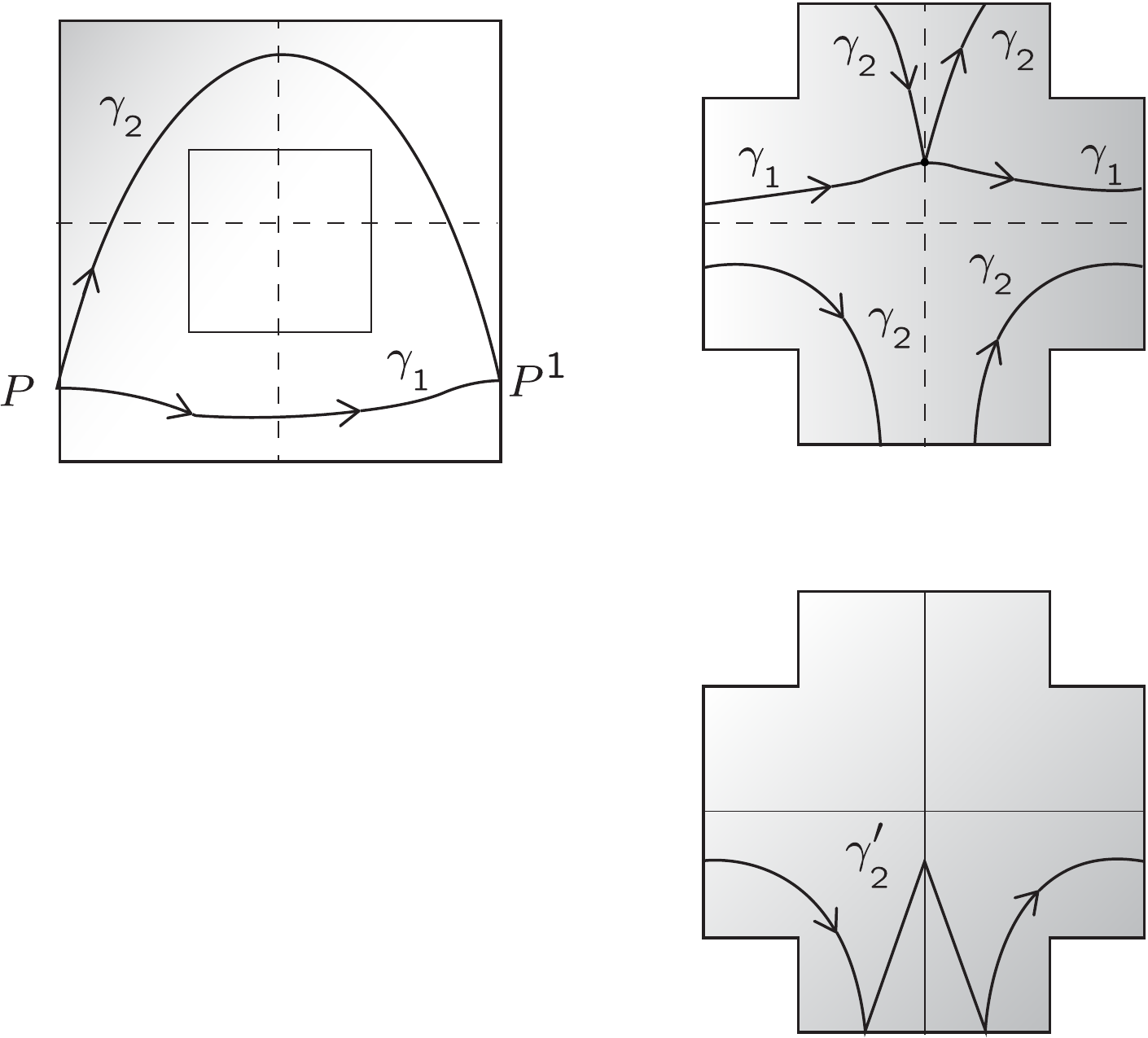} 
\end{center}
\caption{\small  The curves $\gamma_1$ and $\gamma_2$ on the torus are shown on the Swiss cross.  $\gamma_1$ is a standard curve on the Swiss cross but $\gamma_2$ is not continuous.
The curve $\gamma'_2$ of length equal to that of $\gamma_2$
is continuous and standard.}
\label{ffaa1}
\end{figure}

Consider the curves on the torus
in the class $C_1$.  As they are drawn on the Swiss cross
they take two possible forms, as illustrated in Figure~\ref{ffaa1}.
One type of curve, exemplified by $\gamma_1$ is simply
a curve going from the left edge to the right edge, and is
certainly long enough.  The second type of curve is not a continuous curve on the Swiss cross and is exemplified
by $\gamma_2$, which is homologous to $\gamma_1$.  This curve, while going from left to right on the cross, would be continuous
if we used the identifications of the top and bottom edges, as shown
in the figure to the right.  Because
of the isometry, however, such a curve has the same length
as the curve $\gamma'_2$ shown below, where we moved
a piece of the curve to make it continuous on the cross.
As shown, $\gamma'_2$  is a standard curve going
from left to right and must be long enough, showing that 
$\gamma_2$ is long enough.  Any curve of the second type can
be shown (using the isometries) to have the same length as a
curve that goes from left to right. Since both types of curves work
and exactly the same argument
works for the curves in the class $C_2$, the metric induced 
on the torus is indeed admissible.

Now assume the extremal metric on the torus has area $A_1 < A_0$.  Being extremal, the metric must have the isometries
of the torus.   That metric copied into the Swiss cross 
would also have area $A_1 < A_0$ and 
the isometries of the Swiss cross.
Moreover, since this metric is admissible for the torus
the would-be closed curves of the Swiss cross would be good.
Having the isometries, this metric would solve also the 
Swiss cross problem, in contradiction with our 
original statement that the minimal-area metric on the
Swiss cross had area $A_0$.  \hfill $\square$

\section{The programs for the torus with a boundary} 
\label{sec:programs}

We  formulate the primal program
for the torus with a boundary.   We use symmetry arguments to simplify the ansatz for the calibrations and reduce the  work to a fundamental domain one-fourth
the size of the original surface.  We then do a similar analysis for the dual
program, taking our time to obtain a lower bound for the extremal
area using a simple ansatz for the relevant $\varphi$ functions.
We conclude by finding conformal maps of the fundamental domain
 in which the corner point
is mapped to a regular
point on the boundary (the pentagon frame) or is 
mapped to infinity (the strip frame).    

\subsection{The primal program}
\label{sec:primalprogram}

\begin{figure}[!ht]
\leavevmode
\begin{center}
\epsfysize=7.5cm
\epsfbox{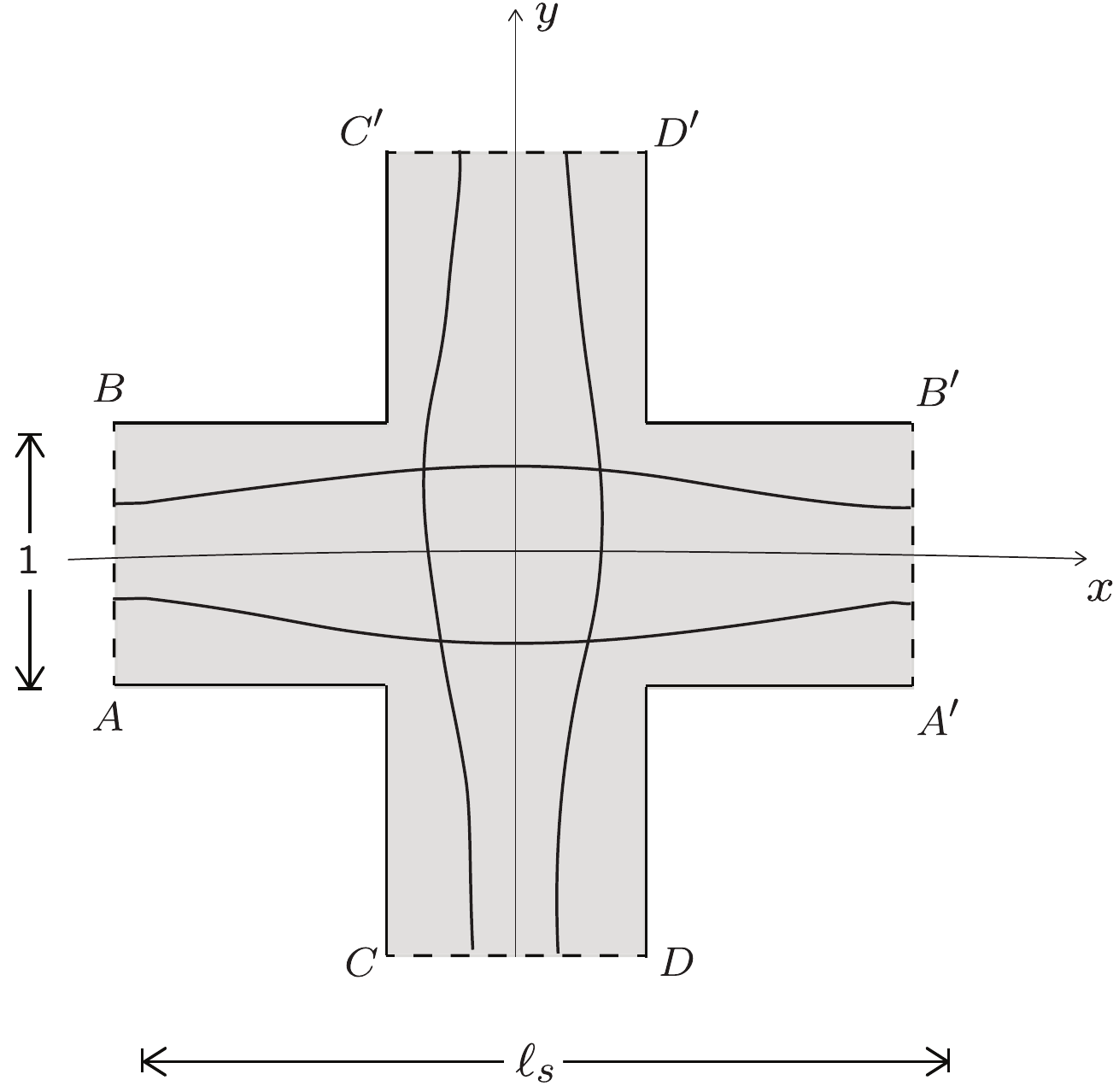}
\end{center}
\caption{\small  Swiss cross presentation of the torus with
a boundary. The vertical edges $AB$ and $A'B'$ are identified.
The horizontal edges $CD$ and $C'D'$ are also identified.
Reflections about the $x$ and $y$ axes map saturating
geodesics to themselves or to other saturating geodesics.}
\label{ff187}
\end{figure}

We now set up the program for the torus with a boundary,
presented as a Swiss cross with identified edges, as shown in Figure~\ref{ff187}.
Let us work in the convention where $a=1$ and the horizontal (and vertical) 
parameter length $\ell_s$ is also the systole.
We must deal with two homology classes: the class $C_1$ corresponding
to curves traveling horizontally and the class $C_2$ corresponding to curves
traveling vertically.   Moreover in this surface we have 
a basis of real closed one-forms spanned by $dx$ and $dy$.  We consider
now  two calibrations, that is two closed one-forms $u^1$ and $u^2$,
corresponding to the classes $C_1$ and $C_2$ respectively.

Following (\ref{calibration-ansatz}) we write the following ansatz
for the calibrations
\begin{equation}
\begin{split}
u^1 \ = \ & \  c_1 dx  + c_2 dy + d\phi^1 \,, \\
u^2 \ = \ & \  c'_1 dx  + c'_2 dy  +  d\phi^2 \,, 
\end{split}
\end{equation}
where $\phi^1$ and $\phi^2$ are functions on the torus that we 
have to determine and  $c_1, c_2, c'_1,c'_2$ are constants also
to be determined.   The exact one-forms $d\phi^1$ and $d\phi^2$
do not contribute to  integrals over closed cycles and 
therefore the period conditions 
\begin{equation}
\int_{C_1} u^1 \ = \ \ell_s \,,  \quad \hbox{and} \quad
\int_{C_2} u^u \ = \ \ell_s \,,
\end{equation} 
determine $c_1 =1$ and $c'_2 =1$ so that 
\begin{equation}
\begin{split}
u^1 \ = \ & \ \phantom{c_1} dx  + c_2 dy + d\phi^1 \,, \\
u^2 \ = \ & \  c'_1 dx  + \phantom{c'_2} dy  +  d\phi^2 \,.  
\end{split}
\end{equation}
We now use the symmetries of the surface to constrain
further the undetermined coefficients and the form 
of the functions $\phi^1$ and $\phi^2$.
Locating the center of the toroidal Swiss cross at the origin of the
$(x,y)$ plane we have a $\mathbb{Z}_2 \times \mathbb{Z}_2$
symmetry generated by the transformations
\be
\mathbb{Z}_2 \times \mathbb{Z}_2: \qquad 
(x, y) \to (-x , y)  \quad \hbox{and} \quad  (x, y) \to (x, -y) 
\ee
The first $\mathbb{Z}_2$ flips the sign of $x$ and the second
$\mathbb{Z}_2$ flips the sign of $y$. 
The extremal metric must be invariant under these symmetries
and therefore the calibrations, which constrain the metric via
$\Omega \geq |u^\alpha|_0^2$,  will transform into themselves
up to signs. For an arbitrary object $\chi$ we write
\be
\chi \sim (\epsilon_1 , \epsilon_2)\,, 
\ee
with $\epsilon_1, \epsilon_2$ equal to plus or minus one, 
if $\chi$ transforms as $\chi \to \epsilon_1 \chi$ under
the first $\mathbb{Z}_2$ and as $\chi\to \epsilon_2 \chi$ under
the second $\mathbb{Z}_2$.  Since $u^1$ contains  
$dx$ and $u^2$ contains $dy$ we must have
\be
\label{calib-trans}
u^1 \sim ( -\, ,\, +) \,, \qquad  u^2 \sim ( + \, , \, - ) \,. 
\ee
It now follows from this condition that $c_2 = c'_1 =0$ and therefore
\begin{equation}
\begin{split}
u^1 \ = \ & \  dx   + d\phi^1 \,, \\
u^2 \ = \ & \  dy  +  d\phi^2 \,.  
\end{split}
\end{equation}
The symmetries constrain the derivatives
of the functions $\phi^1$ and $\phi^2$.  Writing
\begin{equation}
\begin{split}
u^1 \ = \ & \  dx   + \partial_x \phi^1\, dx  + \partial_y \phi^1 dy \,, \\
u^2 \ = \ & \  dy  +  \partial_x \phi^2\, dx  + \partial_y \phi^2 dy \,, 
\end{split}
\end{equation}
the transformations (\ref{calib-trans}) now imply that 
\be
\label{calib-der-trans}
\begin{split}
\partial_x \phi^1 \sim ( +\, ,\, +) \,, \qquad  &\partial_y \phi^1 
\sim ( - \, , \, - ) \,, \\
\partial_x \phi^2 \sim ( -\, ,\, -) \,, \qquad  &\partial_y \phi^2 \sim ( + \, , \, + ) \,. 
\end{split}
\ee
We will see shortly that such results are consistent with our expectations.
There are additional symmetries of the surface, generated by a 
counterclockwise rotation $R$ of $90^\circ$.   This acts as
$R: (x,y) \to (-y,x)$.  Composing this with the first $\mathbb{Z}_2$
we have a symmetry 
\be
R'= \mathbb{Z}_2 \circ R : (x,y) \to (y,x) \,.
\ee
Under this symmetry we have that $R'$ exchanges the calibrations:
\be
R': \ u^1 \to u^2 \,,   \ \ \ u^2 \to u^1 \,. 
\ee
One quickly verifies that this implies that 
\be
\label{exch-deriv-phi}
\partial_x \phi^2(x,y) =  \partial_y \phi^1 (y,x)  \quad \hbox{and} \quad 
\partial_y \phi^2(x,y) =  \partial_x \phi^1 (y,x)  \,. 
\ee
A short calculation using these relations and integrating the
functions starting from the origin shows that 
\be
\label{eroihue}
\phi^2(x,y) - \phi^2 (0,0)  \ = \ \phi^1 (y,x) - \phi^1(0,0) \,. 
\ee
This means that, up to a constant, $\phi^2$ is determined by $\phi^1$.
This is reasonable; one could hope for the simpler $R' : \phi^1 \to \phi^2$ but the constant ambiguity is present 
because both $\phi^1$ and
$\phi^2$ appear acted by exterior derivatives.  

\begin{figure}[!ht]
\leavevmode
\begin{center}
\epsfysize=6.5cm
\epsfbox{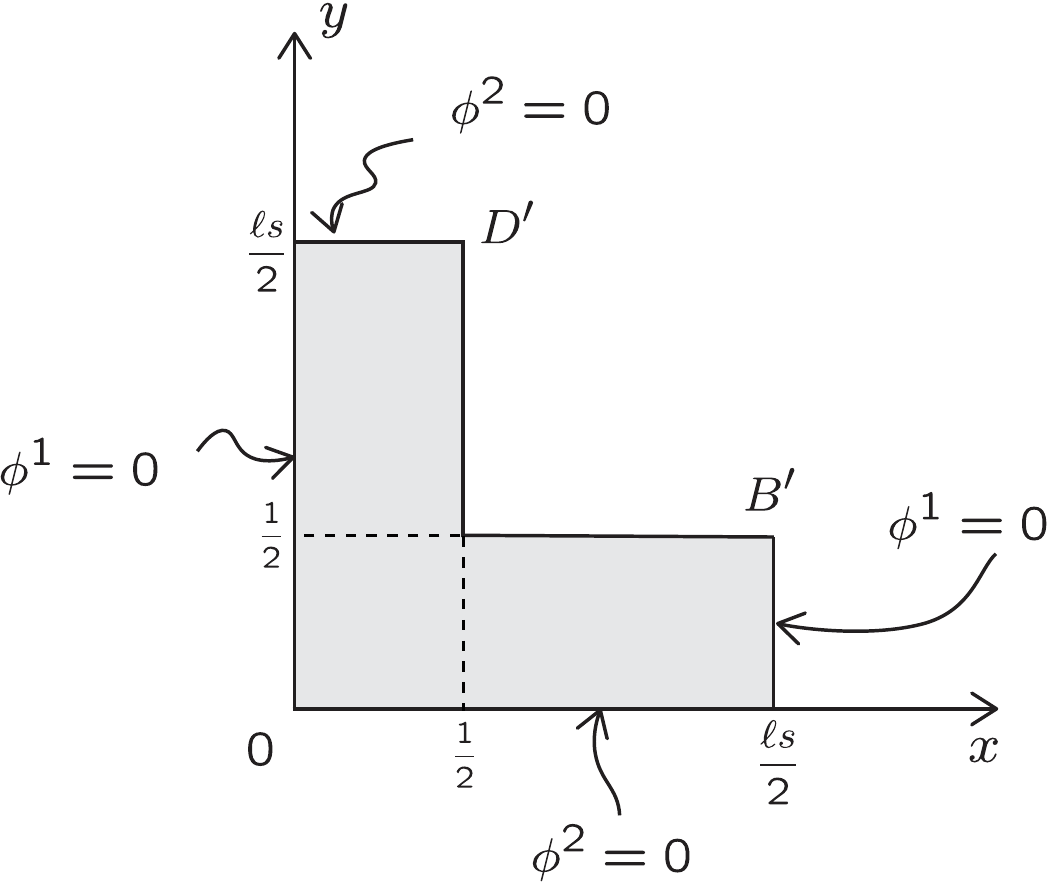}
\end{center}
\caption{\small  A fundamental domain for the functions $\phi^1$ 
and $\phi^2$.}
\label{ff876}
\end{figure}

We can learn a bit more about the functions $\phi^1$ and $\phi^2$
from the symmetry conditions.  Consider first the vertical edges. 
Since $\partial_y\phi^1 \sim (-, -)$,   
it is odd under $(x,y)\to (-x,y)$ relating two points on the vertical edge that are
identified.  Thus $\partial_y\phi^1=0$ on the vertical edges
 and $\phi^1$ is constant along the
vertical edges.  Since the value of $\phi^1$ over the surface can be changed
by an overall constant without any effect, we can choose to make $\phi^1$ vanish at the vertical edges.  An identical argument holds for $\phi^2$ and
we thus conclude that
\begin{equation}
\phi^1 = 0 \, \ \hbox{at the vertical edges} \,,  \qquad 
\phi^2 = 0 \, \ \hbox{at the horizontal edges} \,. 
\end{equation}
The invariance of $\partial_x \phi^1\sim (+,+)$   
under $x$ flips   
together with the value
$\phi^1=0$ at the vertical edges implies that $\phi^1$ is in fact odd
under $x$ flips  (think of integrating $\partial_x \phi^1$ 
along $x$ from the left edge up to a point and compare with the
integral from the $x$-flipped point up to the right edge).  In particular
the value of $\phi^1$ on the $y$ axis is zero.  Pick now two points on the
$y$ axis 
related by reflection about the $x$ axis. Since $\partial_x \phi^1$ is even
under $y$ flips, integrating horizontally $\partial_x \phi^1$
from these two points shows that $\phi^1$ is even under $y$ flips.
All in all, and including the analogous results for $\phi^2$, we have 
\begin{equation}
\label{zeroes-of-trivial-forms}
\begin{split}
\phi^1  & \ \sim (- \,,\, + )  
\hbox{  and is zero on  vertical edges \hskip1pt and the $y$ axis,} \\
\phi^2  & \ \sim (+ \,,\, - )      
\hbox{  and is zero
on horizontal edges and the $x$ axis}. 
\end{split}
\end{equation}
This means that we can use as a fundamental domain 
the part of the surface lying in the first quadrant $x, y \geq 0$.  If 
$\phi^1$ and $\phi^2$ are known there, they can be extended 
accordingly to the full surface.  The region is shown in Figure~\ref{ff876}.

Finally, noting that (\ref{zeroes-of-trivial-forms}) implies that 
$\phi^1(0,0) = \phi^2 (0,0) =0$
we now have that (\ref{eroihue}) gives us the simple relation
\begin{equation}
\label{one-two-idem}
\phi^2 (x, y) \ = \ \phi^1 (y, x) \,.
\end{equation}
If we know $\phi^1$ on the fundamental domain,
we know $\phi^2$ as well.  Effectively we just have one unknown and it can be taken to be $\phi^1$.  

The fiducial metric $g^0_{\mu\nu}$ on the surface is taken to be the constant unit metric: 
\begin{equation}
 g^0_{\mu\nu} dx^\mu dx^\nu = dx^2+ dy^2\,. 
\end{equation}
The program, using (\ref{thirdprogram})  then becomes
\begin{equation}\label{thirdprogram-vm99}
\begin{split}
&\text{Minimize }\, \int_M dx dy \,\Omega\, \quad \hbox{over}  
\ \Omega,\phi^1 , \phi^2 \\
& \hbox{subject to:} \ \ \ 
 \bigl|u^1|_0^2\,   -\Omega\le0, \\
& \hskip60pt   \bigl|u^2|_0^2\,   -\Omega\le0\,.
\end{split}
\end{equation}
Here the fiducial norms are easily calculated:
\begin{equation}
\begin{split}
\bigl|u^1|_0^2\ = \ \bigl|dx+d\phi^1\bigr|_0^2 \ = \ & \ \bigl( 1 + \partial_x \phi^1  \bigr)^2 + \bigl( \partial_y \phi^1  \bigr)^2 \,, \\
\bigl|u^2|_0^2\ = \ \bigl|dy+d\phi^2\bigr|_0^2 \ = \ & \  \bigl( \partial_x \phi^2 \bigr)^2  + \bigl( 1 + \partial_y \phi^2 \bigr)^2 \,. 
\end{split}
\end{equation}
At this point the program is explicitly defined.  Any trial functions
$\phi^1$ and $\phi^2$ on the torus will give an {\em upper bound}
on the minimal area.  Indeed, for any choice of $\phi^1$ and $\phi^2$
we compute at every point on the surface the two norms $\bigl|u^1|_0^2$ 
and $\bigl|u^2|_0^2$  and set
$\Omega$ equal to largest one.  We integrate $\Omega$ over the surface
and that gives the bound.  In searching for the extremal metric, however,
we can use the symmetry constraints discussed above
and restrict ourselves to the fundamental domain.

We can give some intuition for the 
constraints we derived above. 
Consider again the Swiss cross in Figure \ref{ff187}.  
The $\mathbb{Z}_2 \times \mathbb{Z}_2$ invariance
of the  
extremal metric implies that systolic geodesics in $C_1$ 
are invariant under reflections about the $y$ axis and systolic
geodesics in $C_2$ are
invariant under  reflections about the $x$ axis.   
Additionally, reflections about the $x$ axis  map
systolic geodesics in $C_1$ to systolic geodesics in $C_1$ and reflections 
about the $y$ axis map systolic geodesics in $C_2$ to systolic geodesics in $C_2$.  The systolic geodesics must hit orthogonally 
the edges that are to be identified, otherwise the reflection invariance would imply a kink on the geodesic at that point. 

As stated in section~\ref{sec:homology}, on a systolic geodesic in the class $C$, with calibration $u$,  the geodesic  is tangent to the vector $\hat u$ obtained from the one-form $u$ using the metric.  The components of the vector are then
\begin{equation}
\hat u^x =  g^{xx} u_x  = \frac{u_x}\Omega \,, \quad \hat u^y = g^{yy} u_y =  \frac{u_y}\Omega\,, 
\end{equation}  
meaning that the tangent direction is simply defined by $(u_x, u_y)$. 
For the saturating geodesics in $C_1$ the tangent direction is
\begin{equation}
\hat u^1  \sim  \bigl( \,  1 + \partial_x \phi^1 \,,   \, \partial_y \phi^1 \bigr) \,.
\end{equation}
At the vertical edges the previously derived condition
$\partial_y \phi^1=0$ means the 
systolic geodesics hit the vertical edges orthogonally, as expected.   
Looking at the tangents to saturating geodesics along $C_1$ is is clear that
$\partial_x \phi^1$ is even under reflections about the $x$ axis and the $y$ axis, while $\partial_y \phi^1$  is odd under reflections about the $x$ axis and the $y$ axis.  This is exactly as anticipated on the first line of~(\ref{calib-der-trans}).

We have discussed the ansatz for the calibrations and how symmetries
simplify the construction of the program.  A numerical solution, however,
requires discretization.  For the benefit of interested readers, we present
those details in appendix A. 

\subsection{The dual program}

\begin{figure}[!ht]
\leavevmode
\begin{center}
\epsfysize=9.0cm
\epsfbox{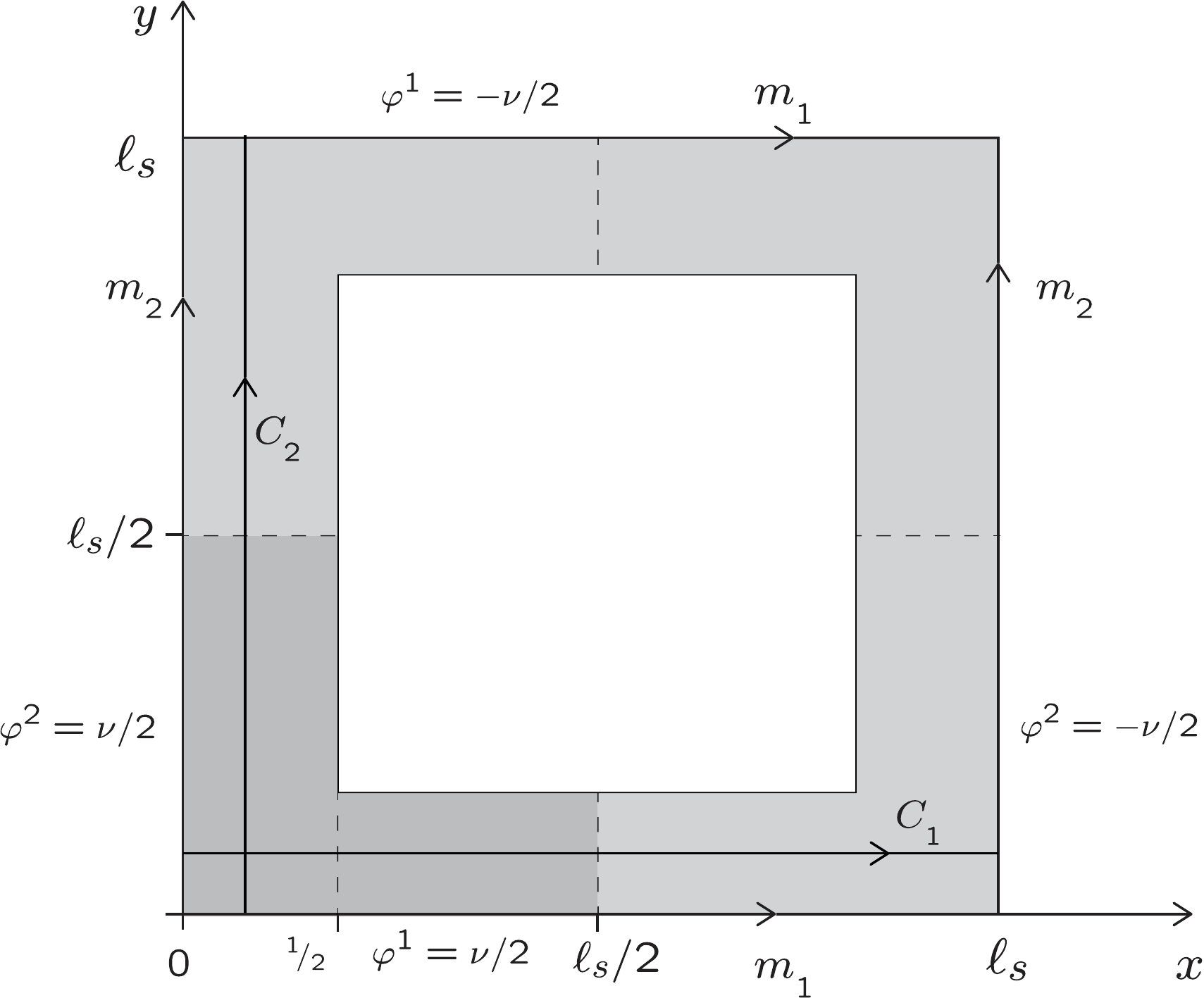}
\end{center}
\caption{\small  The torus with a boundary with the elements
needed for the dual program. Shown are curves in the homology
classes $C_1$ and $C_2$.  The function $\varphi^1$ has a discontinuity
across $m_1$ and the function $\varphi^2$ has a discontinuity across
$m_2$.}
\label{vmIL96}
\end{figure}

We now consider the use of our dual program to study the
torus with a boundary.  In this section, as in the previous one,
we will set up the program for the case of a torus with systole 
$\ell_s$ and $a=1$.  
This corresponds to the Swiss
cross of area $2\ell_s -1$ for the fiducial metric $ds^2 = dx^2 + dy^2$.
The torus, shown in Figure~\ref{vmIL96} as a region on the $(x,y)$ plane, displays 
a horizontal  curve in the homology
class $C_1$ and a vertical curve in the homology class $C_2$.  
We are constraining the length of closed
curves in these homology classes.

Having two homology classes the 
program requires two functions $\varphi^1$ and $\varphi^2$
defined all over the torus.  Following the conditions in
the dual program (\ref{thirddual})  these functions $\varphi^\alpha$ $(\alpha =1,2)$ must have discontinuities $\nu^\alpha$ across some
arbitrarily selected curves $m_\alpha \in C_\alpha$.   
For $\varphi^1$ we choose
$m_1$ to be the horizontal segment $x\in [0,\ell_s], y=0$ which is
also the segment $x\in [0,\ell_s], y=\ell_s$.   For $\varphi^2$ we choose
$m_2$ to be the vertical segment $x=0, y\in [0,\ell_s] $ which is
also the segment $x=\ell_s, y\in [0,\ell_s]$.  

Due to the symmetry of the square torus and  since the two functions are on the same footing, their discontinuities must be exactly the same:
\be
\nu^1 = \nu^2 = \nu \,. 
\ee
 We can set up the discontinuous functions symmetrically:  as $y\to 0^+$
 on the horizontal line $m_1$ we take $\varphi^1 \to \nu/2$ while
 as $y\to \ell_s^-$ on the copy of $m_1$ we have $\varphi^1 \to -\nu/2$.
 Doing similarly for $\varphi^2$ we have:
 \be
 \label{phi1phi2dc} 
 \begin{split}
 \varphi^1 ( x \in [0,\ell_s], y=0) \ = \ & \ \ \   \nu/2\,,  \\
 \varphi^1 ( x \in [0,\ell_s], y=\ell_s) \ = \ & - \nu/2\,,\\[1.0ex]
 \varphi^2 ( x =0,  y \in [0,\ell_s]) \ = \ & \ \ \  \nu/2\,,  \\
 \varphi^2 ( x =\ell_s,  y\in [0,\ell_s]) \ = \ & - \nu/2\,.   
 \end{split}
 \ee
These values are indicated on Figure~\ref{vmIL96}. 
As noted in section~\ref{sec:homology}, we need not be careful
about the orientation of the homology cycles nor about the sign of the
jump. Since we have a torus, the two functions also satisfy the periodicity conditions
  \be
  \label{phi1phi2pc}
 \begin{split}
 \varphi^1 ( x=0,  y \in [0,\ell_s]) \ = \ & \ \varphi^1 ( x=\ell_s,  y \in [0,\ell_s]),   \\
\varphi^2 ( x\in[0,\ell_s],   y=0) \ = \ & \ \varphi^2 ( x \in [0,\ell_s], y=\ell_s)\,.   
 \end{split}
 \ee
Apart from the discontinuities, as stated in the program (\ref{thirddual})
the functions must vanish on the boundary of the surface, thus 
\be
\label{phi1phi2bc}
\varphi^1|_{\partial M}=\varphi^2|_{\partial M} = 0\,,
\ee
where $\partial M$, shown in Figure~\ref{vmIL96},  is the inner square.   

The  dual program 
(\ref{thirddual}) applied to our torus with a boundary becomes 
\begin{equation}\label{thirddualSC}
\text{Maximize}\phantom{\Biggl(} 
\  \Bigl[\, 4\nu\,\ell_s -\int_M
dx dy \,\Bigl(|d\varphi^1|_0+ |d\varphi^2|_0\Bigr)^2\ \Bigr]  \ \     \hbox{over}  \ \  
     \varphi^1, \varphi^2,\nu \,,\ \   
\end{equation}
subject to the discontinuity conditions, periodicity conditions,
and boundary conditions given in  (\ref{phi1phi2dc}), (\ref{phi1phi2pc}),
and (\ref{phi1phi2bc}), respectively. For any set of trial functions $\varphi^1$ and $\varphi^2$ satisfying these conditions the above
maximum gives a lower bound for the minimal area.

In searching
for the optimum, we can use symmetry to constrain the 
functions. 
First, the symmetry $(x,y)\to (y,x)$  of the torus implies 
that one function is determined in terms of the other:
\be
\label{one-from-the-other}
\varphi^2 (x,y) = \varphi^1 (y,x) \,. 
\ee
This relation is respected by the assignment of 
discontinuities on $m_1$ and $m_2$. 
Moreover, the torus is symmetric under reflection
about the vertical line $x=\ell_s/2$ and under reflection about
the horizontal line $y=\ell_s/2$.   The function $\varphi^1$ must
be symmetric under the first 
and antisymmetric under the second:
\be
\varphi^1 (x, y) =  \varphi^1 (\ell_s-x, y)  \,, \qquad
\varphi^1 (x, y) = -\varphi^1 (x, \ell_s-y) \,.
\ee
Indeed, the values of $\varphi^1$ along $m_1$ make this the
only possible choice.  The antisymmetry implies that $\varphi^1$
vanishes for $y=\ell_s/2$:
\be
\varphi^1 (x, y=\ell_s/2) = 0 \,. 
\ee
Given these symmetries, the program
can be formulated restricting the torus to a fundamental
domain one-fourth the size of the surface and shown shaded 
in Figure~\ref{vmIL96}.   

As for the primal, we discretize the fundamental domain
and define the function $\varphi^1$ on all lattice points,
using the discontinuity conditions and the boundary conditions. 
The function $\varphi^2$ can then be read from the values of
$\varphi^1$.  At the center of each plaquette one can evaluate
the fiducial norms $|d\varphi^1|_0$ and $|d\varphi^2|_0$.  
The objective is now constructed by adding the contributions
from all plaquettes and the optimum is found by searching over
positive values of $\nu$ and all functions $\varphi^1$ with the
associated discontinuity.

 \begin{figure}[!ht]
\leavevmode
\begin{center}
\epsfysize=6.0cm
\epsfbox{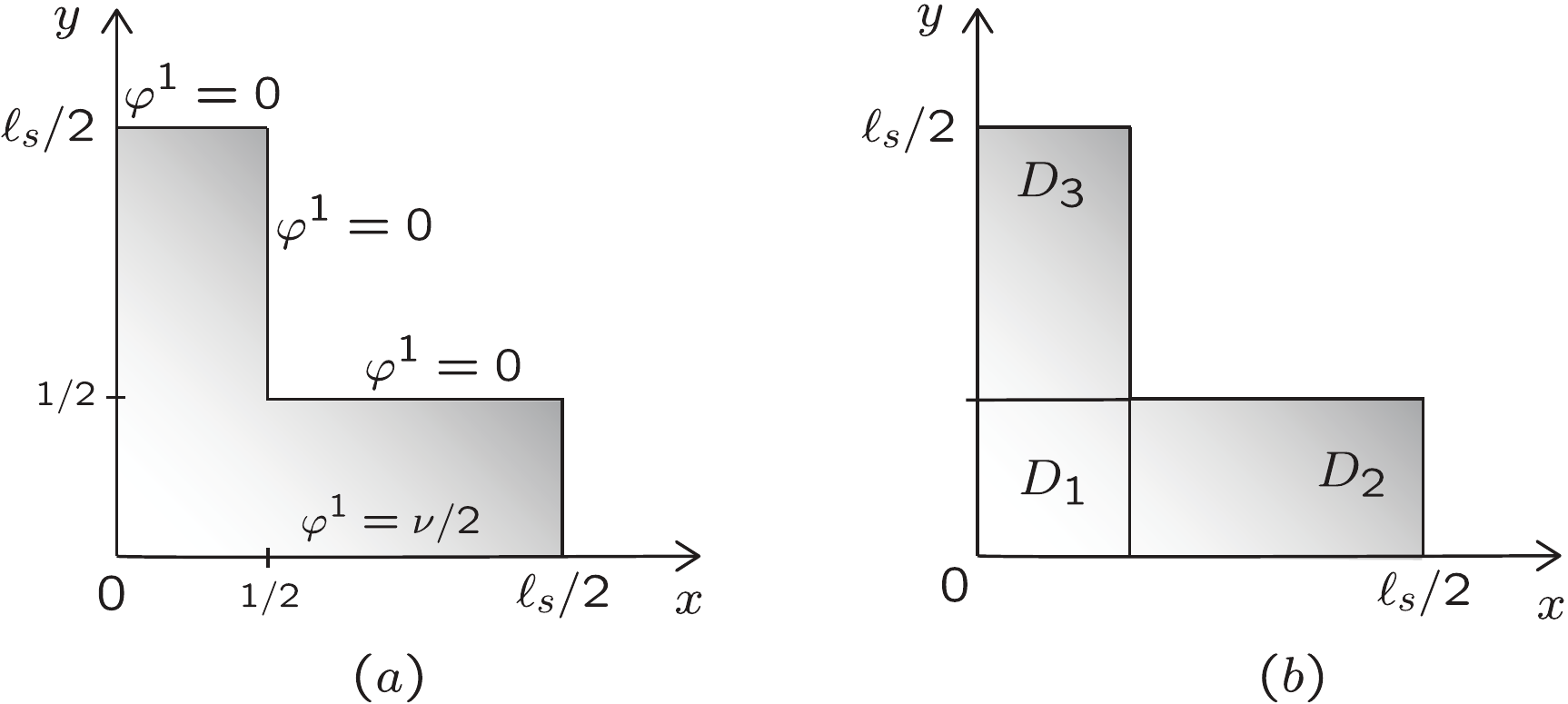}
\end{center}
\caption{\small  (a) The fundamental domain on the torus relevant for the dual program.  The values of $\varphi^1$ are
prescribed over all but two of the segments in the boundary of the
fundamental domain.  Over the other two segments $\varphi^1$
is free to vary.  (b) Defining regions $D_1,D_2$, and $D_3$ over
the fundamental domain.}
\label{iljn}
\end{figure}

\begin{figure}[!ht]
\leavevmode
\begin{center}
\epsfysize=5.5cm
\epsfbox{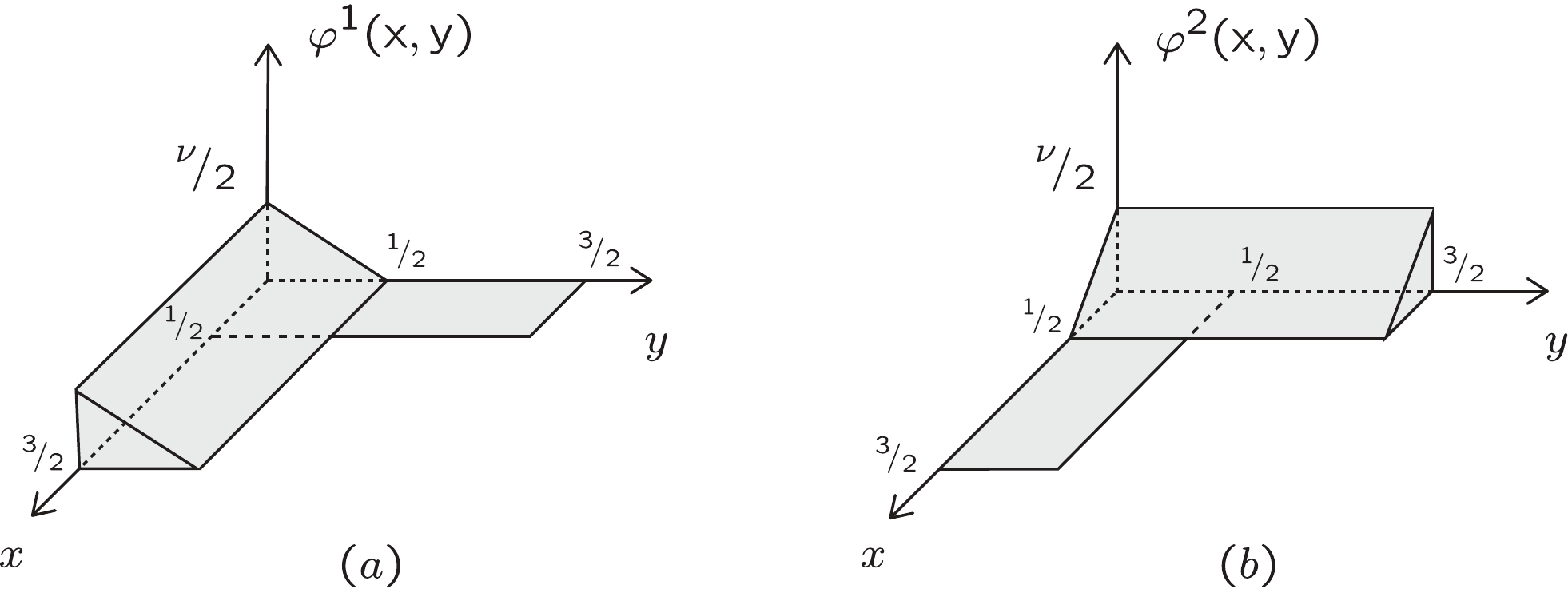}
\end{center}
\caption{\small (a) An ansatz for $\varphi^1$ consistent with the discontinuity conditions and the boundary conditions. (b) The corresponding form of $\varphi^2$ obtained from $\varphi^1$ using
the symmetry of the problem. }
\label{ilcl}
\end{figure}

To illustrate the formalism and obtain a lower bound on the
minimal area, we evaluate the objective using a simple ansatz for
$\varphi^1$ when the discontinuity $\nu$ is given.
For this we focus on the fundamental domain shown in Figure~\ref{iljn} (a).  Consistent with the indicated
boundary conditions we construct a function $\varphi^1$, which
along with $\varphi^2$, is plotted   in Figure~\ref{ilcl}.   These functions
are
\be
\varphi^1(x,y) = \begin{cases} \nu \bigl( \tfrac{1}{2}-y\bigr) \,, \quad  0 < y < \tfrac{1}{2} \, \\  \ \ 0 \,,  \hskip39pt
  \tfrac{1}{2} < y < \tfrac{\ell_s}{2}  \end{cases} \quad \hbox{and} \quad  
  \varphi^2(x,y) = \begin{cases} \nu \bigl( \tfrac{1}{2}-x\bigr) \,, \quad  0 < x < \tfrac{1}{2} \, \\  \ \ 0 \,,  \hskip39pt
  \tfrac{1}{2} < x < \tfrac{\ell_s}{2} \, 
\end{cases} \, . 
\ee
It follows that
\be
d\varphi^1(x,y) = \begin{cases} -\nu dy \,, \quad  0 < y < \tfrac{1}{2} \, \\  \ \ 0 \,,  \hskip24pt
  \tfrac{1}{2} < y < \tfrac{\ell_s}{2}  \end{cases} \quad \hbox{and} \quad  
  \varphi^2(x,y) = \begin{cases} -\nu dx \,, \quad  0 < x < \tfrac{1}{2} \, \\  \ \ 0 \,,  \hskip24pt
  \tfrac{1}{2} < x < \tfrac{\ell_s}{2} \, 
\end{cases} \, . 
\ee
Introducing the subregions $D_1, D_2$, and $D_3$  shown
on Figure~\ref{iljn}(b), we have 
\be
|d\varphi^1|_0 + |d\varphi^2|_0 = \begin{cases}  2\nu \,, \quad (x,y) \in D_1\,, \\
\ \nu \,,  \quad (x,y) \in D_2\,, \\
\ \nu \,,  \quad (x,y) \in D_3\, .\end{cases}
\ee

\begin{figure}[!ht]
\leavevmode
\begin{center}
\epsfysize=6.0cm
\epsfbox{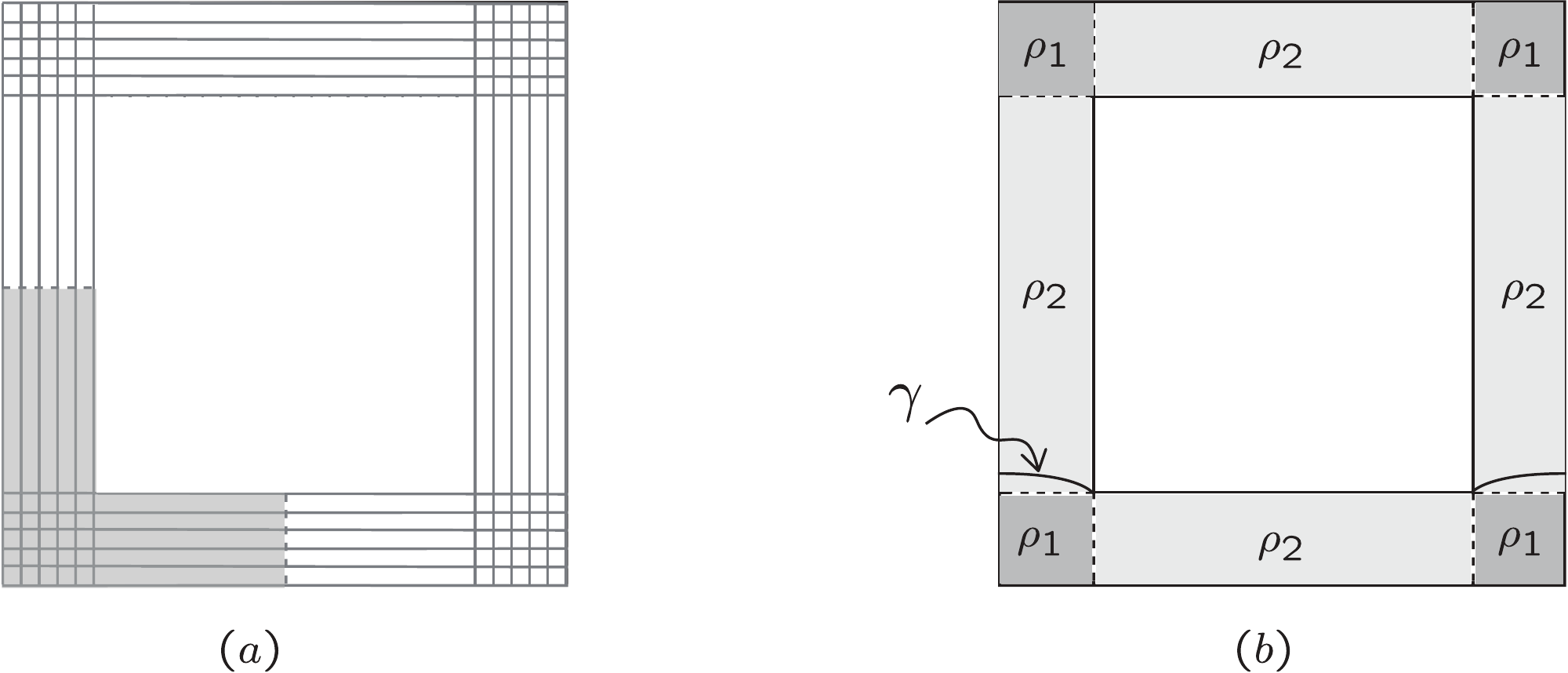}
\end{center}
\caption{\small (a) The curves of constant $\varphi^1$ over the region
where $d\varphi^1 \not=0$ together with the curves of constant $\varphi^2$ over the region where $d\varphi^2 \not=0$.  All these
curves have length three. (b) The two-constant 
metric associated with the ansatz. With $ds= \rho|dz|$ the metric
takes values $\rho_1$ and $\rho_2$ with $\rho_1 = 2\rho_2$. 
The curve $\gamma$ has length less than $\ell_s$.}
\label{ilmk}
\end{figure}

We can now evaluate 
\be
\int_M \bigl(|d\varphi^1|_0 + |d\varphi^2|_0\bigr)^2\,  dx dy =  4 \Bigl( 4\nu^2 \cdot \tfrac{1}{4} + \nu^2 \tfrac{1}{4} (\ell_s-1)  + \nu^2 \tfrac{1}{4} (\ell_s-1)\Bigr) = 2\nu^2 (1+ \ell_s) \,,
\ee
where we multiplied by four the result of the integral over the fundamental domain, itself broken into subregions $D_1, D_2$, and $D_3$. Looking 
back at the objective in (\ref{thirddualSC}) we now have 
\begin{equation}\label{thirdvmdual}
\hbox{Optimum} \ = \  \text{Maximum}
\,  \Bigl[\,4\nu\ell_s  -2 \nu^2(1+ \ell_s) \Bigr]  \ \     \hbox{over}  \ \  \nu \,.\ \   
\end{equation}
The maximum is attained for $\nu_*$ 
\be
\nu_* \ = \ \frac{\ell_s}{1+\ell_s}  \quad \Rightarrow \quad  \hbox{Optimum} = \frac{2 \ell_s^2}{1+\ell_s}\,.
\ee
This means that the minimal area $A$ on the torus with a boundary
is constrained to obey
\be
A_* \  \geq  \ \frac{2 \ell_s^2}{1+\ell_s}\  \,. 
\ee
For $\ell_s=3$ one recovers the familiar lower bound 
$A\geq \tfrac{9}{2}$~\cite{marnerides}.  When the dual program is solved exactly and we have
an optimum the systolic geodesics
are identified as the lines of constant $\varphi^\alpha$ in regions
where $d\varphi^\alpha \not=0$.  We can look at these curves
for the optimum of our ansatz.  We have $d\varphi^1\not=0$ in the region
$D_1 \cup D_2$, and $d\varphi^2\not=0$ in the region $D_1 \cup D_3$.
The lines of constant for $\varphi^1$ and 
$\varphi^2$ (over the regions where their gradients 
are non vanishing on the full torus) are shown in Figure~\ref{ilmk}(a).  
The region $D_1$ 
is covered by two sets of curves, the regions $D_2$ and $D_3$
by a single set of curves. 
The associated metric can also be read from (\ref{solution22}) and
reads
\be
\label{solution212}
\Omega = \Bigl(|d\varphi^1|_0 + |d\varphi^2|_0\Bigr)^2 
\ = \  \begin{cases}  4\nu_*^2\,, \quad (x,y) \in D_1\,, \\
\ \nu_*^2 \,,  \quad (x,y) \in D_2\,, \\
\ \nu^2_* \,,  \quad (x,y) \in D_3\, .\end{cases}
\ee
Recalling that $\Omega = \rho^2$ where $ds = \rho |dz|$, this is
a two-constant metric: $\rho_1=2\nu_*$ on region $D_1$ and $\rho_2= \nu_*$
in regions $D_2$ and $D_3$.  We see that the metric is higher in the
region with two sets of constant $\varphi$ curves and lower in the
regions with one set of constant $\varphi$ curves. Indeed, it makes
sense to raise the metric in the region where more geodesics 
can gain length. 
 This two-constant metric
(see Figure~\ref{ilmk}(b)) was obtained
by Marnerides~\cite{marnerides} who used Beurling's criterion to 
show that it is the minimal-area metric for a problem where we
only constrain the strictly horizontal and strictly vertical curves
on the torus.  One can easily check that on this metric all 
these curves, shown in Figure~\ref{ilmk}(a) have length equal to $\ell_s$.  
Of course, $\ell_s$ is not the systole for this metric, as
there are non-contractible curves shorter than $\ell_s$. 
One such curve $\gamma$ is
shown  in Figure~\ref{ilmk}(b): making use of the discontinuity of the metric, it bends slightly to avoid the region where the metric is $\rho_1$.

\subsection{Using other conformal frames}
\label{sec:conformal-frames}

As it will become clear in the following section, the form of 
the optimal metric near and at the corners
of the central square of the Swiss cross is difficult to assess from the
numerical data.  While the metric seems to diverge as one approaches
these `central corners',  the behavior of the curvature is less clear.  
As we travel along the Swiss cross boundary in the counterclockwise
direction, the central corners are points where there is a 
turning angle of $-\pi/2$. The turning angle is 
defined as that rotation angle
angle of magnitude less than or equal to $\pi$ taking
the tangent to the curve right before the critical point to the tangent to
the curve right after the critical point.  The rotation angle is measured
as positive in the counterclockwise direction and as negative in the
clockwise direction.  As opposed to the central corners, the other corners
of the Swiss cross all have turning angles of~$\pi/2$.

In this subsection we give two alternative presentations
for the Swiss cross, for convenience chosen to have $\ell_s=2$. 
They are both conformal maps of a portion
of the Swiss cross, an L-shaped region with a single central corner, into 
polygons where the central corner has become an ordinary point or a point
at infinity. These polygons are chosen because they are amenable to discretization.
We will give two presentations.  One is a pentagon with the
central corner mapped to the middle of one edge, and the
other is a semi-infinite strip with the central corner pushed to infinity.

\subsubsection{Pentagon}

Let us consider first the pentagon presentation. 
As we will see in the next section, the fact that this frame straightens out the boundary at the corner greatly clarifies the behavior of the metric in the vicinity of this point.

\begin{figure}[!ht]
\leavevmode
\begin{center}
\epsfysize=10.5cm
\epsfbox{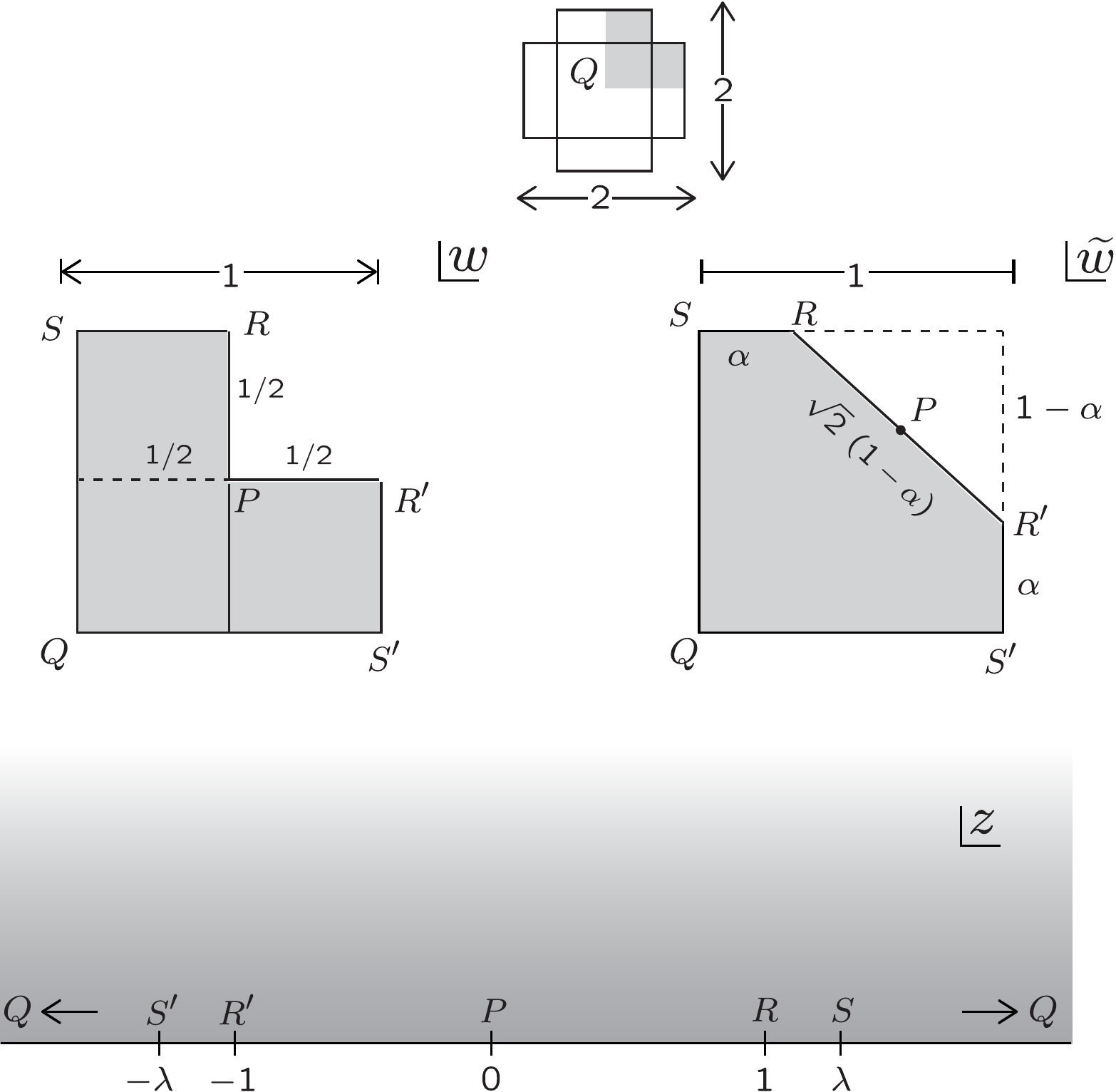}
\end{center}
\caption{\small  Mapping conformally an L-shaped region of an $\ell_s=2$ 
 Swiss cross (upper left) to a pentagonal region (upper right) where the central corner $P$  becomes a regular point on the boundary. 
 Both figures can be mapped to the
same upper half plane $z$ when $\lambda = 1.154701$.  The value of $\alpha$ 
defining the pentagon turns out to be $\alpha=0.308963$.}
\label{pent27}
\end{figure}

In Figure \ref{pent27}, top left, we show
the L-shaped subregion of the Swiss cross as a domain in a $w$ plane (with the full Swiss cross in the
little figure above).  The pentagonal region, shown to the right as a domain
in the $\tilde w$ plane, is expected
to be conformally equivalent to the L-region for some choice of the 
parameter $\alpha$.  In the conformal map corners labeled with the same letters go into
each other.  Our goal is to find the value of $\alpha$.  We will demonstrate
the map exists and calculate $\alpha$
by mapping both shapes into the same $z$ upper-half plane
with some marked points. 

 We first map
the L-shaped region in the $w$ plane  into  the $z$ upper half plane with the central corner $P$ mapped to $z=0$, the $R,R'$ corners mapped to $\pm 1$,
the $S,S'$ corners mapped to $\pm \lambda$, and $Q$ mapped to infinity.
Using the standard rules for turning angles in the Schwarz-Christoffel 
map, we have
\be
{dw\over dz} \ = \ N \frac{\sqrt{ z}}{\sqrt{(1-z^2) (\lambda^2 - z^2) }} \,,
\ee
where $N$ and $\lambda$ are unknown parameters. 
The value of $\lambda$ is determined by the condition
that on the L-shaped region the side $PR$ (mapped to $z\in [0,1]$) has the same length as
the side $RS$ (mapped to $z\in [1,\lambda]$).  This condition gives
\be
\int_0^1 dz  {\sqrt{ z} \over \sqrt{(1-z^2) (\lambda^2 - z^2)} } =  
\int_1^\lambda  {\sqrt{ z}\over \sqrt{(z^2-1) (\lambda^2 - z^2)} }\,.
\ee
Given our purposes a numerical solution for $\lambda$ is all we need.
We find, to excellent accuracy,
\be
\label{lambda-value}
\lambda = 1.154\, 701\,.
\ee
With this value the ratio of the  length of the two edges differs
from one by less than $10^{-6}$.

The next step is mapping the pentagon in the $\tilde w$ plane to the upper
half plane as well.  Again, we must map  the image $P$ of the
central corner to $z=0$, the $R, R'$ corners to $\pm 1$,
the $ S, S'$ corners to $\pm \lambda$ and $Q$ mapped to infinity. 
This time the conformal map takes the form 
\be
{d\tilde w \over dz}   \ = \  {N\over \sqrt{(\lambda^2 - z^2) \,\sqrt{1-z^2}\, }}\,.
\ee
For the L region and the pentagon to be conformally equivalent  the value
of $\lambda$ here must be that in (\ref{lambda-value}). 
Note that, as required, there is no turning point at $z=0$, corresponding to the
central corner.    The unknown in the pentagon is the length $\alpha$
of the $ST$ and  $S'R'$ edges.  By simple geometry
we find that the length of the $PR$ segment is $(1-\alpha)/\sqrt{2}$.
This requires 
\be
 {\alpha\over {1-\alpha\over \sqrt{2}}}    
 \ = \
 {\int_1^\lambda   {1\over \sqrt{(\lambda^2 - z^2) \,\sqrt{z^2-1}\, }}  \over 
\int_0^1   {1\over \sqrt{(\lambda^2 - z^2) \,\sqrt{1-z^2}\, }}  }  \ = \ 0.632\,295 \,, 
\ee
where the last equality follows by numerical evaluation of the two integrals 
with the known value of~$\lambda$.   We now find  
\be
\alpha \simeq  0.308\, 963  \, ,  \quad \hbox{and} \quad  R'R = \sqrt{2} (1- \alpha ) \simeq 0.977\, 274\,,
\ee
where the second value is the length, on the pentagon, 
of the side that now contains the central point.  This completes the determination
of the map and makes it possible to write the primal and dual programs
on the pentagonal region.

\subsubsection{Strip}

Let us now turn to the map of the L-shaped region to a semi-infinite strip. We include this discussion mainly for completeness and because it could be useful in future investigations. The results we present in the rest of the paper were not obtained using the strip frame, so the reader should feel free to skip this subsection.

\begin{figure}[!ht]
\leavevmode
\begin{center}
\epsfysize=10.5cm
\epsfbox{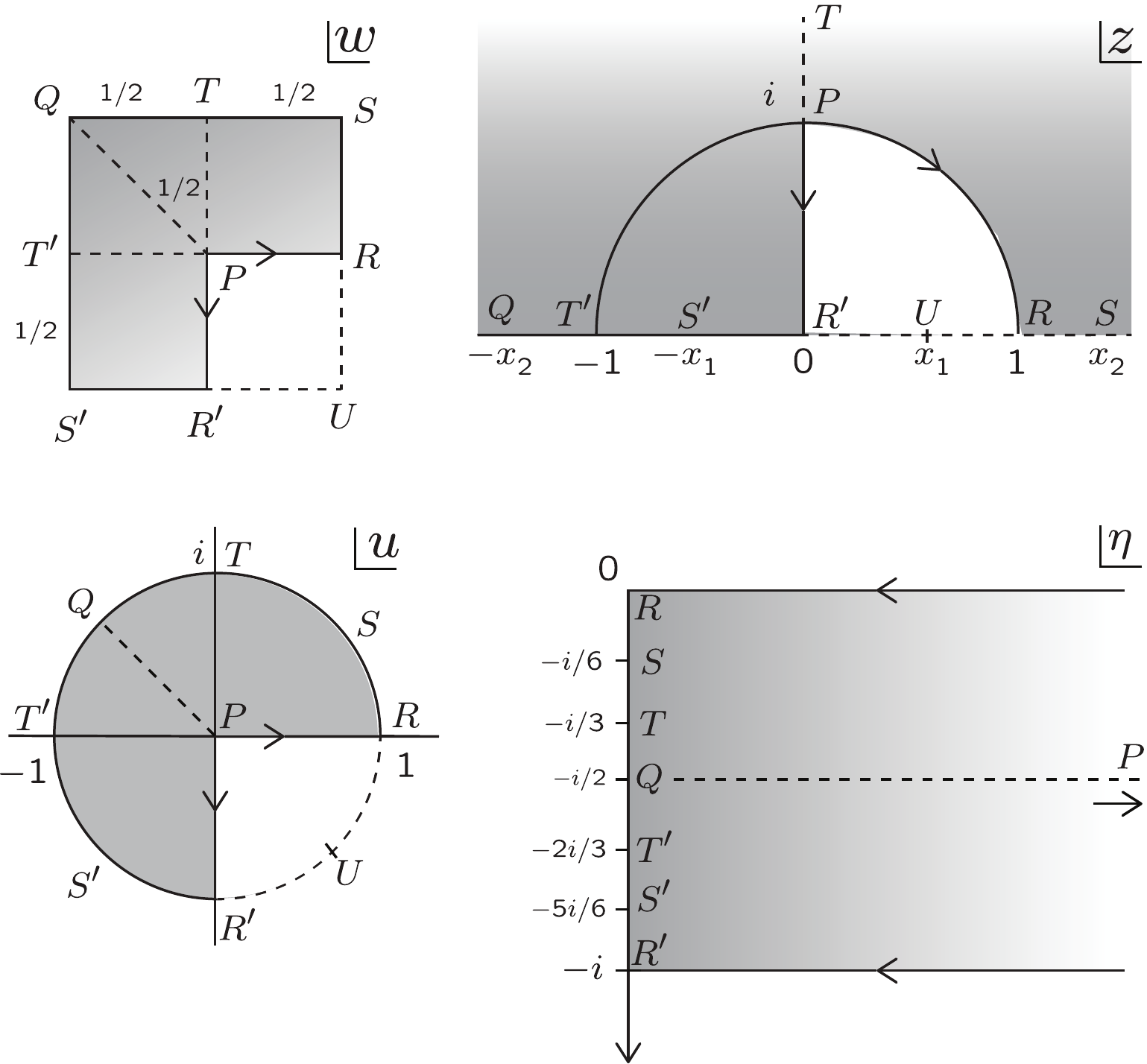}
\end{center}
\caption{\small  Mapping conformally an L-shaped region of an $\ell_s=2$ 
 Swiss cross (upper left) to a semi-infinite strip (lower right) where the central corner $P$  is pushed to infinity.  Symmetry considerations show that the
 map to the $u$u-plane unit disk must exist.  The map is constructed via
 the $z$ upper half-plane.  Finally a logarithmic map takes the region in the
 unit disk into the strip in the $\eta$ plane.}
\label{strip-p}
\end{figure}

Our task becomes much easier if we complete the L-shaped region into
a square, as shown in Figure~\ref{strip-p}, upper left, with a new corner point $U$.  This square can be mapped into the interior of a unit disk $|u| \leq 1$, 
with the four corners
and the four edge midpoints mapped into the boundary of the disk
in such manner that the symmetries of the square
are respected.  As shown in Figure~\ref{strip-p}, the central
corner $P$ ends at the center of the circle,  the vertical and horizontal 
lines passing through the middle of the square become vertical and horizontal
lines going through the origin of the unit disk and,  finally, because of symmetry,
the corners of the square go into points on the unit disk with argument $\pm 45^\circ$ and $\pm 135^\circ$.    Note that the L-shaped region occupies the interior of the unit disk with 
\be 
\arg (u) \in [0, \tfrac{3\pi}{2}\, ]\,.
\ee

The map from the L-shaped
region in the $w$ plane to the $u$ disk can be constructed
with the help of intermediate maps to a $z$ upper-half plane.  
Choosing to map the $u=0$ central corner $P$ to $z=i$, the  point $R'$, at $u=-i$  to $z=0$,
and the point $T$ at $u=i$ to $z=\infty$, we have
\be
z \ = \ -i \, {u+i \over u-i} \, \ \ \  \hbox{or} \  \ \  u \ = \ -i  \, {1 + iz\over 1-iz} \,. 
\ee 
Note that in this map the real points $T'$ and $R$ on the circle are mapped to
plus and minus one, respectively.  Moreover, one quickly verifies that
the points $U$ and $S'$ are mapped to $x_1$ and $-x_1$, respectively,
while the points $S$ and $Q$ are mapped to $x_2$ and $-x_2$, respectively,
where
\be
x_1 \ = \ \sqrt{2}-1   \,,   
\quad x_2 \ = \ \sqrt{2} +1  \,.  
\ee
Having constructed the map from $|u|\leq 1$ to the upper half plane, we
also need the map from the square into the upper half plane $z = f(w)$.
This is a Schwarz-Christoffel map that takes the form
\be
{dw\over dz} \ = \ {N\over \sqrt{(z^2 - x_1^2 ) (z^2 - x_2^2)}} \,, \quad 
N \simeq  0.762\,759\, 78\,, 
\ee
where the constant $N$ was determined from the condition that
the square containing the L-shaped region have the correct size. 
We now have the map $z= f(w)$ from the L-shaped region to the upper
half plane, and the map $u = g(z)$ from the upper half plane to the
$u$ unit disk.  Finally we map that the image of the L-shaped 
region in the $u$ disk to a semi-infinite strip in the $\eta$ plane.
This is achieved by
\be
 \eta \ = \ -{2\over 3\pi}  \, \ln u \,.
\ee
The image of the L-shaped region is now the semi-infinite strip ${\cal S}$
defined by 
\be
{\cal S} \ = \ \{ \eta \ \bigl| \ \hbox{Im} (\eta ) \in [0, -1] , \ \hbox{Re} (\eta) \in [0, \infty]
\}  \,. 
\ee
The central corner at $u=0$ has been mapped to the end of the strip, at 
$\eta = -\tfrac{i}{2} + \infty$, aligned with the center $P'$ of the Swiss 
cross, at $\eta = -\tfrac{i}{2}$.  The position of the varius additional corners
are indicated in Figure~\ref{strip-p}.

\section{The results for the extremal metric}
\label{sec:results}

We begin this section by discussing the results of the primal and 
dual programs for a Swiss cross with $\ell_s=2$ ($a=1$).  The
area can be found to six significant digits.  The patterns of systolic
geodesics  as well as the curvature of the surface are discussed.  
Special attention is given to the (central) corner points where the
extremal metric is found to diverge in a way to make them
regular points on the boundary.  The global 
picture of the surface is shown in Figure~\ref{fig:embedding}. 
We then study the moduli space of the Swiss cross, or torus with boundary,
finding that the extremal length conformal invariant $\lambda$ is, to a good
approximation, linearly related to a parameter $t$ describing the shape of
the rectangles that define the Swiss cross.  In the moduli space we find two
critical points.  One occurs as the positively curved central dome
reaches the edges of the Swiss cross.  The second occurs when, in the torus picture,
the length of the boundary becomes equal to the systole.  The latter
one is relevant to the study of the torus with a puncture. 

\subsection{Area, metric, and curvature}
\label{sec:metric}

In this subsection, we will explore the extremal metric for the Swiss cross (or equivalently the torus with a boundary) obtained from solving the primal and dual programs described in the previous section. 
For concreteness, throughout this subsection we will focus on the case $\ell_s=2$
with $a=1$.  For this surface, the $\rho=1$ metric is admissible and has 
an area $A =3$. 
We will plot all quantities on a fundamental domain of the $\mathbb{Z}_2\times\mathbb{Z}_2$ symmetry, in both the original frame (referred to as the ``L'' frame) and the frame in which the L is mapped to a pentagon\footnote{To discretize the pentagon frame we used 38 plaquettes on the short side and 123 on the long side.  This is a good approximation since 
$38/123 \approx 0.308\, 943$ is pretty close to value $\alpha  = 0.308\, 963$, 
required for conformal equivalence with the $\ell_s=2$ surface.  Indeed, with this lattice, we find an extremal area of  $2.806\hskip1pt 874$, very close to the 
value $A$ in (\ref{extremalarea1}).} (referred to as the ``pentagon'' frame; see Figure\ \ref{pent27}).
The data we present are from the highest-resolution runs of the primal and dual programs in both frames.  
Based on the runs in the L frame, we estimate the extremal area $A$ as  
\begin{equation}\label{extremalarea1}
A = 2.806\hskip 1pt970(1)\,.
\end{equation}
The quoted error estimate of $10^{-6}$ is derived from the convergence of the results as a function of the lattice resolution; see appendix \ref{sec:convergence} for details.  As explained there, the error appears to decrease like $(\hbox{res})^{-3}$, and the results of the primal and dual are surprisingly close
to each other at each value of the resolution.  The extremal area is
about $6.4\%$ lower than the area $A=3$ of the $\rho=1$ metric. 

\begin{figure}[!ht]
\leavevmode
\begin{center}
\epsfysize=8cm
\epsfbox{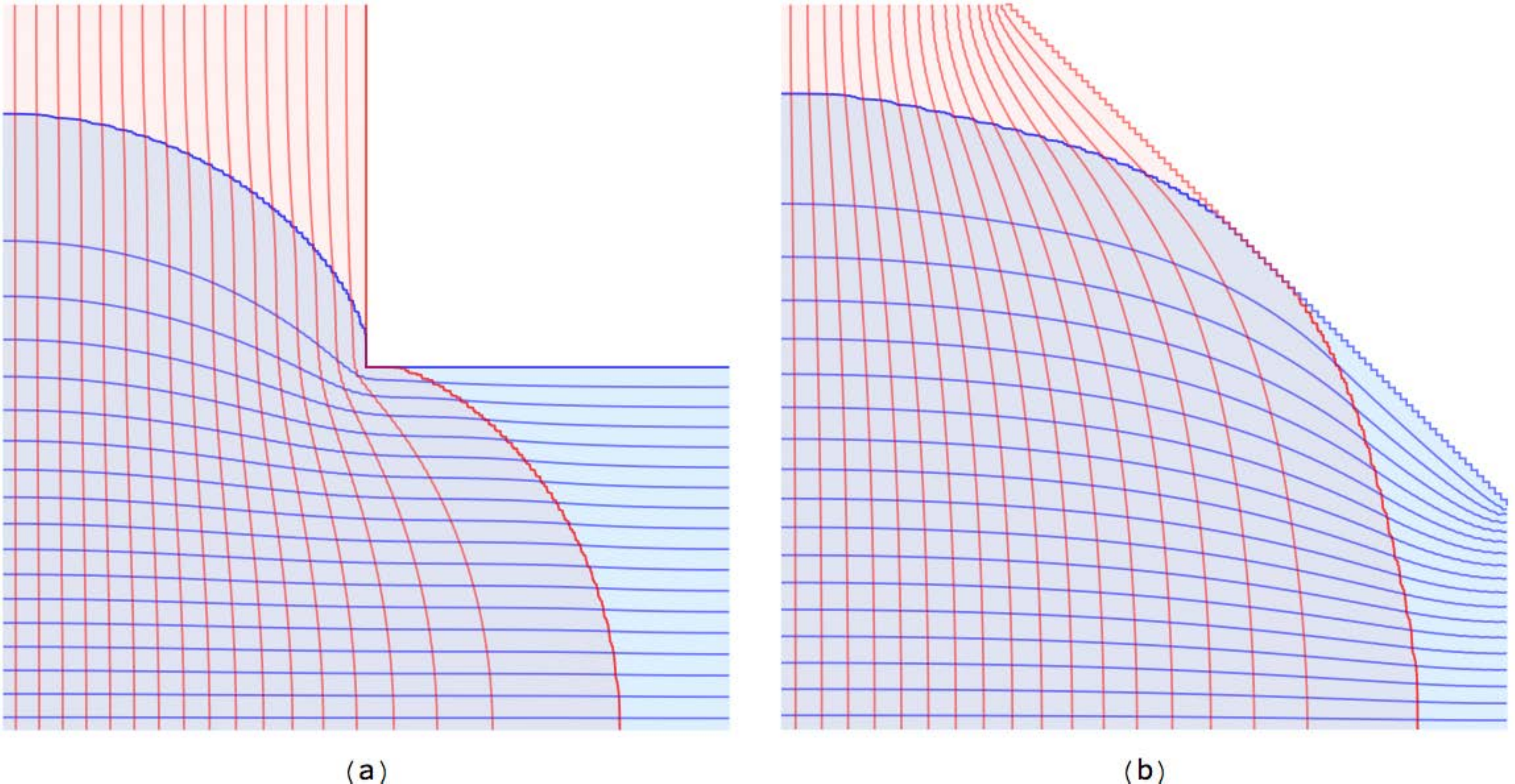}
\end{center}
\caption{\small 
Systolic geodesics on the fundamental domain of the $\ell_s=2$ Swiss cross, drawn in the (a) L frame and (b) pentagon frame. The blue curves are 1-geodesics and the red curves 2-geodesics. The light blue region contains only 1-geodesics, the pink region only 2-geodesics, and the purple region both.  
}
\label{fig:sgs}
\end{figure}

\begin{figure}[!ht]
\leavevmode
\begin{center}
\epsfysize=6cm
\epsfbox{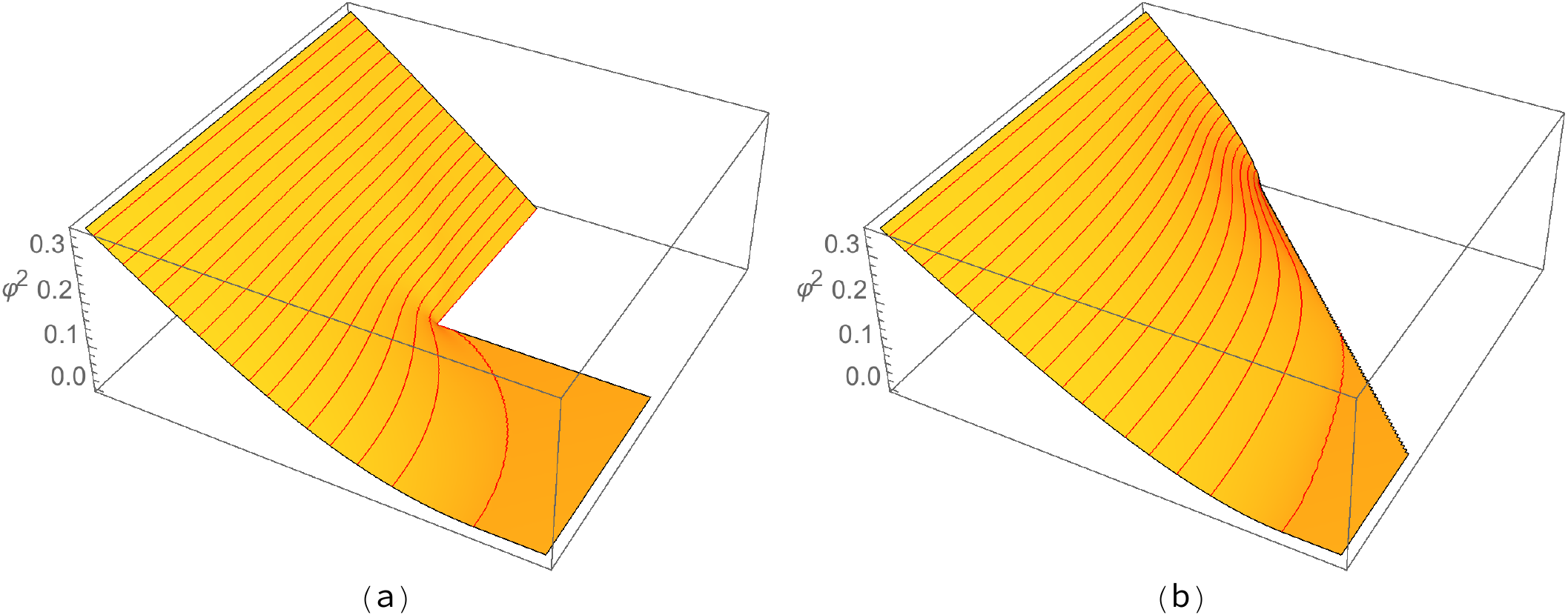}
\end{center}
\caption{\small 
The optimal configuration of the function $\varphi^2$ appearing in the dual program, drawn in the (a) L frame and (b) pentagon frame. The region where the gradient of $\varphi^2$ is non-zero defines the red and purple regions of Figure\ \ref{fig:sgs}.  The level sets of $\varphi^2$ are the 2-geodesics shown in red.  
}
\label{fig:psiplot}
\end{figure}

The systolic geodesics are plotted in Figure\ \ref{fig:sgs}. These are obtained as the level sets of the functions $\varphi^1$ and $\varphi^2$ of the dual program, and are drawn as
blue and red curves, respectively.  
 The function $\varphi^2$ is plotted in Figure\ \ref{fig:psiplot}. We see from Figure\ \ref{fig:sgs}  that there is a region foliated by 1-geodesics (in light blue), another foliated by 2-geodesics (pink), and a third foliated by both (purple). We will denote the last region as $U_2$. If we follow the uppermost 1-geodesic from the left in the L frame, it appears to meet the vertical boundary tangentially, before making a sharp $90^\circ$ left turn counterclockwise and and following the horizontal boundary; similarly for the rightmost 2-geodesic. In the pentagon frame, these last geodesics  join the diagonal boundary of the pentagon tangentially at its midpoint.

\begin{figure}[!ht]
\leavevmode
\begin{center}
\epsfysize=20cm
\epsfbox{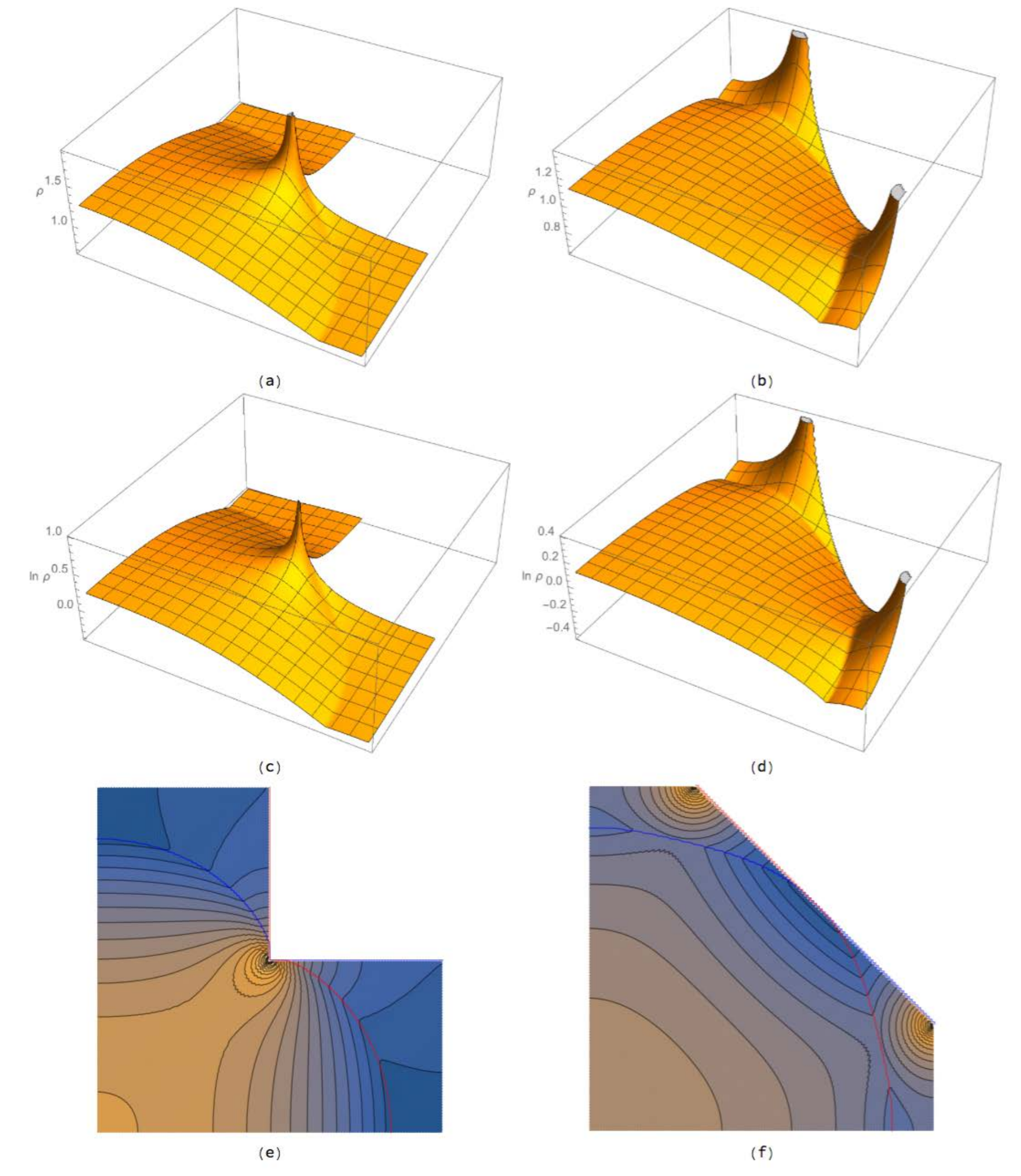}
\end{center}
\caption{\small 
The line element $\rho=\sqrt\Omega$ in the (a) L and (b) pentagon frames. $\ln\rho$ in the (c) L and (d) pentagon frames. Contours plots of $\ln\rho$ in the (e) L  and (f) pentagon frames.  Both in the (e) and (f) figures, the uppermost 1-geodesic and rightmost 2-geodesic are also shown.
}
\label{fig:rhoplots}
\end{figure}

\begin{figure}[!ht]
\leavevmode
\begin{center}
\epsfysize=5cm
\epsfbox{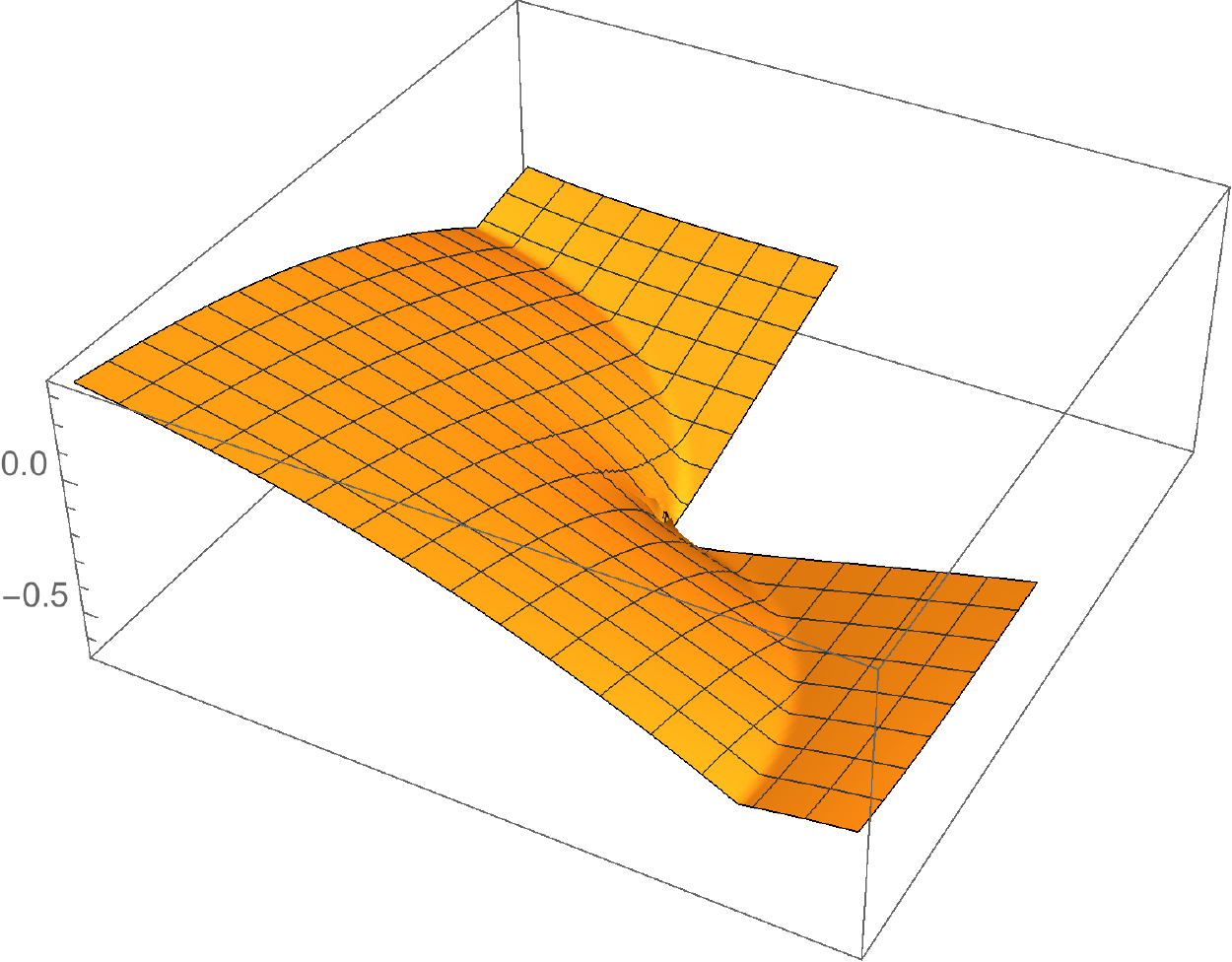}
\end{center}
\caption{\small 
The quantity $\ln\rho_w-\ln(|w-w_P|^{-1/3})$ in the L frame, where $w_P$ is the coordinate of the inner corner $P$. The finiteness of this difference at $P$ confirms the prediction \eqref{rhowpred}.
}
\label{fig:rhominus}
\end{figure}

Next, we study the extremal metric, or more precisely its line element $\rho=\sqrt\Omega$. Figure\ \ref{fig:rhoplots} shows $\rho$,  $\ln\rho$, and a contour plot of $\ln\rho$ in both the L and pentagon frames. Of course, $\rho$ is not a scalar, so it gets rescaled by the factor $|d\tilde w/dw|$ between the two frames, where $w$ is the L coordinate and $\tilde w$ is the pentagon coordinate (see Figure\ \ref{pent27}):
\begin{equation}
\label{rhowtrhow}
\rho_w = \left|\frac{d\tilde w}{dw}\right|\rho_{\tilde w}\,.
\end{equation}
In the L frame, $\rho$ exhibits a spike at the inner corner; however, it remains finite at the corresponding point in the pentagon frame, the center of the diagonal boundary (labelled $P$ in Figure\ \ref{pent27}). Near that point, the 
conformal map takes the form 
\begin{equation}
\tilde w-\tilde w_P \sim (w-w_P)^{2/3}\,,
\end{equation}
where $w_P$ and $\tilde w_P$ are the $w$- and $\tilde w$-coordinates of $P$ respectively. Hence
\begin{equation}
\left|\frac{d\tilde w}{d w}\right| \sim |w-w_P|^{-1/3}\,.
\end{equation}
Given the finiteness of $\rho_{\tilde w}$ near $P$,
equation (\ref{rhowtrhow}) gives the following prediction for the behavior of $\rho_w$ near $P$:
\begin{equation}\label{rhowpred}
\rho_w \sim |w-w_P|^{-1/3}\,.
\end{equation}
To check that our $\rho_w$ is consistent with the prediction \eqref{rhowpred}, in Figure\ \ref{fig:rhominus} we plot the difference $\ln\rho_w-\ln(|w-w_P|^{-1/3})$; indeed, it is clearly bounded near the corner. We can say that the minimal-area problem wants the boundary to be smooth; by giving the metric the spike \eqref{rhowpred}, it straightens out the corner that is present in the L coordinate system.

The same effect leads to the spikes in $\rho_{\tilde w}$ at the two corners of the pentagon. The pentagon is the $\mathbb{Z}_2\times\mathbb{Z}_2$ fundamental domain of a torus with a diamond (i.e.\ a square with sides at $45^\circ$) removed. The corners of the diamond are at the points labelled $R$ and $R'$ in Figure \ref{pent27}. To straighten out those corners, the minimal-area problem makes the metric diverge like $|\tilde w-\tilde w_R|^{-1/3}$ and $|\tilde w-\tilde w_{R'}|^{-1/3}$. On the other hand, the boundary of the torus is already smooth at the points $R,R'$ in the L frame, so the metric remains bounded there. In a coordinate system in which the boundary has no corners---for example, a torus with a disk removed---the metric would be bounded everywhere.

\begin{figure}[!ht]
\leavevmode
\begin{center}
\epsfysize=15cm
\epsfbox{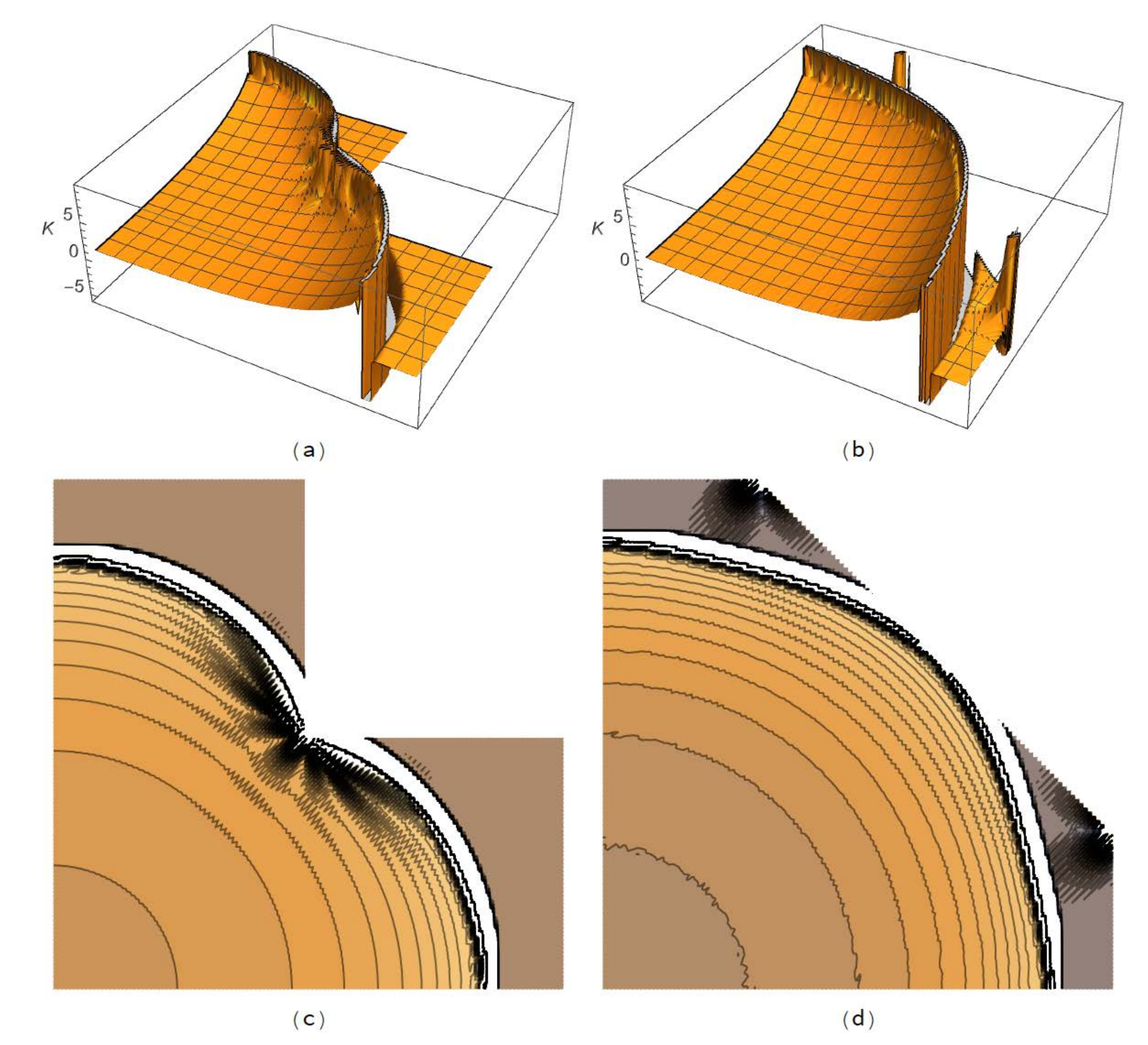}
\end{center}
\caption{\small 
Gaussian curvature $K$ in the (a) L frame and (b) pentagon frame. Contour plots of $K$ in the (c) L frame an (d) pentagon frame. The computation of $K$ by finite differencing based on \eqref{K} amplifies noise in the underlying values of $\rho$, leading to artifacts such as the spikes at the corners of the pentagon. Nonetheless, certain features are evident, such as the positive curvature in $U_2$ that increases as its boundary is approached, the deep valley on its boundary representing the negative delta-function, and the flatness of the wings (most clearly seen in the L frame).
}
\label{fig:curvature}
\end{figure}

Another notable feature of $\rho$ in both frames, apparent in the plots of Figure\ \ref{fig:rhoplots}, is the ``dome'' occupying the region $U_2$ and bounded by a crease coinciding with the rightmost and uppermost systolic geodesics. We compute the Gaussian curvature $K$ using the formula
\begin{equation}\label{K}
K = -\frac1{\rho^2}\nabla_0^2\ln\rho\,,
\end{equation}
where $\nabla_0^2$ is the Laplacian in the fiducial metric. We find that $K$ is positive in $U_2$, increases as we approach the boundary of $U_2$, and has a negative delta-function along its boundary, indicating a line curvature singularity. This can be confirmed by computing $K$ numerically, as shown in Figure\ \ref{fig:curvature}.   Our numerical results seem to indicate that the curvature 
$K$ remains bounded as the boundary of $U_2$ is approached; however, we cannot make a definitive claim about this.  We also see, as expected, that the curvature $K$ vanishes in the regions foliated by just one band of geodesics.

\begin{figure}[!ht]
\leavevmode
\begin{center}
\epsfysize=5cm
\epsfbox{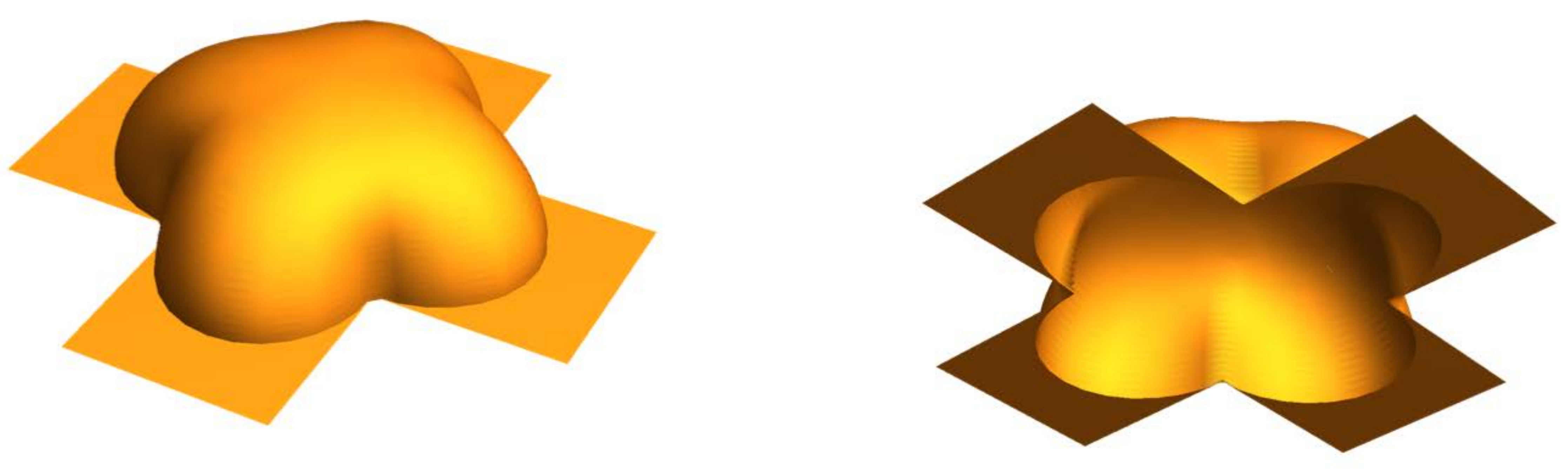}
\end{center}
\caption{\small 
Top and bottom views of a qualitative embedding diagram illustrating the global structure of the minimal-area metric on the Swiss cross.}
\label{fig:embedding}
\end{figure}

Based on this data, we now put together a global picture of the minimal-area metric on the Swiss cross. For this purpose, we return to the full Swiss cross, not just one $\mathbb{Z}_2\times\mathbb{Z}_2$ fundamental domain. It consists of a positively-curved dome glued to four flat wings. Each wing is a rectangle with a half-disk removed. As can be seen in Figure\ \ref{fig:sgs}(b), the curves that bound the dome $U_2$ are tangent where they meet. (In the L frame shown in Figure\ \ref{fig:sgs}(a), they appear to meet at a right angle; however, as emphasized above, the metric is singular at the corner where they meet, and we want a picture where the metric does not diverge in this region.) There are four such curves, and together they form a complete geodesic circle. Therefore, by the Gauss-Bonnet theorem, the Gaussian curvature integrates to $2\pi$ over the dome; the dome is essentially a hemisphere, albeit not with constant positive curvature. There is a delta-function of negative curvature along the curve where the dome is glued to each wing. This curve has different extrinsic curvatures on its two sides; on the dome side it has vanishing extrinsic curvature while on the wing side it has negative extrinsic curvature, with total turning angle $\pi$ (like a semicircle). Therefore the Gaussian curvature integrates to $-\pi$ over each gluing curve. This is consistent with the fact that the total Gaussian curvature over any of the two  bands of systolic geodesics vanishes, since each such band crosses the full dome as well as two gluing curves. An embedding picture of the whole Swiss cross is shown in Figure \ref{fig:embedding}. This is not a quantitatively accurate picture of the metric, but shows the important qualitative features. In particular, the dome is pictured as a hemisphere with four points on its boundary ``pinched in'' to leave $90^\circ$ concave angles, allowing the wings to be glued in at the corners; again, the boundary is regular at these corners.

\subsection{Moduli space of the torus with a boundary}

In this section we investigate Swiss crosses---or tori with a boundary---of different shapes.  We identify a simple parameter, called $t$, that
specifies the shape, and attempt to relate it to the extremal length
conformal invariant $\lambda$ that is a simple function of the extremal area.
Our numerical analysis suggests a rather surprising linear relation between
$\lambda$ and $t$, but careful consideration shows that the relation does not 
hold exactly.  We also show how one can use the extremal metric on one
surface to construct, by simple amputation, the extremal metric on a surface
of different shape.   

\begin{figure}[!ht]
\leavevmode
\begin{center}
\epsfysize=5.5cm
\epsfbox{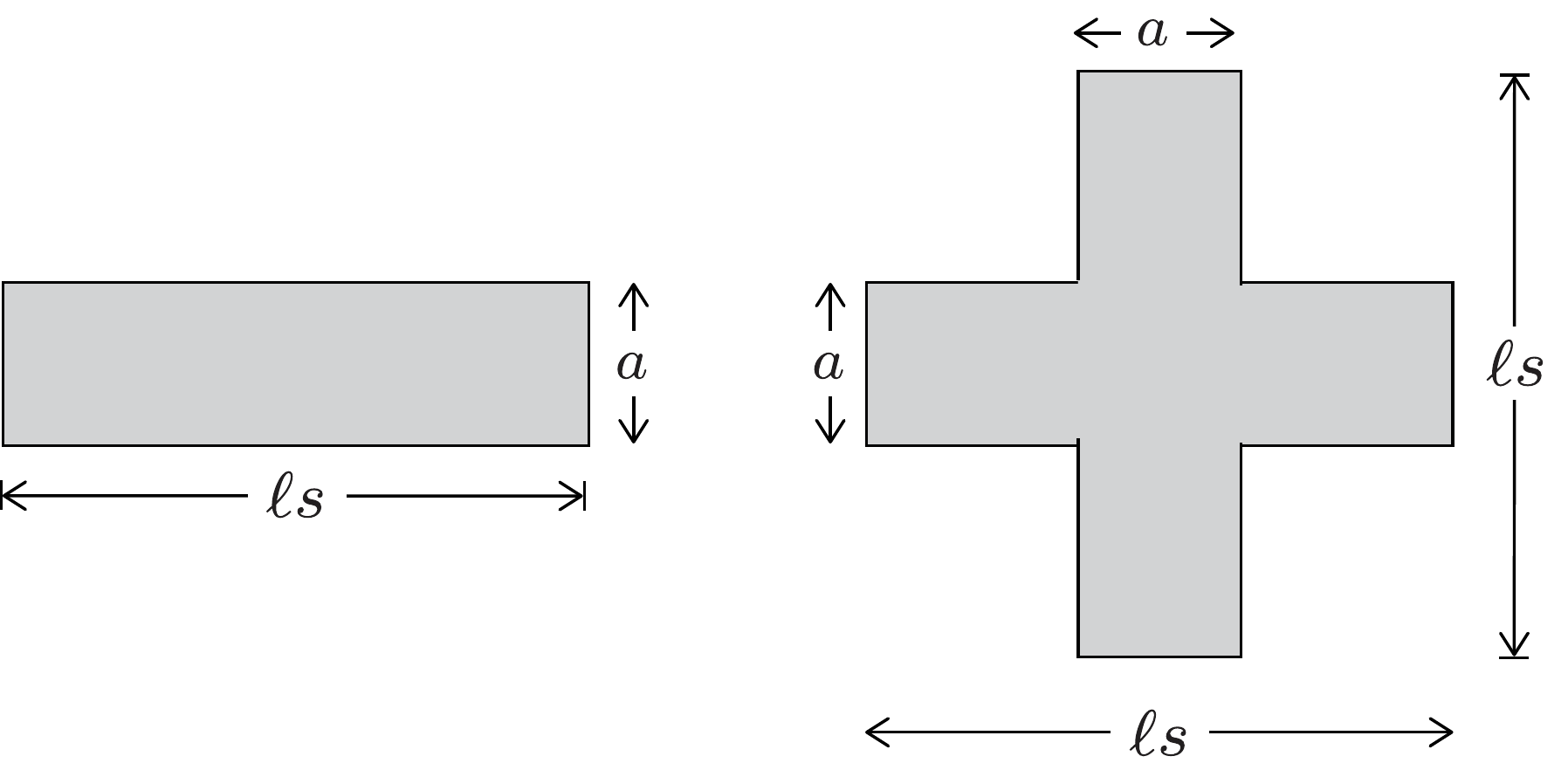}
\end{center}
\caption{\small  The Swiss cross is the part of the plane covered
by two identical rectangles, crossing orthogonally and symmetrically.  
The modulus of the rectangle provides a modulus for the torus with
a boundary.}
\label{ff27}
\end{figure}

We suggested in section \ref{sec:constrthesurf} 
that with fixed $\ell_s$ the parameter $a$ would be a modulus.  In the Swiss cross
presentation of the torus (Figure~\ref{ff1}), the parameter $a$ is the length of the vertical and the horizontal edges that are identified.   Alternatively, we can think
of the Swiss cross as the part of the $z$ plane covered by  two rectangles, each of length $\ell_s$ and width $a$, one placed horizontally and the other vertically,
with coincident centers, as shown in Figure~\ref{ff27}.

Clearly, for fixed $\ell_s$ we can take $a \in (0, \ell_s)$.  
As $a\to 0$ the boundary of the torus grows until it eats up the  
whole torus.  In the limit $a \to \ell_s$ the boundary of the torus is becoming
infinitesimally small.   Alternatively,  for fixed $a$ we
can take $\ell_s \in (a, \infty)$.

We can use the modulus of each of the rectangles as a tentative modulus
for the torus with a boundary (or the Swiss cross).  
The modulus of a rectangle is the ratio of the two sides. 
We thus define the modulus $t$:
\begin{equation}
t \ \equiv \  {\ell_s\over  a} \, \in  \bigl( 1 , \infty \bigr) \,. 
\end{equation}
$t$ is a modulus if we
can find a conformal invariant $\lambda$ of the surface such that
$t \not= t'$ implies $\lambda \not= \lambda'$.  As we explain now,
the requisite
conformal invariant, called extremal length, can be constructed using data from 
the minimal-area metric.

The extremal length $\lambda$~\cite{ahlfors}
is defined  for a surface with a set of curves $\Gamma$:
\begin{equation}
\lambda (\Gamma) \ \equiv \  \sup_\rho \Bigl( {L^2(\Gamma) \over A(\rho)}\Bigr)\,, \quad  L(\Gamma) \ \equiv \ \inf_{\gamma \in \Gamma} L( \gamma, \rho)  \, .  
\end{equation}
For a given choice of metric $\rho$ the length $L(\Gamma)$ is the length of the 
shortest curve $\gamma$ in the set $\Gamma$,  and $A(\rho)$ is the area of the surface in that metric.  The ratio $L^2(\Gamma) /A(\rho)$ is manifestly
invariant under scaling of the metric by an arbitrary constant.  After computing this ratio one searches over all possible metrics to find the {\em largest}
possible value for the ratio.   The extremal length is determined by the minimal
area metric as follows.  Each metric $\rho$ can be rescaled
so that $L(\Gamma) = \ell_s$, where $\ell_s$ is some arbitrary
length,  now the length of the shortest curve in $\Gamma$.  We then have 
\begin{equation}
\lambda (\Gamma) \ =  \  \sup_\rho \Bigl( {\ell_s^2  \over A(\rho)}\Bigr)\,.    
\end{equation}
Since the supremum is achieved by having $A(\rho)$ as small as possible, 
$\lambda$ is the ratio of $\ell_s$-squared divided
by the area $A$ of the least-area metric for which the shortest
curve in $\Gamma$ is longer than or equal to $\ell_s$:
\begin{equation}
\lambda \ =  \   {\ell_s^2  \over A} \,.   
\end{equation}
For our problem the surface is the torus with a boundary
 and  $\Gamma$ is the set of
curves in the two homology classes specified earlier.  
Since the shape of the torus with 
a boundary is fully determined
by the parameter $t$, the conformal invariant $\lambda$ must ultimately be a function of $t$:
\begin{equation}
\lambda(t)  \ =  \   {\ell_s^2  \over A(t)} \,.   
\end{equation}
For a torus defined by some value of $t$, the area $A(t)$ of the
 minimal-area metric determines the
extremal length $\lambda(t)$. 
Recall that for the extremal metric we also have, from (\ref{area-heights}),  
\begin{equation}
A \ = \ 2 \ell_s  \nu  \,,
\end{equation}
where $\nu$ is the optimal common parameter for the bands of geodesics.
It is also, as explained around (\ref{identify-nu-height}), the length in the extremal metric 
of the terminal edges of the Swiss cross
presentation, where 
there is only one foliation. 
 These edges have parameter length $a$ 
in our construction of the surface.   From the last two equations the conformal invariant $\lambda(t)$ can be given in terms of~$\nu$:
\begin{equation}
\label{lambdanu}
\lambda(t)  \ = \ {\ell_s \over 2 \nu}  \,,   \quad \hbox{and}  \quad   t = {\ell_s \over a}  \in (1, \infty)\,.  
\end{equation}
With $\nu$ the length of the edge $a$ of the Swiss cross, we expect
$\nu \to a$ as $\ell_s \to \infty$ because $\rho \to 1$ on the arms of the
cross.  Our goal now is to determine the function $\lambda (t)$ for $t\in (1, \infty)$.
We will see that $\lambda$ is a monotonic function of $t$, meaning that $t$ is 
indeed a modulus. Moreover, we find that to first approximation
the relation between $\lambda$ and $t$ is surprisingly simple.  This relation,
however, is not exact.

To determine  $\lambda(t)$ we solve for the 
extremal area for tori with several values of $t$ and then attempt to guess
a relationship.   For these calculations we take $a=1$ and vary $\ell_s$.
As a result 
\be
a = 1  \  \ \to   \  t= \ell_s\,, 
\ee
the modulus equals the systole.  In the table below we computed the extremal
area $A$ and the modulus $\lambda$ for integer values of $t$ from $2$ to $11$.
\begin{center}
 \begin{tabular}{|| c |c| c||} 
 \hline
 $t$  & $A$ & $\lambda(t) $ \\ [0.5ex] 
 \hline\hline
  2  & 2.80697 & 1.4250 \\
 \hline
3 & 4.67514 & 1.9251 \\
 \hline
 4  & 6.5977 & 2.4251 \\
 \hline
 5  & 8.5466 & 2.9251 \\
 \hline
 6  & 10.511 & 3.4252 \\
 \hline 
 7  & 12.484 & 3.9252 \\
 \hline
 8  & 14.463 & 4.4251 \\
 \hline
 9  & 16.446 & 4.9251 \\
 \hline
 10  & 18.433 & 5.4251 \\
 \hline
 11  & 20.422 & 5.9251 \\
 \hline
\end{tabular}
\end{center}

\begin{figure}[!ht]
\leavevmode
\begin{center}
\epsfysize=7.0cm
\epsfbox{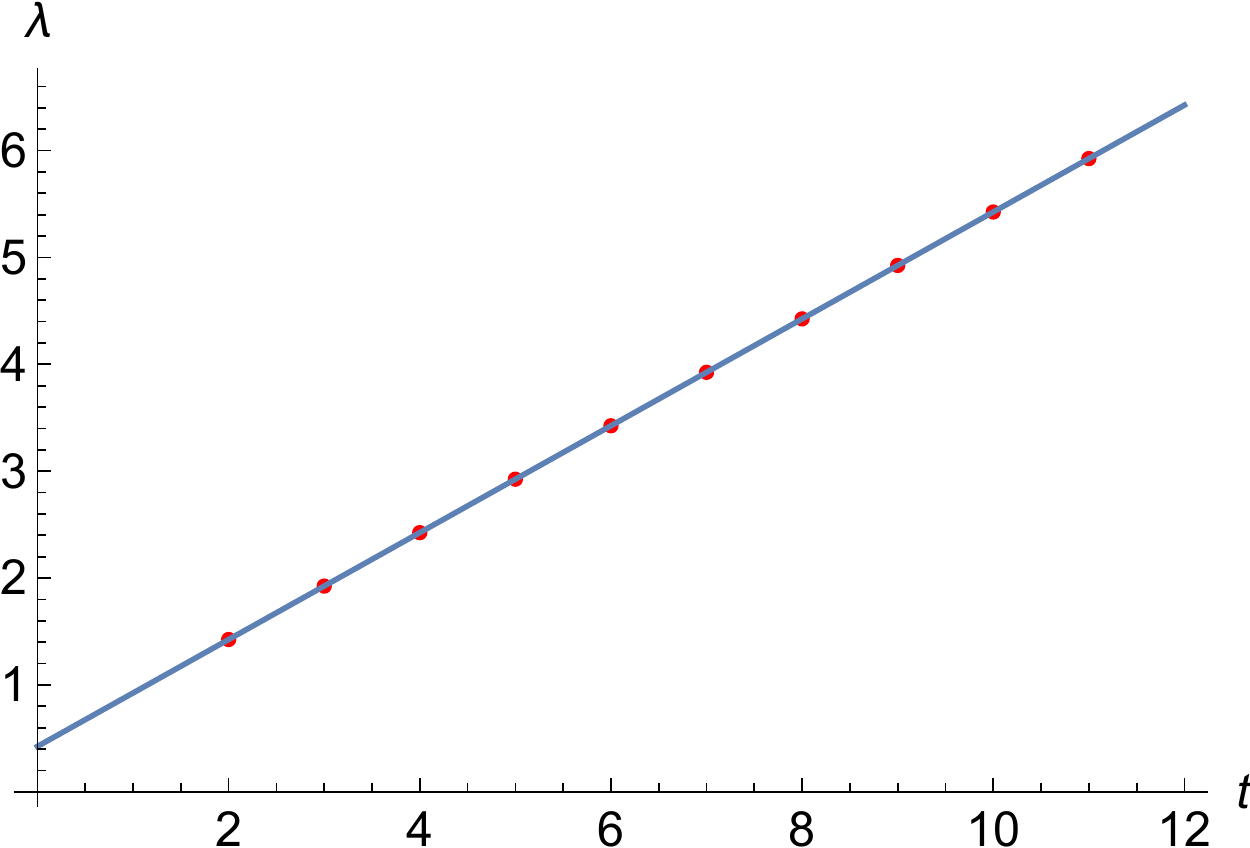}
\end{center}
\caption{\small  The extremal length $\lambda$ as a function of the 
parameter $t$ taking integer values from 2 to 11.  The blue line
is the linear fit to the data.}
\label{ffit}	
\end{figure}

The above data is shown in  Figure \ref{ffit}.  A linear fit seems appropriate and gives
\begin{equation}
\label{01-6403}
\lambda(t) \  \simeq \  0.425052 + 0.50001 \,   t \,. 
\end{equation}
This led us to ask: Can it be that we have an exact relation of the form
\be
\label{not-true}
\lambda (t) =\lambda_0  +  \tfrac{t}{2}  \ ? 
\ee 
Here $\lambda_0$ would be a constant
with value of about $0.425$.    First note that this cannot
hold for all $t\in [1,\infty]$.  As $t\to 1$ the Swiss cross becomes a square
with side length equal to one, the minimal-area metric is $\rho=1$ and
the extremal length $\lambda(1) = 1$.  This is certainly not consistent
with (\ref{not-true}).  Still, it is conceivable that for $t$ greater than some
critical value $t_0$ a linear relation could be exact.  The critical value 
$t_0\simeq 1.73$ is such that for $t > t_0$ the edges of the Swiss cross
lie beyond the region with curvature.    Even for $t > t_0$  
we believe that (\ref{not-true}) is not exact.  
Numerical exploration, which we describe now, 
suggests that with the assumption of a linear relation,
the slope is close but not exactly equal to $1/2$.  

For $\ell_s = 2$, the highest resolution data from the dual program   
gives
\be
A (\ell_s=2) \  = \ 2.806\,970\,011 \,,  \quad  |\epsilon| < 4\times 10^{-6}\,,
\ee
where the absolute error $|\epsilon|$ is estimated from the convergence rate
as the lattice size is increased.  In fact, the difference between the above value and the
value obtained from the primal is $3\times 10^{-6}$. 
For $\ell_s = 3$, the highest resolution data from the dual program   
gives
\be
A (\ell_s=3) \  = \ 4.675\, 146\, 916\, \,,  \quad  |\epsilon| < 4\times 10^{-5}\,,
\ee
The estimate here is rather conservative, the error is probably less
than $1\times 10^{-5}$; indeed the difference between the above value
for the area and the value obtained from the primal is $6\times 10^{-7}$.
Given this information we get
\be
\lambda(3) - \lambda (2) =  0.500\, 05 \,\, \pm \, 0.000\, 02 \,.
\ee
This is not consistent with $\lambda(3) - \lambda (2)=0.5$,
as would be predicted by (\ref{not-true}), thus making clear 
that such a simple linear relation between $\lambda$ and $t$ 
does not hold.
   
\begin{figure}[!ht]
\leavevmode
\begin{center}
\epsfysize=4.5cm
\epsfbox{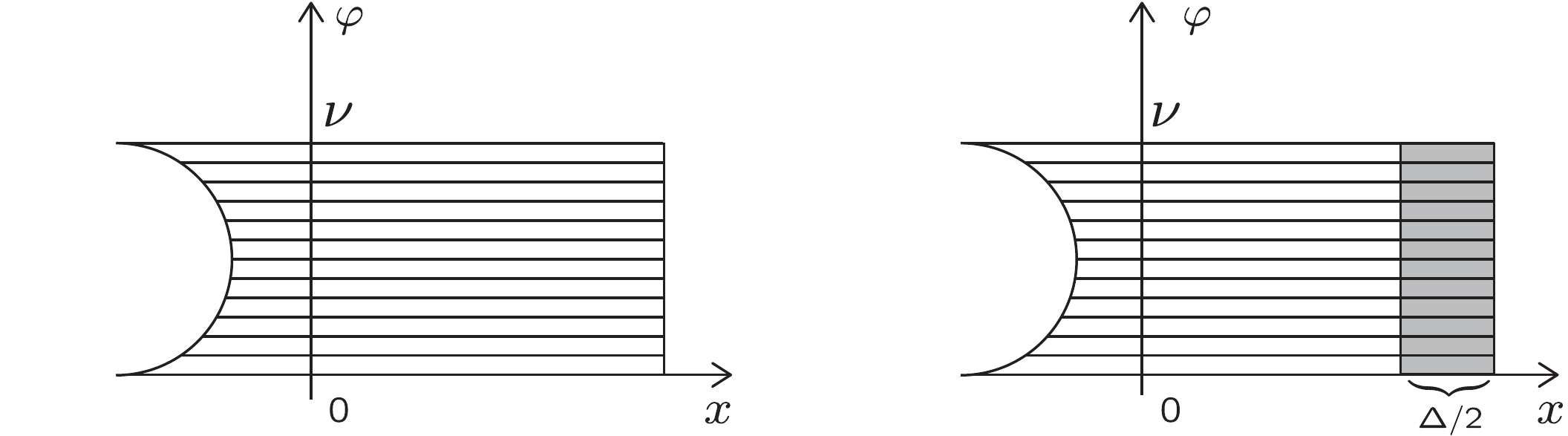}
\end{center}
\caption{\small Left: A region on the Swiss cross where there is a single band of 
geodesics, presented in a $w$-frame where $\rho=1$ and $w= x+ i \varphi$. 
The region is a piece of a rectangle.  Right: Amputating the surface by removing
in each $w$ plane the indicated small rectangle of length $\Delta/2$ and height $\nu$.}
\label{ffamp}
\end{figure}

  We have learned that the minimal-area metric on the Swiss cross 
  is positively curved on the region with a double foliation and
  flat in the four regions with a single foliation (see Figure~\ref{ffjn3}).  
  The two types of regions
  are separated by negative line curvature singularity.    Let us focus 
  on the regions with flat metric.  Having just one band of geodesics
  the metric in that region can be written as
  \be
  ds^2  =  dx^2  + d\varphi^2\,,
  \ee 
  and the region (stretching up to the right vertical edge of the 
  Swiss cross)  is shown to the left in Figure~\ref{ffamp}.  The region is 
  a piece of a rectangle:  the bottom and top horizontal boundaries are
  the curves $\varphi=0$ and $\varphi= \nu$, respectively and
  the geodesics are the (horizontal) curves of constant $\varphi$.  All the geodesics
  are orthogonal to the edge, which must then be a (vertical) line of constant $x$.
 In this region, where we can define the complex coordinate $w= x + i \varphi$, 
 we have $\rho=1$.  In the $z$ complex plane where the Swiss cross
 is defined, the numerical work indicates that the metric on the flat region is not 
 constant.  This means that the map between the $z$-plane flat region and
 the $w$-plane flat region of Figure~\ref{ffamp} is nontrivial. 
 
 The presentation using the $w$ plane, which can be used separately
 for each of the four flat regions of the Swiss cross, allows us to use
 amputation to constructe extremal metrics for Swiss crosses of different moduli.

Let $\Sigma$ denote a Swiss cross with extremal
length $\lambda$ and consider the
$w$ plane presentation of the flat sides of the surface with
the extremal metric. 
Let $\Sigma_\Delta$ denote
the surface obtained by cutting, on each of the four $w$ plane
presentations,  small rectangles of length
$\Delta/2$ and height $\nu$, while keeping the metric
the same (Figure~\ref{ffamp}).  The new surface $\Sigma_\Delta$ 
is still a Swiss cross because it is conformally equivalent to a round 
disk where the edges are invariant under orthogonal reflections.

\noindent
{\em Claim:} The metric induced 
on $\Sigma_\Delta$ 
is of minimal area under the condition
that all curves between the new vertical edges and a
between the new horizontal edges are longer than or equal to
$\ell^\Delta_s = \ell_s - \Delta$.   

\noindent{\em Proof:} First note that
all relevant curves in $\Sigma_\Delta$ have length greater than
or equal to $\ell^\Delta_s$ because if there was a shorter curve adding 
back the removed rectangles we could extend that curve to one
in the original surface $\Sigma$ with length less than $\ell_s$, which
is impossible.
Moreover, if there was a metric on $\Sigma_\Delta$ with less
area and still having all curves longer than or equal to $\ell^\Delta_s$
we could add back the rectangles and obtain a new metric with
less area on $\Sigma$.  \hfill $\square$
 
Using (\ref{lambdanu}) the new surface $\Sigma_\Delta$  has extremal length
\begin{equation}
\lambda_\Delta \ = \ \lambda -  {\Delta \over 2 \nu}\,.
\end{equation}
It is also clear that the new area is still equal to twice the systole
times the edge length:
\begin{equation}
A^\Delta =  A - 4 \tfrac{\Delta}{2} \nu  =  2 \ell_s \nu - 2 \Delta \nu
 = 2 (\ell_s - \Delta ) \nu \,. 
\end{equation}
If we know the $t$ parameter of the initial Swiss cross $\Sigma$, however,
the cutting procedure does not tell us the $t$ parameter of $\Sigma_\Delta$.
Indeed, since the mapping from the $z$ plane to the $w$ plane is nontrivial,
the $w$-plane cuts do not correspond to straight cuts in the $z$-plane
presentation and thus the new value of the parameter $t$ cannot be identified.

\subsection{Special tori in the moduli space} 
\label{sec:spetorinthemodspa}

\begin{figure}[!ht]
\leavevmode
\begin{center}
\epsfysize=6.5cm
\epsfbox{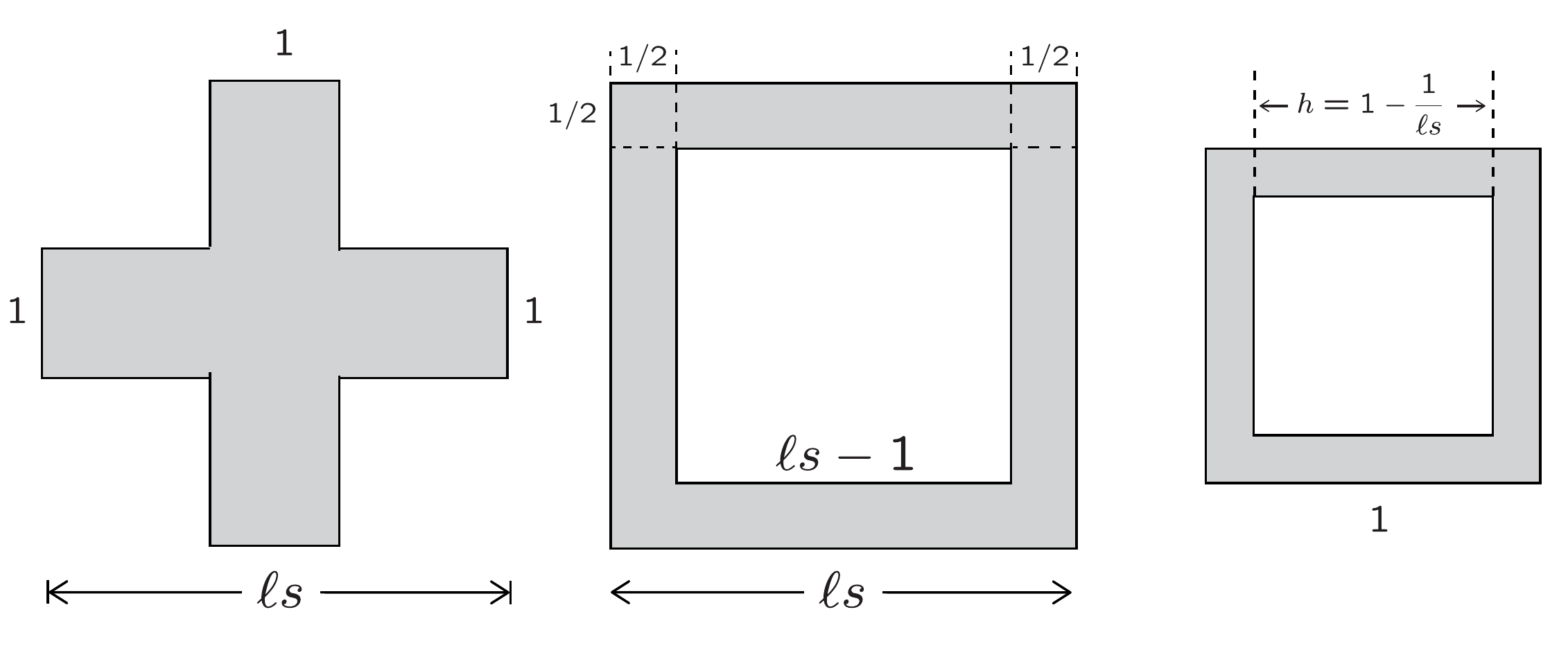}\hskip15pt
\end{center}
\caption{\small  A Swiss cross can be described in terms of $\ell_s$ (with $a=1$) or in terms of the size $h$ of the edge of the boundary when the systole is 
set equal to one. The right-most figure is obtained by scaling
down by $\ell_s$ all lengths in the center figure.}
\label{ells-to-h}
\end{figure}

Setting $a=1$,  the torus with a boundary, or Swiss cross, is determined by the parameter $\ell_s$, which we take to be equal
to the systole in the minimal-area problem.   
There is another parameterization of the various
surfaces that is useful when we want to keep the systole constant
and equal to one.  In that case we view the surface as a torus with a square
boundary and the parameter, called $h$, is the length of the edge
of the square boundary.  The relation between $\ell_s$ and $h$ with $a=1$
is illustrated in Figure~\ref{ells-to-h}: 
\be
\label{hfromls}
h \, = \, 1 - {1\over \ell_s} \,.
\ee 
The Swiss cross with $\ell_s=3$ corresponds to $h= 2/3$ while the
Swiss cross with $\ell_s=2$, which we used to describe much of our results,
corresponds to $h=1/2$, a boundary edge half as large as the systole.
In this section we will identify  two critical values of $h$ 
at which the
minimal-area metric undergoes interesting changes.

\begin{figure}[!ht]
\leavevmode
\begin{center}
\epsfysize=6.5cm
\epsfbox{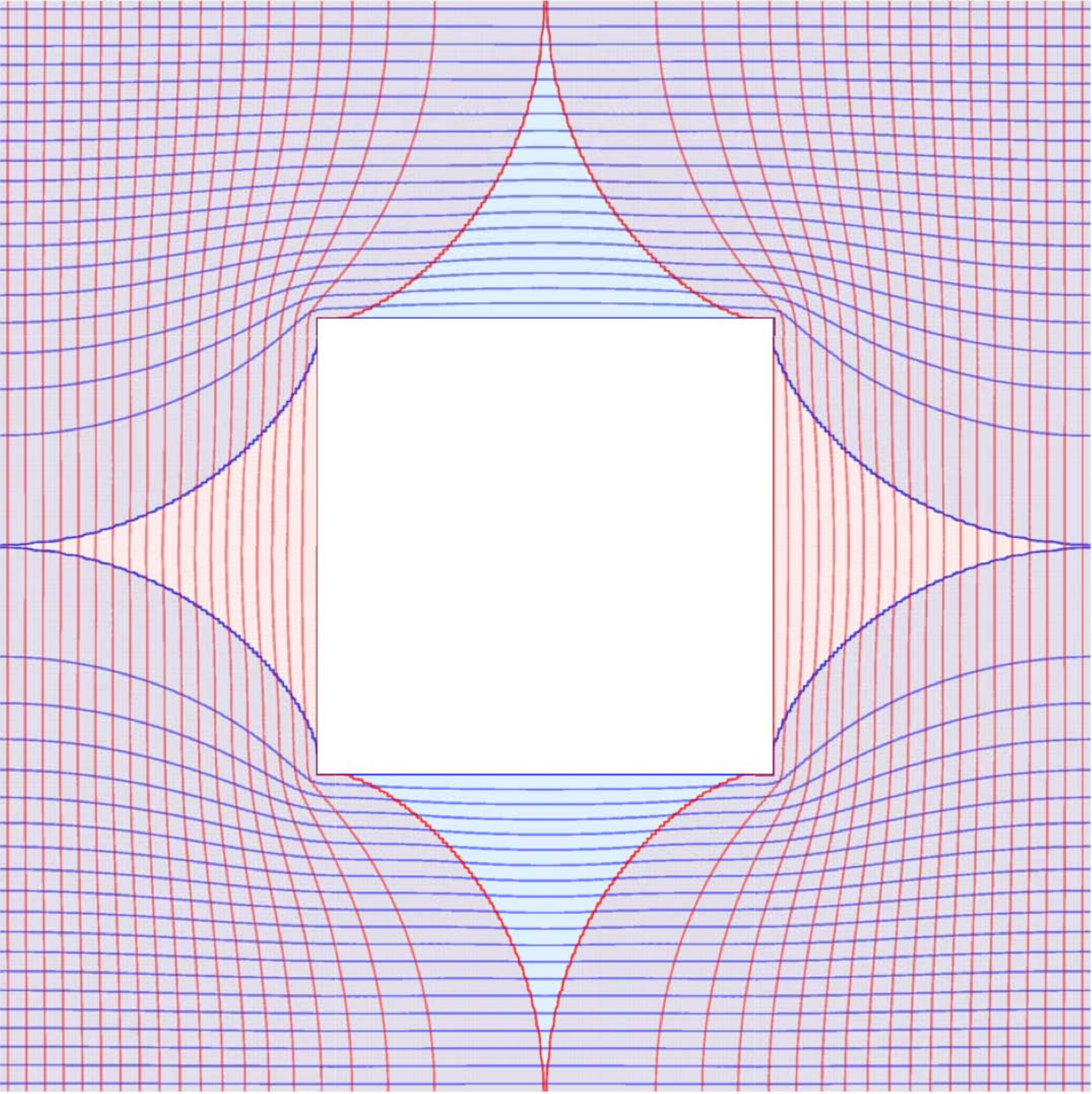}
\epsfysize=6.5cm
\hspace{1cm}
\epsfbox{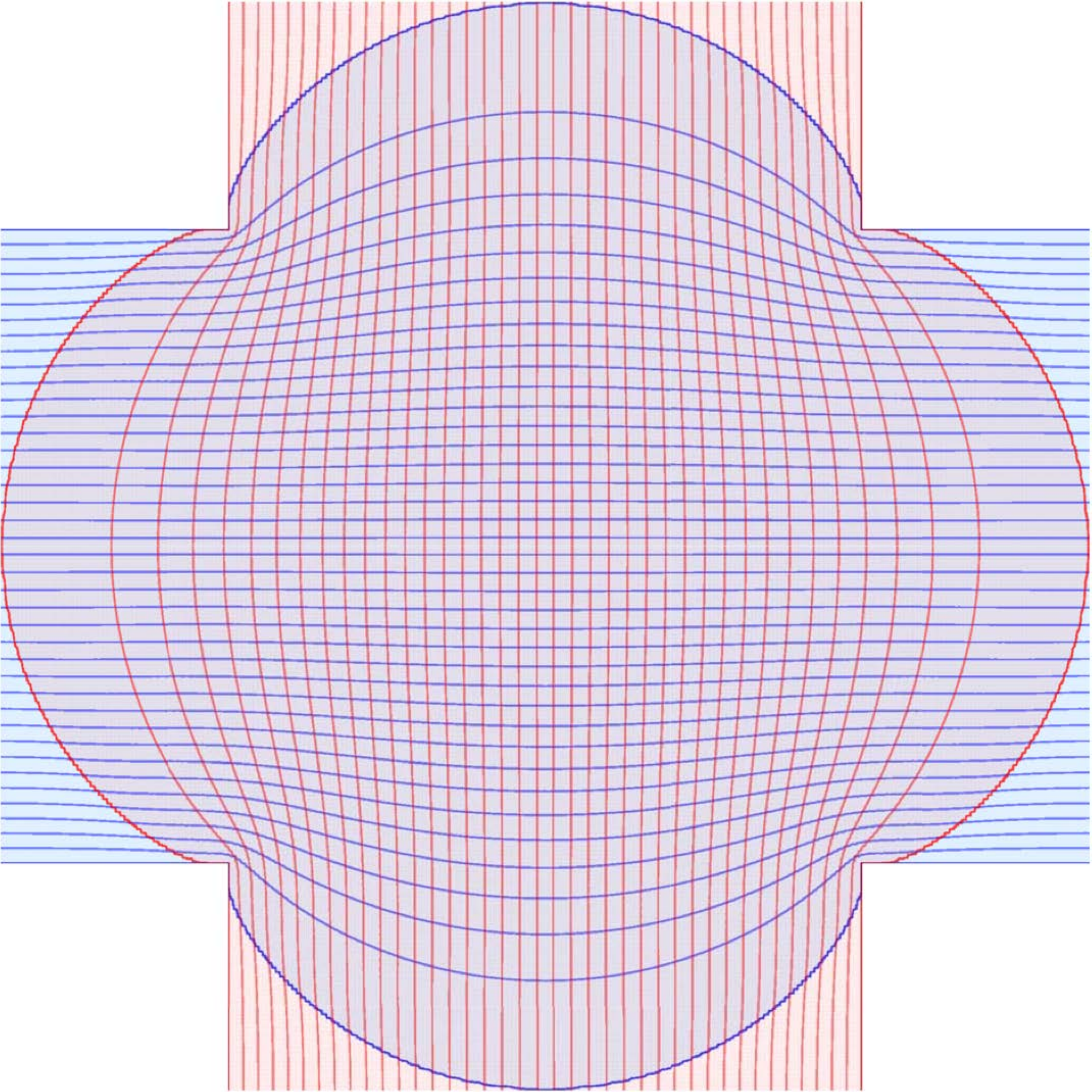}
\end{center}
\caption{\small  
Systolic geodesics for the torus (left) or Swiss cross (right) with 
$h= 54/129\simeq 0.4186$. This is close to the critical value $h^{(1)} \simeq 0.4201$ for which the dome touches the edge of the Swiss cross, or touches itself in the
torus picture.}
\label{fig:just-touch}
\end{figure}

As we have observed with the $h=2/3$ and $h=1/2$ solutions,
the metric in the Swiss cross picture 
has a central ``dome'' with positive Gaussian curvature bounded
by negative curvature line singularities.  The dome is the region covered
by two bands of systolic geodesics. As we reduce $h$ further,
at some point the line curvature singularities touch the edges of the Swiss cross.
This defines a critical value of $h$ that we have estimated at 
\be\label{ells1}
  h^{(1)} \simeq  0.4201  \,. 
\ee
This critical value was found by analyzing the value of the gradient
of the $\varphi$ function associated with the geodesics that are about to hit
the edge.  As we recall, a non-zero gradient at a point 
implies there is a systolic geodesic at the point.  We find  that as
$h$ approaches its critical value from below, 
the {\em gradient} at the edge midpoint decreases linearly to a
very good approximation. Beyond the critical value the gradient
becomes zero, 
indicating that the geodesics do not reach the end.  With
a few values of $h$ near the critical point, we estimated the critical
value by linear extrapolation.  The geodesics for $h=54/129$, just below $h^{(1)}$, are shown in Figure~\ref{fig:just-touch}.

The second critical point occurs for a value $h^{(2)}$ 
that defines a surface relevant to the minimal 
area problem of the
square torus with a puncture.   This problem is discussed in detail in the 
following section, but we note here a couple of useful facts.
In the torus-with-boundary problem studied so far, we have only constrained
the lengths of the curves in {\em two} homology cycles.  The closed curves
around the boundary are not constrained.  In a theory of closed and open
strings, closed curves homotopic to a boundary are also required to 
satisfy the length condition.   Since curves homotopic to the boundary
are homologically trivial, our methods require modifications that are
 explained in the  following section. For now assume we can impose this extra condition.
 
\begin{figure}[!ht]
\leavevmode
\begin{center}
\epsfysize=7.3cm
\epsfbox{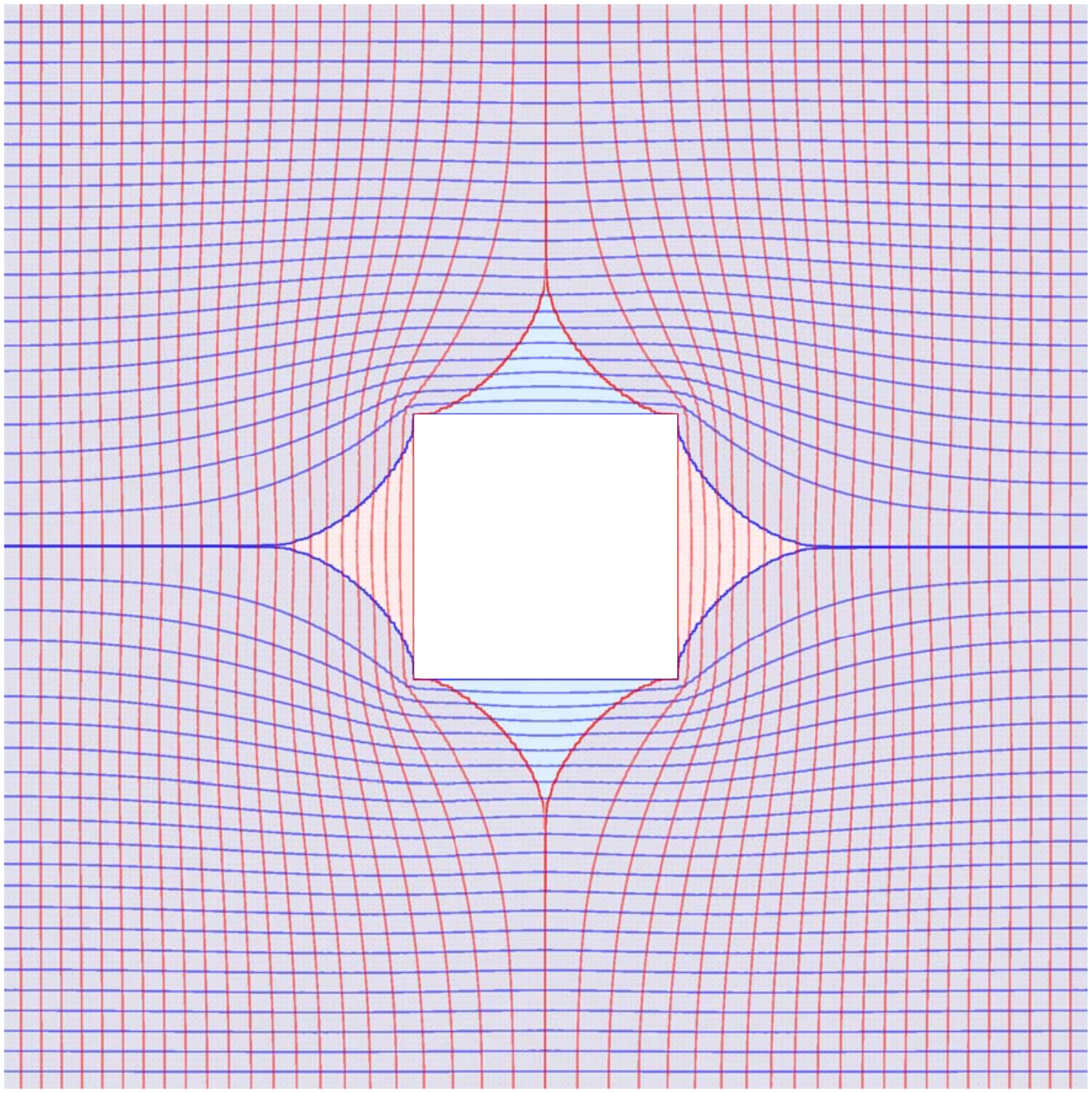}
\hspace{1cm}
\epsfysize=7.3cm
\epsfbox{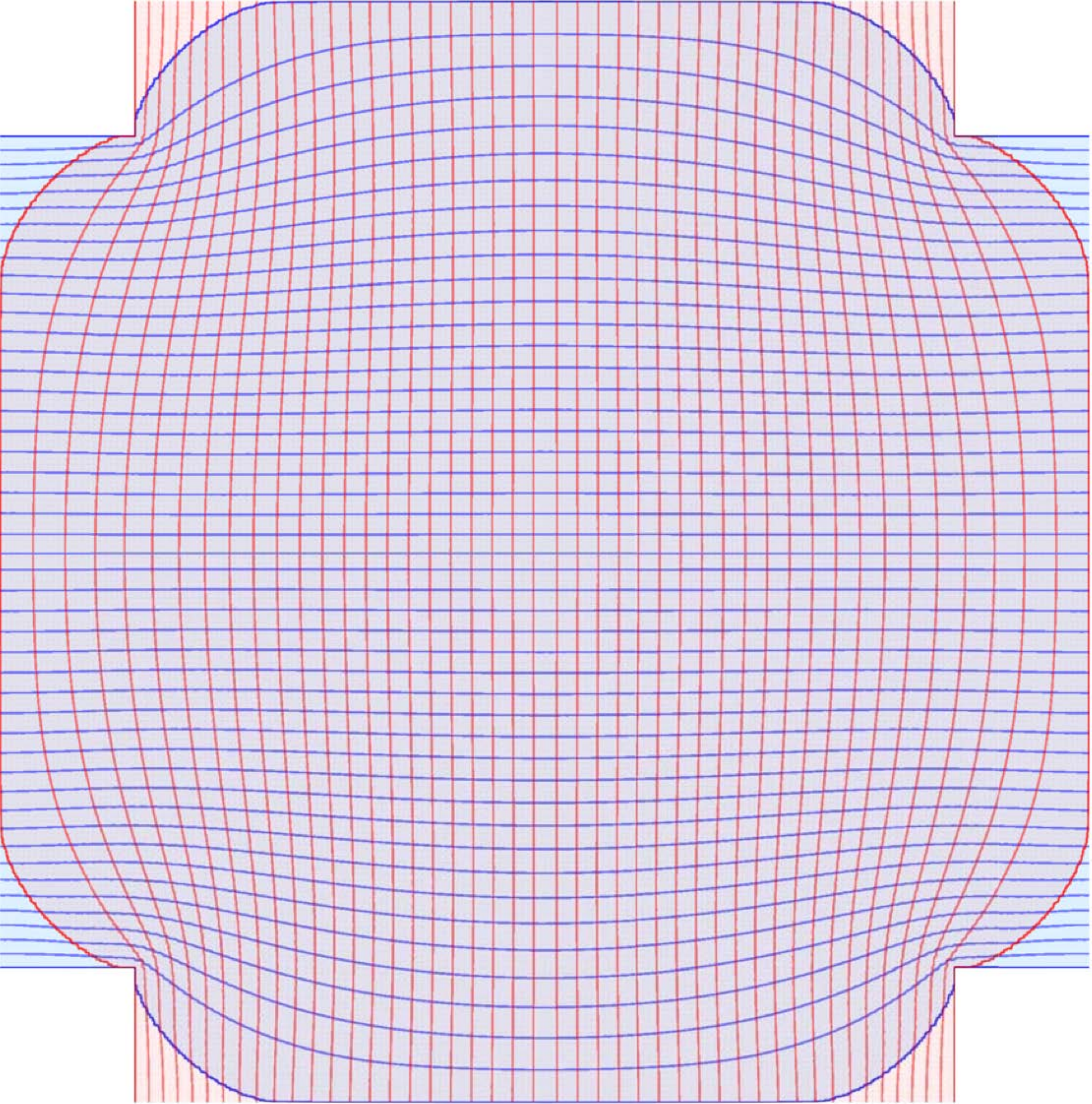}
\end{center}
\caption{\small  
Systolic geodesics for the torus (left) or Swiss cross (right) with $h=23/94$. This
is, to a good approximation, the surface for which the length $L$ of the boundary is in fact equal to the 
systole~1.
}
\label{fig:punct-torus-fig}
\end{figure}

We claim that for $h> h^{(2)}$, in the minimal-area metric with  {\em two} length conditions, the length $L$ of the boundary is in 
fact greater than the systole $1$.  Even more, all curves homotopic to the boundary are longer than $1$.   
This means that the solution of the two-condition problem
is in fact a solution for the problem with an extra length condition on  the curves homotopic to the boundary.  As $h$ approaches $h^{(2)}$ from above, $L$ approaches 1 from above.  
An analysis performed in 
section~\ref{sec:PTresults} implies that 
$L=1$ for $h=h^{(2)}$ where
\be\label{ells2}
h^{(2)} = 0.24469\,. 
\ee 
Numerical analysis requires
a rational representation of $h$ and we can take $h = 23/94 \simeq 0.24468$.  The geodesics for this torus are shown in Figure \ref{fig:punct-torus-fig}.  Note that the boundary is surrounded by
 regions of flat metric with a single band of geodesics.  It is curious that $h^{(2)}\sim 1/4$,  the value for which the flat 
$\rho=1$ metric makes the length of the boundary
equal to the systole.

For $h < h^{(2)}$ the two-condition minimal
 area metric has curves homotopic to the
 boundary that are shorter than $1$, and this metric is not
 a solution for the string field theory problem.
 If we imposed an additional  length condition on curves homotopic to the boundary
 the minimal-area metric would have larger area and it would have a new
 foliation by curves homotopic to the boundary.  Thus the value $h^{(2)}$
 is a critical point in the moduli space:  the solution of both problems is the
 same for $h\geq h^{(2)}$  and the solution
 is different for $h< h^{(2)}$.

More detailed exploration reveals a new feature
occurring below the value of the {\em first} critical point.
We find that for $h< h^{(1)}$ the extremal
metric on the two-foliation dome 
features a region with {\em negative curvature} in addition to the region
with positive curvature.   
As the dome collides with itself, the affected portion of its boundary becomes smooth; the line curvature is gone. Negative curvature 
spreads out becoming bulk curvature.   Although a more detailed
description will be given in the following section, 
given that $h^{(2)} < h^{(1)}$, the surface displayed in 
Figure \ref{fig:punct-torus-fig} has both positive and negative bulk 
curvature.  The negative curvature in the Swiss cross presentation (right)
exists near the four edges, on each one on a neighborhood of the piece 
of the  `last' systolic geodesic that runs along the edge.  In the torus picture these segments are the central vertical curve
(red)  and the central horizontal curve (blue).     The bulk negative
curvature exists in some neighborhood of those curves.

\sectiono{The once-punctured torus}
\label{once-p-torusvm}

In this section we determine the minimal-area metric for a square torus
with a puncture.  This is perhaps the simplest unknown metric in closed
string field theory.  This problem is also a natural next step after
our study of the torus with a boundary.  Indeed, we may view the
punctured square torus as the limit of the square torus with a square
boundary considered so far, when the size of the
boundary goes to zero.  
For the once-punctured torus, curves homotopic to the puncture are
of nontrivial homotopy and must be constrained in
the minimal-area problem of closed string field theory.   Such curves, however,
are homologically trivial and one cannot have a calibration $u$ with nonzero
integral along them.  In terms of the conserved flow associated to
a closed form,  a nonzero integral implies a net flow across any curve surrounding the puncture.  But such a flow, sourced at the puncture, has nowhere to go on the surface because there is no possible sink.

We will use a device proposed in \cite{headrick-zwiebach} to deal with this complication.  We consider some covering surface
$(\tilde M, p)$ of the original surface $M$, where $p: \tilde M \to M$ 
is  the projection map.
In $\tilde M$ the curves surrounding the puncture become homologically
nontrivial.  The desired calibrations on the covering surface $\tilde M$ are such
that at points $\tilde x$ that are projected down to $x \in M$ the 
one-form $u(\tilde x)$ is equal to  $u(x)$ or $-u(x)$.   Much of the discussion in 
\cite{headrick-zwiebach} was couched in terms of a double cover, which
in principle suffices to deal with the situation.  In that double cover
of the punctured torus, we have two punctures and the
curves surrounding the punctures are homologically nontrivial,
allowing a calibration to exist.  In terms of flows, one
puncture can act as a source while the other will act as a sink.  We will
discuss below the double cover briefly. It turns out, however, that
a fourfold cover is far more convenient.  The double cover breaks some of the discrete symmetries of the original punctured square torus.  
On the other hand the fourfold cover breaks none of the symmetries.
In the fourfold cover we have a torus with four punctures; two will be sources
and two will be sinks.  The calibration is indeed the same, up to signs, on points
that project to the same point on the original surface $M$.

We will first discuss the formulation of the programs and then our results.

\subsection{Programs for the  punctured  square torus}
\label{sec:puncturedtorusprograms}

For the construction of the requisite one-forms on the double and fourfold cover of the torus, we will use the Weierstrass
zeta function $\zeta (z)$ defined in terms of the {\em half periods} $\omega_1$ and $\omega_2$ of a torus $z\sim z + 2\omega_1$ and $z \sim z + 2\omega_2$.  
This zeta function is odd
\be
\zeta (z) \ = \ - \zeta (-z)\,,
\ee
and it has a single pole at $z=0$ in the fundamental domain of the torus
\be
\zeta (z) = \, {1\over z}  + \hbox{regular}\,,  \qquad \hbox{near}  \ z=0\,.
\ee 
The function is quasi-periodic on the torus:
\be
\begin{split}
\zeta (z + 2\omega_1) \ = \ &  \zeta (z) +  2 \eta_1\,, \\
\zeta (z + 2\omega_2) \ = \ &  \zeta (z) +  2 \eta_2\,, \\
\end{split}
\ee
with constants $\eta_1 = \zeta (\omega_1)$ and $\eta_2 = \zeta(\omega_2)$.
While the dependence of the zeta function on the half periods is usually omitted,
some authors use $\zeta (z ; \omega_1, \omega_2)$ or $\zeta (z , g_2, g_3)$,
with $g_2(\omega_1, \omega_2)$ and $g_3(\omega_1, \omega_2)$ some functions of $\omega_1$ and $\omega_2$.

For the minimal-area analysis of the punctured torus we fix $\ell_s=1$.
In practice, when we discretize, we deal with the puncture by 
replacing it by a small square hole, thus
going back to the problem of a torus with a boundary.  The curves surrounding 
the boundary are constrained and the size of the boundary will be
reduced.  With the punctured torus being the unit square with identifications,
we take the puncture to be at the center $\lambda$  of the square.  In summary:
\be
\hbox{Punctured torus:}  \quad 
2\omega_1 = 1 \,,  \quad 2 \omega_2 =  i\,,  \quad \lambda= \tfrac{1}{2} + \tfrac{1}{2}i  \,. 
\ee

\begin{figure}[!ht]
\leavevmode
\begin{center}
\epsfysize=7.5cm
\epsfbox{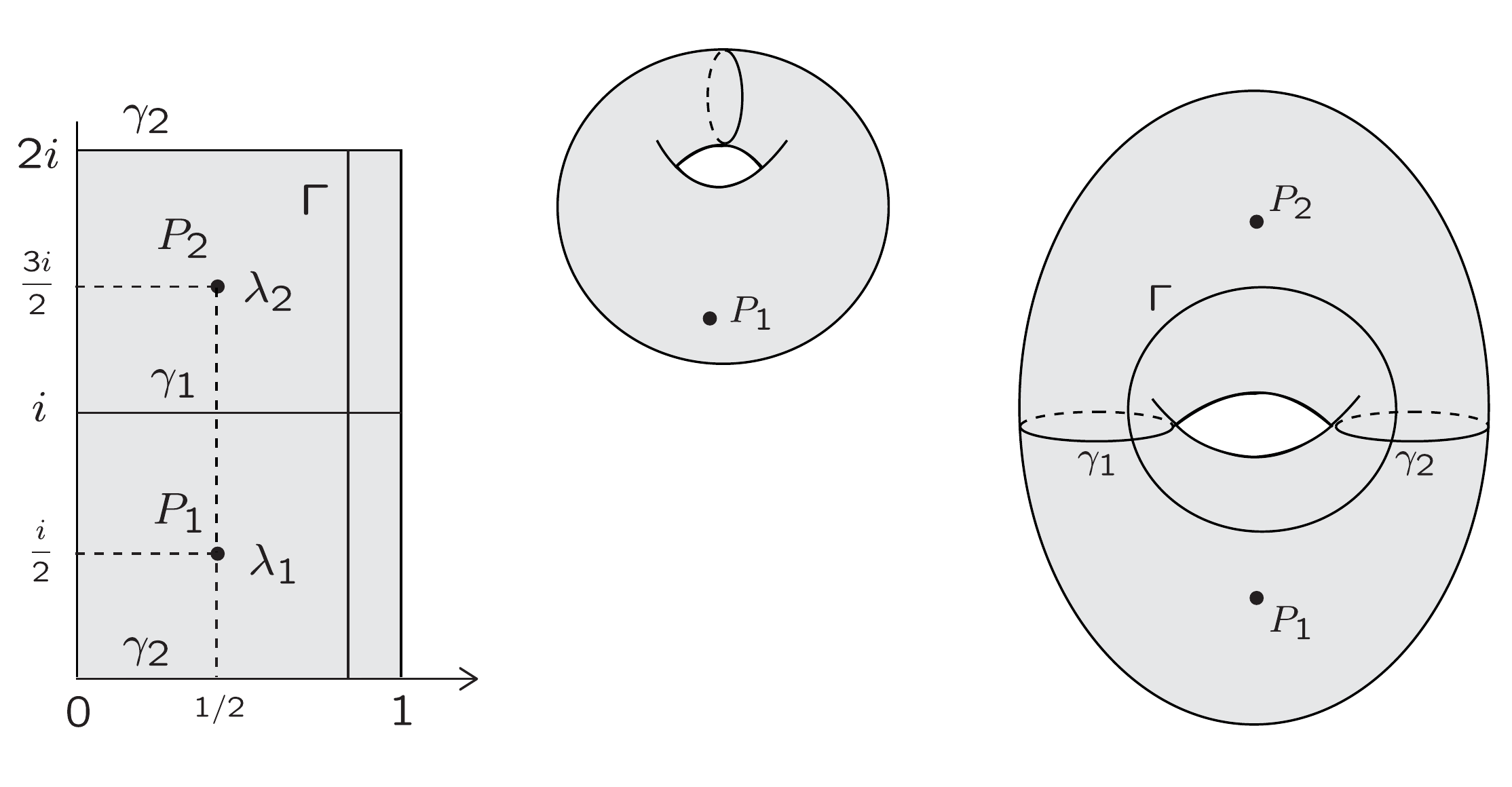}\hskip15pt
\end{center}
\caption{\small  Center: A punctured torus and a cutting curve.  Once the torus is cut, the two new boundaries $\gamma_1$ and $\gamma_2$ are used to glue a second copy of the torus, as shown to the right and to the left.  }
\label{d-cover}
\end{figure}

\subsubsection{Double cover}

The double cover is itself a torus with half periods
and puncture locations $\lambda_1$ and $\lambda_2$ as follows:
\be 
\begin{split}
\hbox{Double cover:} \quad  2\omega_1 = & \  1 \,,  \quad 2 \omega_2 = 2 i\,, \\
\lambda_1 = \ & \ \tfrac{1}{2} +  \tfrac{1}{2} \, i \,, \quad  
\lambda_2 =  \  \tfrac{1}{2} +  \tfrac{3}{2} \, i \,. 
\end{split}
\ee
The double cover is shown in Figure~\ref{d-cover}.  One can imagine cutting
open the original punctured torus (center) along a curve that corresponds
to the identified horizontal segments at $y=0$ and $y= i$ (left).  
A copy of the punctured torus is glued across those curves to obtain the torus shown on the right and left.   Since the punctures must be 
treated like boundaries, it is clear that in the double cover the curves homotopic
to the punctures are, as desired, homologically nontrivial.

\begin{figure}[!ht]
\leavevmode
\begin{center}
\epsfysize=8.3cm
\epsfbox{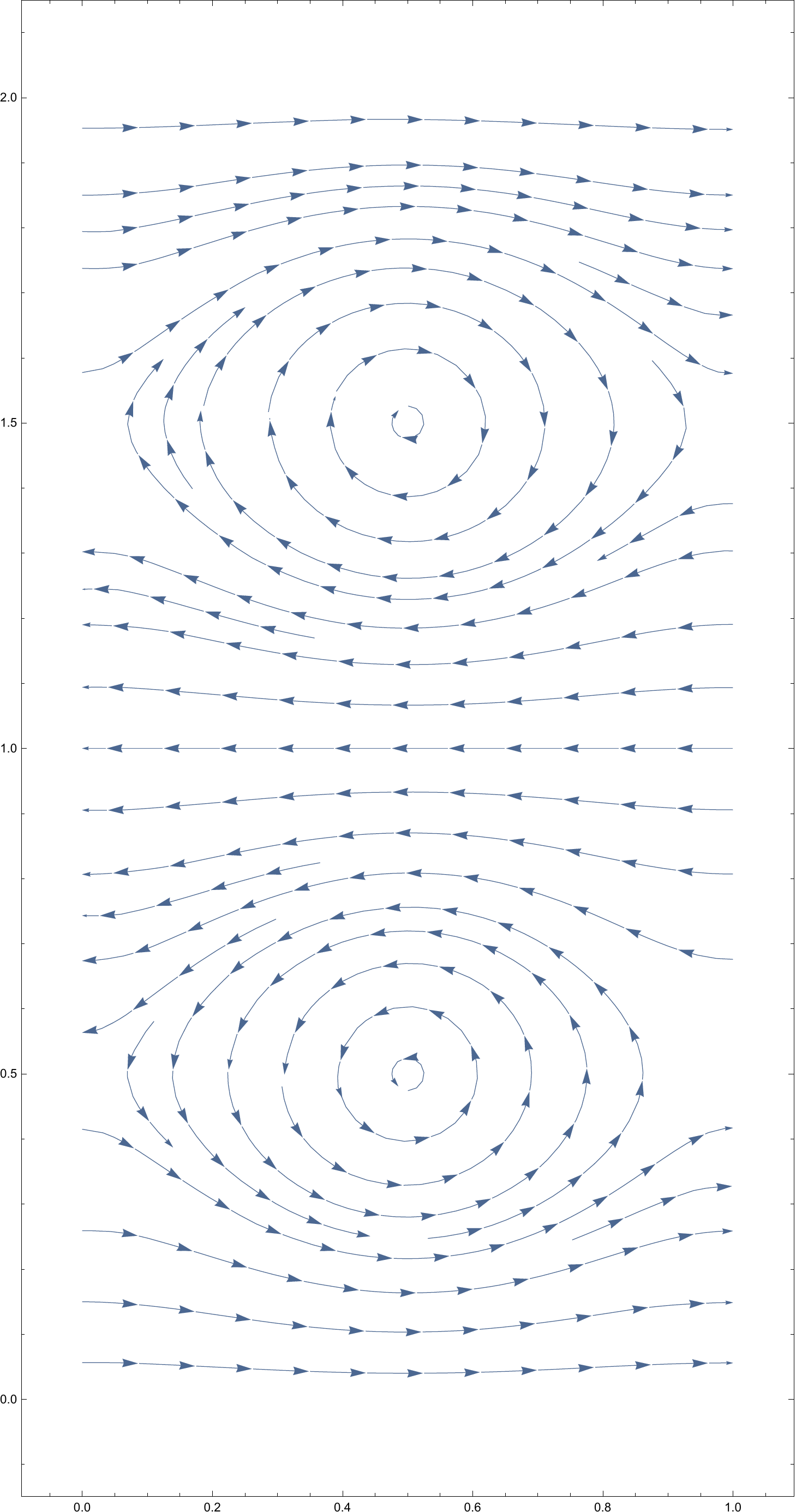}\hskip15pt
\epsfysize=8.3cm
\epsfbox{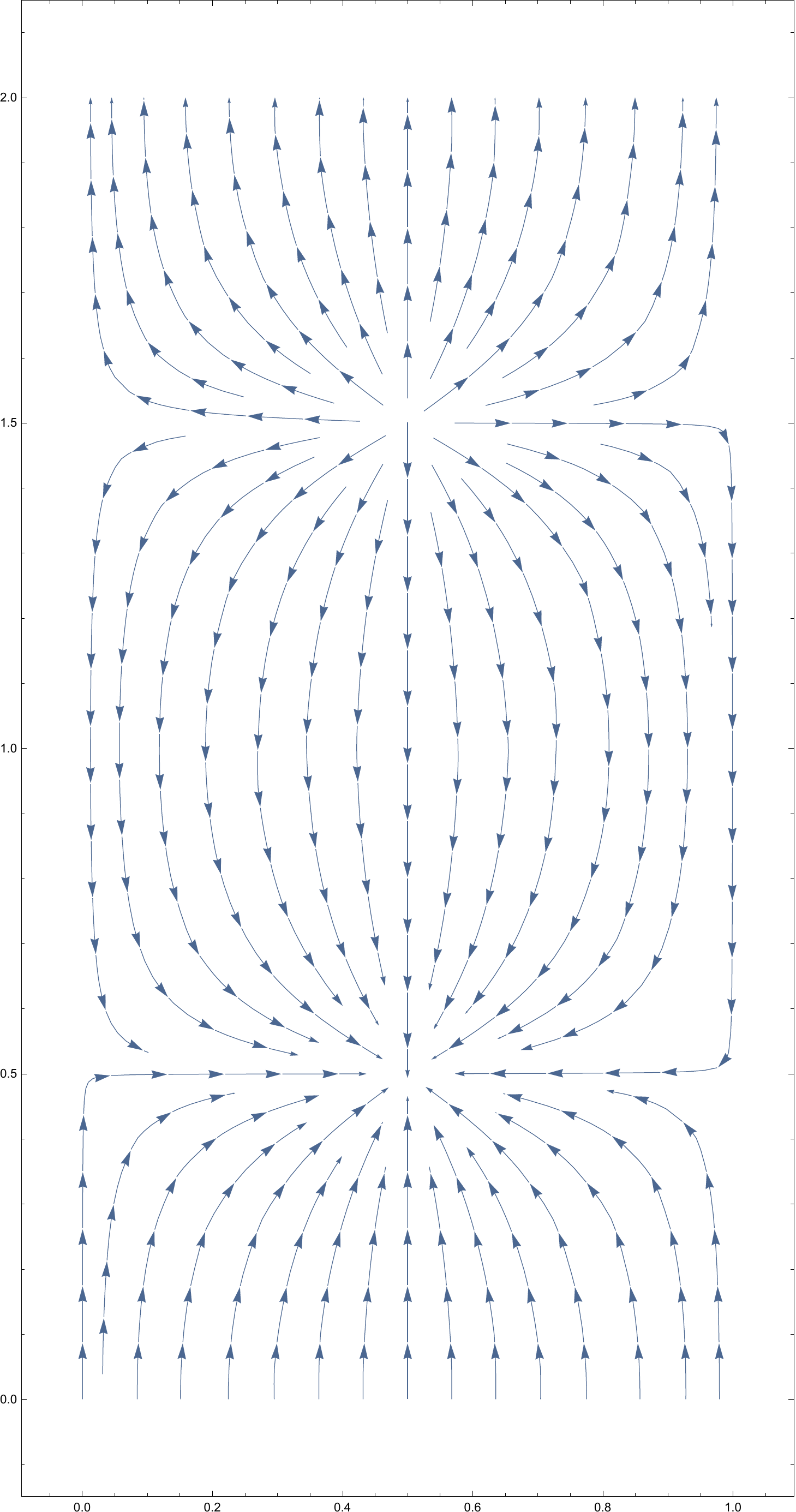}
\end{center}
\caption{\small  The meromorphic abelian differential $\omega_{\rm dc}$ 
for the double cover of the punctured torus is that of a light-cone diagram. 
Its real part is the one-form represented in this figure.   Left: The vector field lines associated with the one-form.   Right:  The vector field lines for  the associated flow.  In this diagram the square symmetry of the original punctured square torus is lost.   On the left one can see the bottom (incoming) string splitting into two strings, one upwards and one downwards, that later merge to form the top (outgoing) string.}
\label{punct-torus-double-cover}
\end{figure}

To construct the primal we need a calibration for each of the homologies
we want to constrain.  We have already discussed 
in section~\ref{sec:primalprogram} the one-forms $u^1$ and $u^2$ for curves that run horizontally or vertically on the original torus.  These need no
modification.
We now need to consider a one-form  $u^3$ to constrain the curves 
that surround the punctures.  Following (\ref{calibration-ansatz}) we write $u^3$ on the double torus
as follows
\be
\label{dc-ansatz-vmgh}
u^3 \ = \  \omega^{(3)}  +  c_1 dx  + c_2 dy  +  d \phi^3 \,.
\ee 
Here $c_1$ and $c_2$ are constants and the last term is an exact
form written in terms of some unknown function $\phi^3$. 
The closed form $\omega^{(3)}$ must have
the correct unit period along curves homologous to the puncture $P_1$
\be
\oint_{P_1}  \omega^{(3)}  \ = \ 1 \,. 
\ee
This means that, necessarily
 \be
\oint_{P_2}  \omega^{(3)}  \ = \ -1 \,. 
\ee
For the program one must determine the form $\omega^{(3)}$ explicitly,  
while the constants $c_1, c_2$ and the function 
$\phi^3$ are determined from the minimization
of the objective.   To find $\omega^{(3)}$ we first construct a meromorphic
abelian one-form $\omega_{\rm dc}$, with `dc' standing for double cover.
The requisite form, we claim, is given by 
\be
\label{dc-canform}
\omega_{\rm dc} (z) \ = \ \Bigl( \zeta (z - \lambda_1) - \zeta(z-\lambda_2) - \zeta (i)\, \Bigr) 
{dz \over 2\pi i }  \,. 
\ee
The first two terms  guarantee exact periodicity on the torus:
$\omega_{\rm dc} (z+1) = \omega_{\rm dc}(z+2i) = \omega_{\rm dc}(z)$.  
Because it has unit residues at the punctures we have
\be
\oint_{\lambda_1}  \omega_{\rm dc}  \ = \ 1 \,, \qquad
\oint_{\lambda_2}  \omega_{\rm dc}  \ = \ -1 \,.
\ee
 The constant last term in (\ref{dc-canform})
ensures the anti-periodicity
\be
\omega_{\rm dc} (z + i ) \ = \ - \omega_{\rm dc} (z)\,,
\ee
where $z+i$ and $z$ are related by the projection in the double cover. 
The form $\omega_{\rm dc}$ satisfies all the constraints we want from
$\omega^{(3)}$ except that it is not a real one-form (a form $f dx + g dy$ with
real $f$ and $g$).  The primal is formulated using real forms.   This is no
complication:  since the periods of $\omega_{\rm dc}$ are real we can 
simply set $\omega^{(3)}$ equal to the real part of the meromorphic
one-form $\omega_{\rm dc}$:
\be
\omega^{(3)} \ \equiv  \hbox{Re} \bigl(  \omega_{\rm dc} \bigr)  \,.  
\ee 
Thus defined, $\omega^{(3)}$ has the right periods, the proper anti-periodicity
and is well defined on the double torus.   It is clear in the ansatz (\ref{dc-ansatz-vmgh})  that $c_1 = c_2= 0 $ since the forms $dx$ and $dy$ do not have the 
requisite anti periodicity under $y \to y + 1$.  Our ansatz then becomes 
\be
\label{dc-ansatz-vmgh-99}
u^3 \ = \  \hbox{Re} \bigl(  \omega_{\rm dc} \bigr)  +  d \phi^3 \,.
\ee 
It is interesting to note that $\omega_{\rm dc}$ is in fact the abelian
differential that defines a one-loop light-cone diagram with one incoming
state and one outgoing state.  Indeed, this one-form only has periods 
around the punctures and around the cycles corresponding to the curves
$\gamma_1$ and $\gamma_2$ of Figure~\ref{d-cover}.  We can visualize
$\omega_{\rm dc}$ by plotting its real part $h_x dx + h_y dy$ 
as vector field lines with tangent $(h_x, h_y)$.  This is shown 
on the left of Figure~\ref{punct-torus-double-cover}.   To the right
we show the associated flow, using the dual form $-h_y dx + h_x dy$.
The bottom puncture is a sink
and the top puncture is a source.   The figures make clear that
the discrete symmetries of the original punctured square torus are partially
broken:  neither the form nor the flow, restricted to the original torus, 
are invariant under rotations by $\pi/2$.   
As we will see now we can get a simpler
formulation of the problem using a fourfold cover.

\begin{figure}[!ht]
\leavevmode
\begin{center}
\epsfysize=8.3cm
\epsfbox{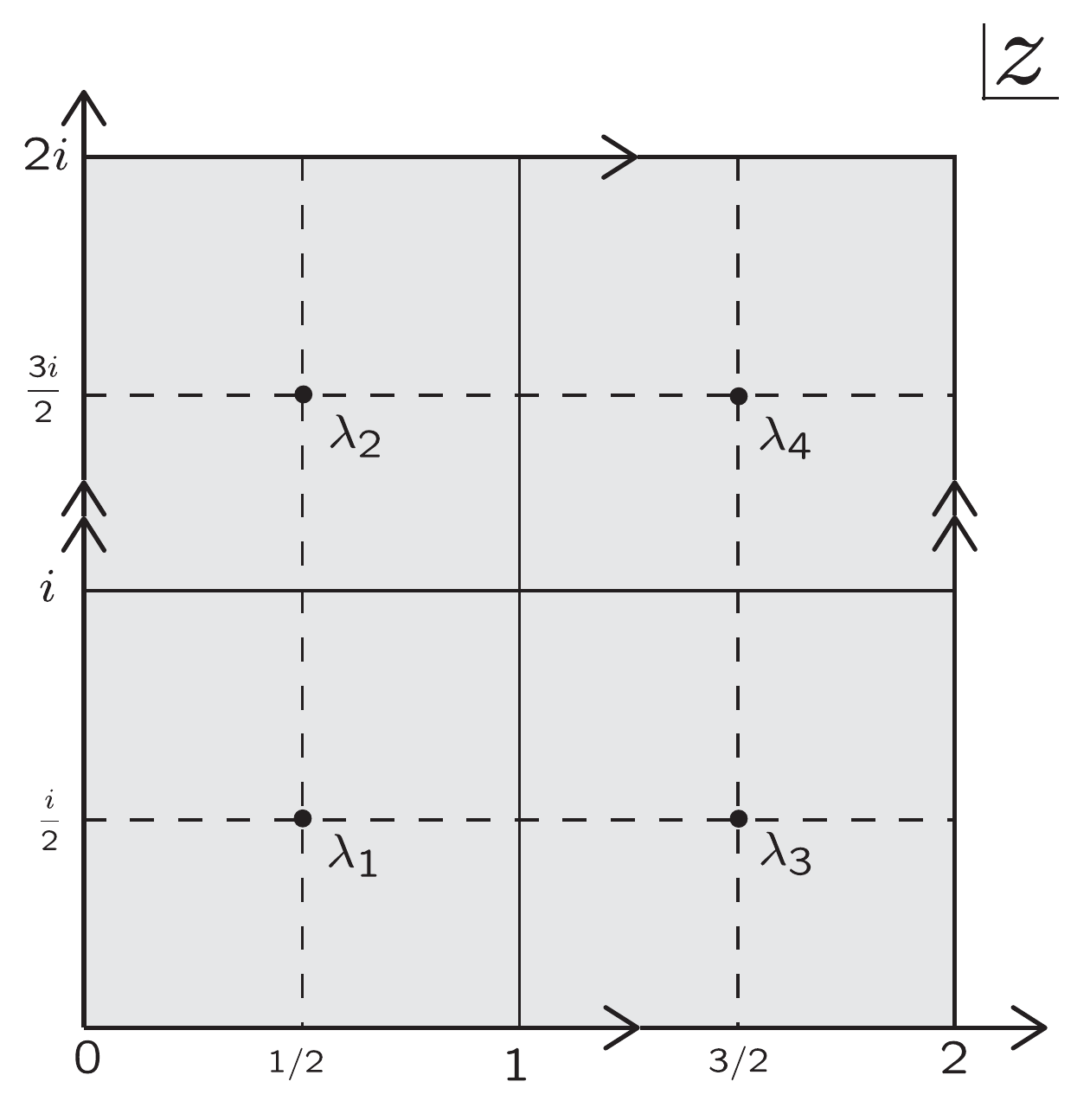}\hskip15pt
\end{center}
\caption{\small  The fourfold cover of the original punctured square torus $z\sim z+1$ and $z \sim z+ i$. 
The result is a torus twice as large, with identifications $z \sim z+2$ and $z \sim z+ 2i$.  There are four punctures, located at $\lambda_1, \lambda_2, \lambda_3, \lambda_4$.}
\label{fc-cover}
\end{figure}

\subsubsection{Fourfold cover}

We will now consider a fourfold cover of the punctured torus.  We cut open
the two cycles corresponding to the opposite edges of the unit square, and
glue three copies to form a single square torus double the size, as shown
in Figure~\ref{fc-cover}.
The fourfold cover of the  punctured square torus is itself a square 
torus with periods $2$ and $2i$ and four punctures located at the centers
$\lambda_i, \, i = 1, \cdots 4$,
of the unit squares:
\be 
\begin{split}
\hbox{Fourfold cover:}  \quad  & 2\omega_1 = 2 \,,  \quad 2 \omega_2 = 2 i\,, \\
\lambda_1 = \ & \ \tfrac{1}{2} +  \tfrac{1}{2} \, i \,, \quad
\lambda_2 =  \  \tfrac{1}{2} +  \tfrac{3}{2} \, i \,,  \\
\lambda_3 = \ & \ \tfrac{3}{2} +  \tfrac{1}{2} \, i \,, \quad 
\lambda_4 =  \  \tfrac{3}{2} +  \tfrac{3}{2} \, i \,.
\end{split}
\ee
A closed one-form on this fourfold cover 
with the correct integrals around the punctures and with the
requisite antisymmetry or symmetry under projection to the
original manifold is build from  a meromorphic abelian differential 
 $\omega_{\rm fc}$, with `fc' for four-cover:
\be
\omega_{\rm fc} (z) \ = \ \Bigl( \zeta (z - \lambda_1) -\zeta (z - \lambda_2) - \zeta(z-\lambda_3)+ \zeta(z-\lambda_4)  \Bigr) 
{dz \over 2\pi i }  \,. 
\ee
This form has the requisite periodicities
\be
\omega_{\rm fc} (z + 2) \ = \  \omega_{\rm fc} (z)\,,\qquad
\omega_{\rm fc} (z + 2i) \ = \  \omega_{\rm fc} (z)\,,
\ee
as well as the proper anti-periodicities
\be
\omega_{\rm fc} (z + 1) \ = \  -\omega_{\rm fc} (z)\,,\qquad
\omega_{\rm fc} (z + i) \ = \  -\omega_{\rm fc} (z)\,,
\ee
and the required integrals around the punctures,
\be
\oint_{\lambda_1}  \omega_{\rm fc}  \ = \ 1 \,, \quad
\oint_{\lambda_2}  \omega_{\rm fc}  \ = \ -1 \,, \quad
\oint_{\lambda_3}  \omega_{\rm fc}  \ = \ -1 \,, \quad
\oint_{\lambda_4}  \omega_{\rm fc}  \ = \ 1 \,.
\ee
Just as we explained for the case of the double cover, we can use
the real part of the meromorphic differential $\omega_{\rm fc}$ as our 
real one form, because it also satisfies all of the requisite properties. 
\begin{figure}[!ht]
\leavevmode
\begin{center}
\epsfysize=6.0cm
\epsfbox{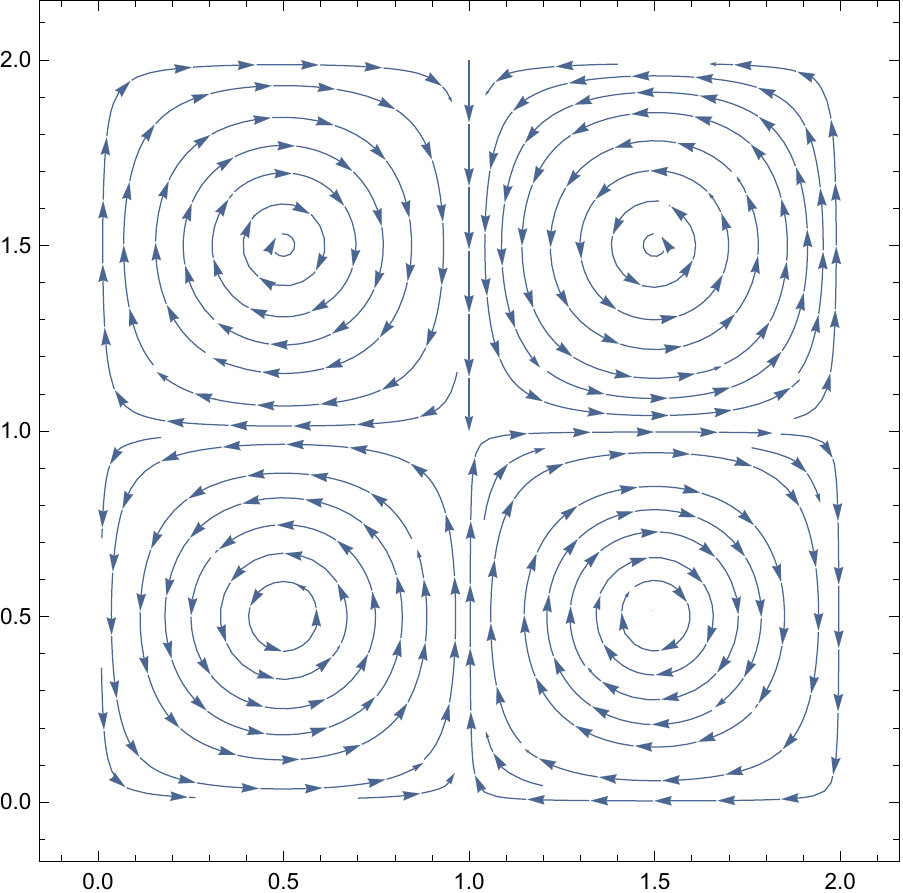}\hskip15pt
\epsfysize=6.0cm
\epsfbox{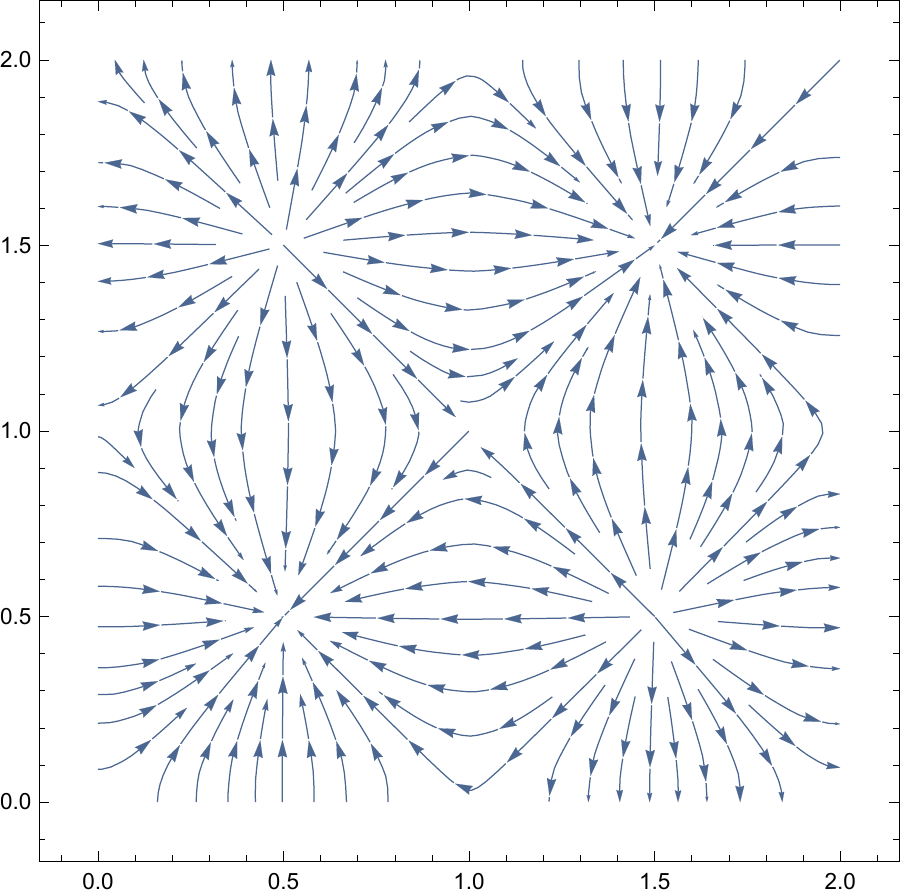}
\end{center}
\caption{\small  The meromorphic abelian differential $\omega_{\rm fc}$ 
for the fourfold cover of the punctured torus. 
Its real part is the one-form represented in this figure.   Left: The vector field lines associated with the one-form.   Right:  The vector field lines for  the associated flow.  The form and the flow, restricted to the original torus preserve the 
square symmetry.  The abelian differential $\omega_{\rm fc}$, squared, defines a Jenkins-Strebel quadratic differential that corresponds to a contact interaction for a torus and four external states.  This is a contact interaction because the union of the development of each of the four strings covers fully the surface. }
\label{punct-torus-fig99}
\end{figure}
The one-form $u^3$ for the fourfold cover is therefore
\be
u^3 \ = \ \hbox{Re} \bigl( \omega_{\rm fc} \bigr)  + d\phi^3 \,. 
\ee 
The abelian differential $\omega_{\rm fc}$ is rather special.  Its square
is a Jenkins-Strebel quadratic differential that defines a one-loop {\em contact} interaction of four strings on a torus.  Indeed the quadratic differential takes the form $\sim (dz)^2/(z-\lambda_i)^2$ near each puncture
$\lambda_i$.
In Figure~\ref{punct-torus-fig99} we show the representation of the one-form 
$\hbox{Re}(\omega_{\rm fc})$ as vector field lines (left) and the representation of the associated flow (right).   The significant advantage of the fourfold cover 
over the double cover, made manifest in the figure,  is that the form $\omega_{\rm fc}$ restricted to the 
original torus has the discrete symmetries of the square.   Using the
fourfold cover, both in the primal and in the dual programs 
we can restrict ourselves to
fundamental domains within the original punctured torus.  In fact, we can focus
on the domain $0 \leq x, y \leq 1/2$, one fourth of the size of the original torus, and even within this domain we can cut it in half using the symmetry 
under $(x, y) \to (y, x)$.

\begin{figure}[!ht]
\leavevmode
\begin{center}
\epsfysize=6.0cm
\epsfbox{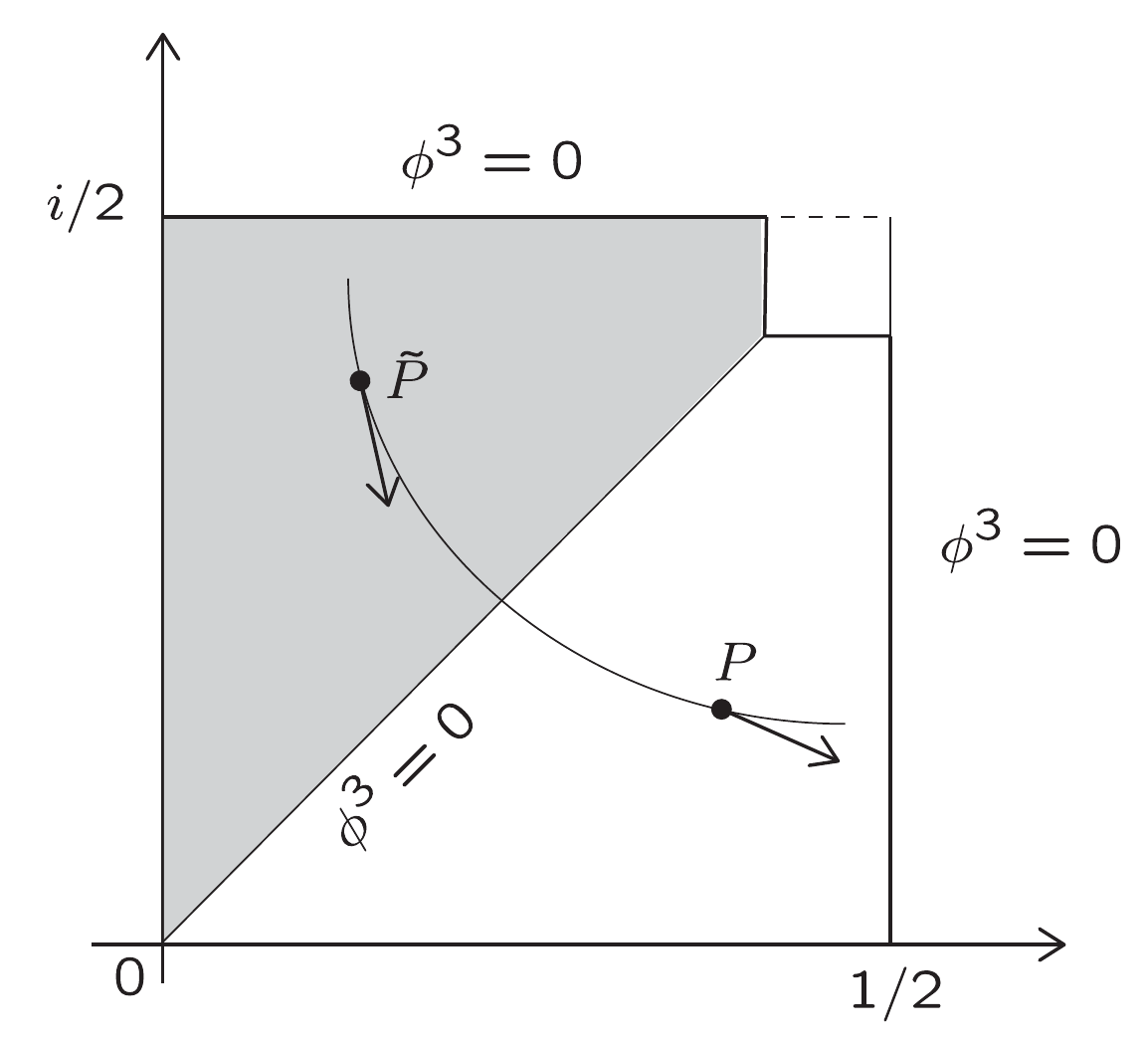}\hspace{1cm}
\epsfysize=6.0cm
\epsfbox{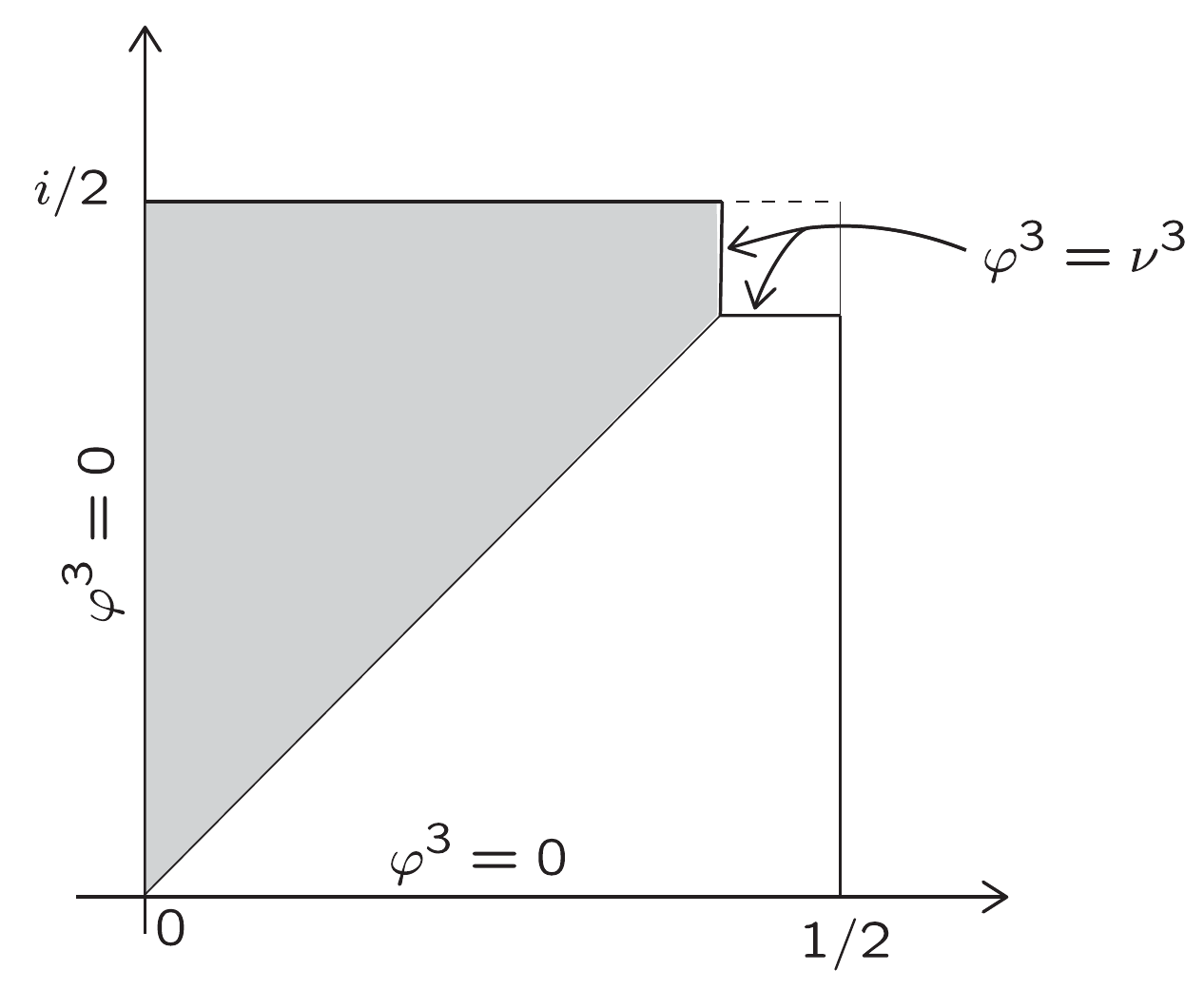}
\end{center}
\caption{\small Left:  The boundary conditions on the function $\phi^3$ defining the trivial form $d\phi^3$ that enters the calibration $u^3$ for curves homotopic to the puncture.  The function $\phi^3$ is antisymmetric about the diagonal.
Right:  The boundary conditions on the function $\varphi^3$ entering the dual.
The function takes the prescribed value $\nu^3$ at the boundary regulating the puncture, and is symmetric about the diagonal. 
 }
\label{phi3-fig}
\end{figure}

Having determined the one-form $\hbox{Re} (\omega_{\rm fc})$ that provides the unit period around the puncture, let us examine
the symmetry constraints on $\phi^3$.  We do this in the context of the
regulated puncture, where a small square hole replaces the puncture.  
The fundamental domain is shown in Figure~\ref{phi3-fig}, left.  
The integral of $d\phi^3$
around the puncture should provide no extra contribution to the period.  
It follows that the integral of $d\phi^3$
along any contour starting anywhere on the horizontal edge $y = 1/2$ and ending
anywhere along the vertical edge $x=1/2$ must vanish because any nonzero 
contribution, by symmetry, would give four times that value for the full integral around the puncture.  This means that $\phi^3$ must be equal to the same constant on both edges.  Without loss of generality
we take $\phi^3 = 0$ along these edges, as shown in the figure.  A further
constraint arises from the $\mathbb{Z}_2$ transformation $(x,y) \to (y,x)$.
Consider the two points $P$ and $\tilde P$ related by such an exchange. The 
$x$ and $y$ components of the form
$u^3$ at those points must be related as follows
\be
u^3_x (\tilde P) \ = \ - u^3_y (P) \,, \quad  u^3_y(\tilde P) = - u^3_x (P) \,.
\ee
One can verify that the form $\hbox{Re}(\omega_{fc})$ has this symmetry.
The trivial form $d\phi^3$ must also have the symmetry, so
\be
\partial_x \phi^3\bigl|_{(y,x)}  \ = \ - \partial_y \phi^3\bigl|_{(x,y)}  \,, \quad  \partial_y \phi^3\bigl|_{(y,x)}  \ = \ - \partial_x \phi^3\bigl|_{(x,y)} \,.
\ee
These relations require that
\be
\phi^3 (y, x) \ = \ - \, \phi^3 (x, y) \,. 
\ee
With this condition, we now learn that $\phi^3=0$ along the diagonal from 
the origin up to the corner of the small boundary and one can work with the
shaded part of the domain (Figure~\ref{phi3-fig}, left).  The value of $\phi^3$ is then free on the vertical
segment of the small boundary and on the vertical segment $x=0, y\in [0,1/2]$.

Let us now turn to the dual program.  Here we also use the fourfold cover,
giving us strong symmetry constraints on the function $\varphi^3$ associated with the curves homologous to the puncture.  Our discussion of the dual implied that the discontinuity $\nu^3$ on the value of $\varphi^3$ across a representative curve of the homology class can be implemented by setting $\phi^3 = \nu^3$ at the boundary of the original torus.  On the fourfold cover, the 
function $\varphi^3$ must also be anti-periodic under
$z \to z+ 1$ and under $z \to z+i$.  It follows that the value of $\varphi^3$ at the
other three boundaries must be either plus or minus $\nu^3$.  
The expected invariance of the systolic
geodesics under reflection about the vertical line $x=1$ of
the covering surface implies that $\varphi^3$ must be antisymmetric under this reflection.  Similarly, $\varphi^3$ must be antisymmetric under reflection
about the horizontal line $y=1$.  It follows that 
$\varphi^3=0$ on those lines, and its anti-periodicity 
implies that $\varphi^3=0$ along the apparent boundaries $x=0,2,$ and $y=0, 2,$ of the fourfold cover. The takeaway of this analysis
is shown on the right part of Figure~\ref{phi3-fig}.  As indicated there,
$\varphi^3 = \nu^3$ on the small boundary, and $\varphi^3=0$ along
the vertical segment $x=0, y\in [0,1/2]$ as well as on the horizontal 
segment $x\in [0,1/2], y=0$.   The function $\varphi^3$ is also symmetric
under the diagonal reflection $(x,y) \to (y,x)$ and therefore it suffices to find out its
value on the shaded region.   These are the boundary conditions to be used in maximizing the dual objective.

\subsection{Results}
\label{sec:PTresults}

We expect the minimal-area metric on the punctured torus to include a semi-infinite cylinder of circumference 1, with the puncture at infinity. It is therefore useful  for numerical purposes  to cut off the cylinder, by considering instead the square torus with a finite square, of edge size $h$, removed. If the hole is small enough,  it will be surrounded by a region foliated by systolic geodesics that wrap the hole, which we call 3-geodesics. In the minimal-area metric, this region will be a finite cylinder. The rest of the torus will be covered by 1-  and 2-geodesics, wrapping the torus horizontally and vertically respectively.  Generic 1- and 2-geodesics cannot intersect the 3-geodesics; the reason is that if they intersected they would have to intersect twice, but systolic geodesics may not intersect more than once. The only exceptions are the ``last'' geodesics of the respective foliations, which may coincide along the mutual boundary of the regions they foliate.

These expectations are borne out by the configurations obtained numerically. Here we present data for the case $h=1/16$ from our highest-resolution solution of the dual program (128 lattice points on a side of the fundamental domain).  The results from solving the primal program are closely consistent. The systolic geodesics are shown in Figure \ref{fig:punc-h1/16-fig}. We see four distinct types of regions: The white annular region is foliated by 3-geodesics (shown in black); the two light-blue triangular regions foliated by 1-geodesics (blue); the two pink triangular regions foliated by 2-geodesics (red); and the purple region foliated by both 1- and 2-geodesics. The last 3-geodesic coincides with segments of the last 1- and the last 2-geodesic; this is where the cylinder is attached to the rest of the torus.

We obtained the following values for $\nu^\alpha$ and the area $A$:
\begin{equation}\label{nuAvals}
\nu^1 = \nu^2 = 0.45795\,,\qquad
\nu^3 = 0.21709\,,\qquad
A = 1.1330\,.
\end{equation}
The errors on these quantities, as estimated from the difference with the results of a lower-resolution run (64 lattice points on a side of the fundamental domain) are roughly $\pm1$ on the last digit. From the discussion in section 5 of \cite{headrick-zwiebach}, we know that the values of $\nu^\alpha$ and $A$ should satisfy $A = \sum\nu^\alpha$ (recall that we have set the systole $\ell_s$ to 1), which is indeed satisfied by the values \eqref{nuAvals}. Furthermore, since the cylinder is foliated by a single band of geodesics, its height and area are both given by $\nu^3$. Therefore the rest of the surface must have area
\begin{equation}\label{nu1vals}
\nu^1+\nu^2 = 0.91589 \,.
\end{equation}

\begin{figure}[!ht]
\leavevmode
\begin{center}
\epsfysize=5.0cm
\epsfbox{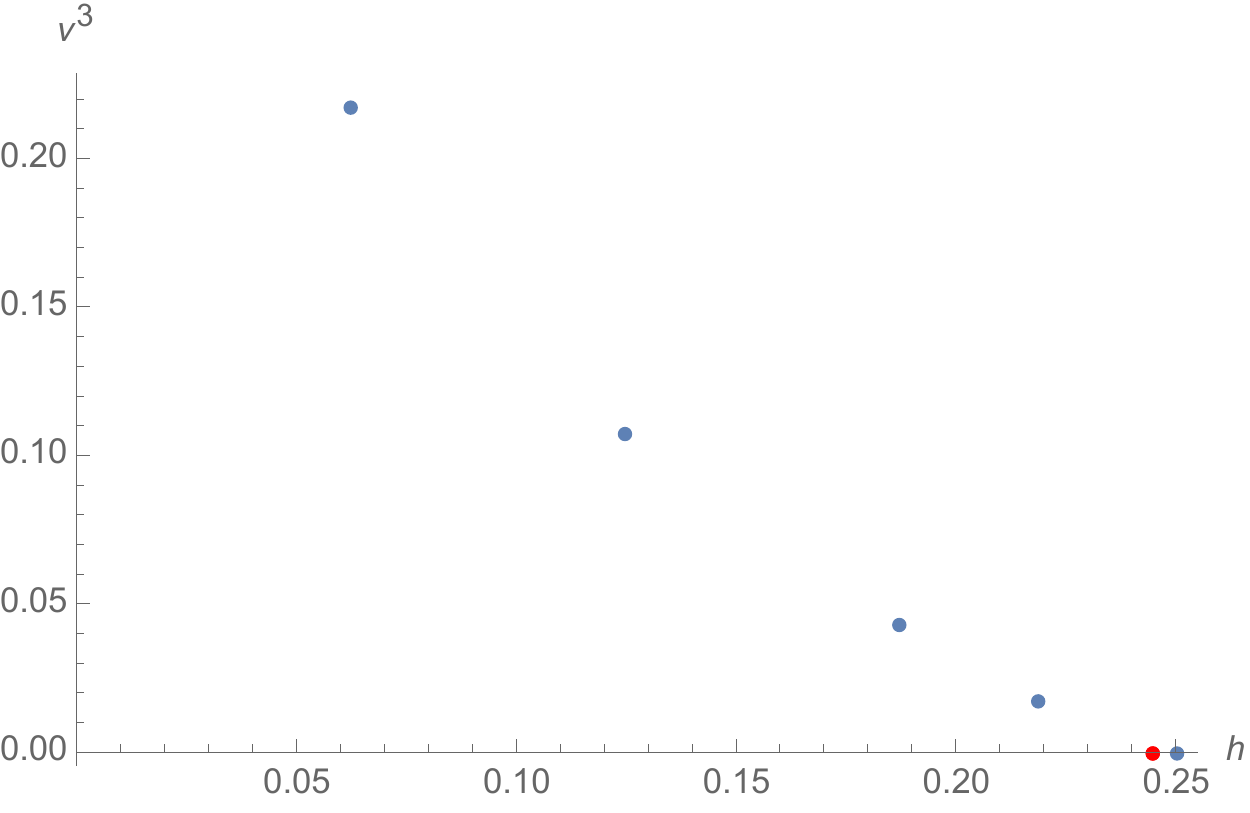}
\end{center}
\caption{\small 
The height $\nu^3$ of the cylinder as a function of the 
size $h$ of the boundary. The blue dots are data points from solving the dual program. The red dot is $h^{(2)}$, defined in subsection \ref{sec:spetorinthemodspa} as the value of $h$ for which, \emph{without} imposing the length constraint on curves wrapping the boundary, the boundary has length 1. As argued in the main text, this coincides with $h_{\nu^3=0}$, the smallest value of $h$ for which $\nu^3$ vanishes when the length constraint \emph{is} imposed.
 }
\label{fig:nu3plot}
\end{figure}

Before describing the geometry of the $h=1/16$ surface, we make a small digression concerning surfaces with different hole sizes. As $h$ increases, the height $\nu^3$ of the cylinder decreases, and at a certain value $h=h_{\nu^3=0}$ it vanishes (see Figure \ref{fig:nu3plot}).  For $h>h_{\nu^3=0}$, there are no 3-geodesics, and the boundary has circumference greater than 1. Therefore these surfaces are the same as the ones obtained in section \ref{sec:results} \emph{without} imposing the length constraint on curves wrapping the hole. At $h=h_{\nu^3=0}$, the boundary circumference is 1. It follows that $h_{\nu^3=0}=h^{(2)}$, the second critical value discussed in subsection \ref{sec:spetorinthemodspa}, and indeed this is consistent with the data obtained for various values of $h$. On the other hand, for $h\le h^{(2)}$, while the cylinder height depends on $h$, the geometry of the rest of the surface---the complement of the cylinder---should be independent of $h$ up to a complex coordinate transformation. In particular, $\nu^{1,2}$ should be independent of $h$. Indeed, we confirmed this prediction for $h=1/8$, 3/16, and 7/32, finding values of $\nu^{1,2}$ equal (within their respective errors) to the one in \eqref{nuAvals}. The value of $h^{(2)}$ quoted in \eqref{ells2} was found by searching the moduli space of tori with boundary (i.e.\ without imposing the length constraint on curves wrapping the hole) for the one with $\nu^{1,2}$ equal to the value in \eqref{nuAvals}.

\begin{figure}[!ht]
\leavevmode
\begin{center}
\epsfysize=7.0cm
\epsfbox{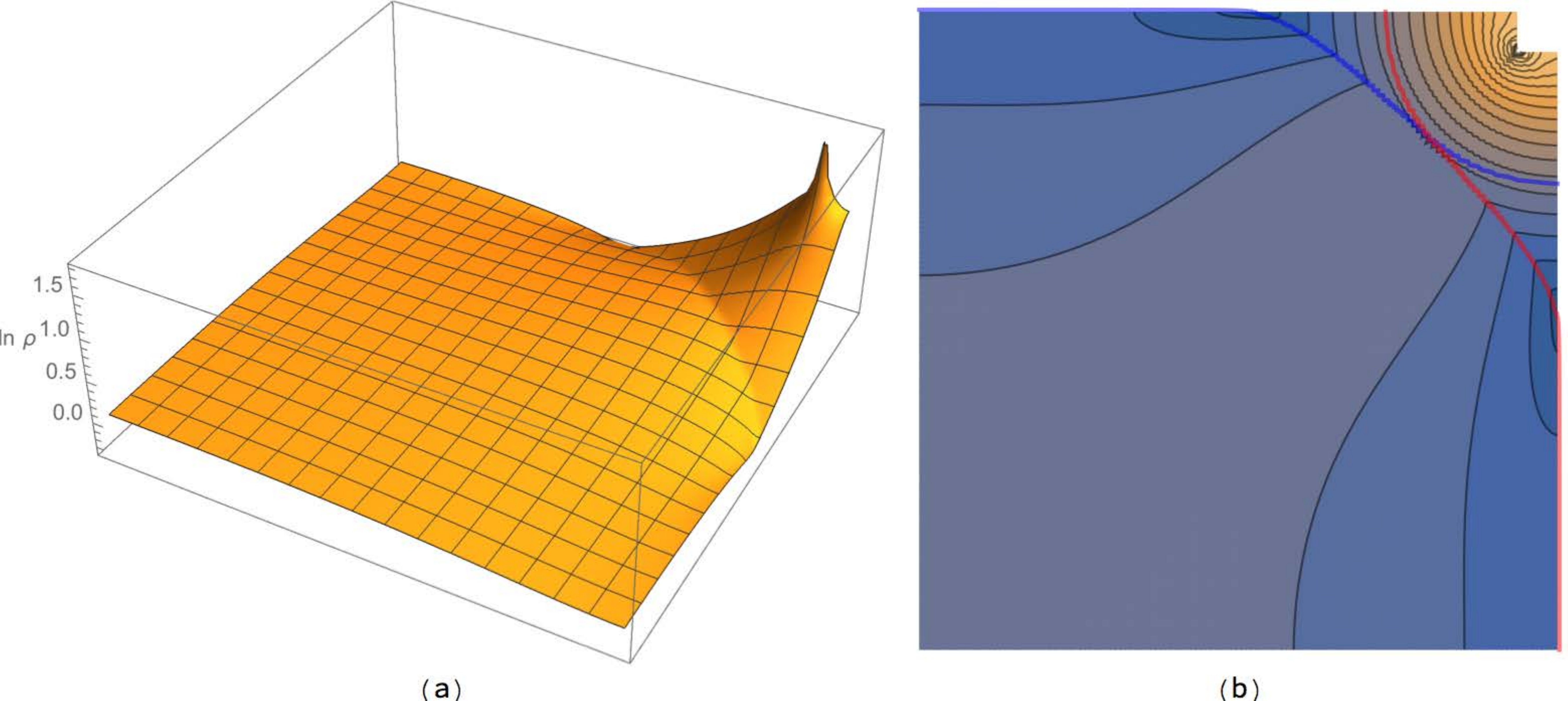}
\end{center}
\caption{\small 
Plot of $\ln\rho$ on the fundamental domain of the $h=1/16$ torus.
 In the contour plot on the right, the last 1- and 2-geodesics are shown in blue and red, respectively.
 }
\label{fig:PTrhoplots}
\end{figure}

\begin{figure}[!ht]
\leavevmode
\begin{center}
\epsfysize=7.0cm
\epsfbox{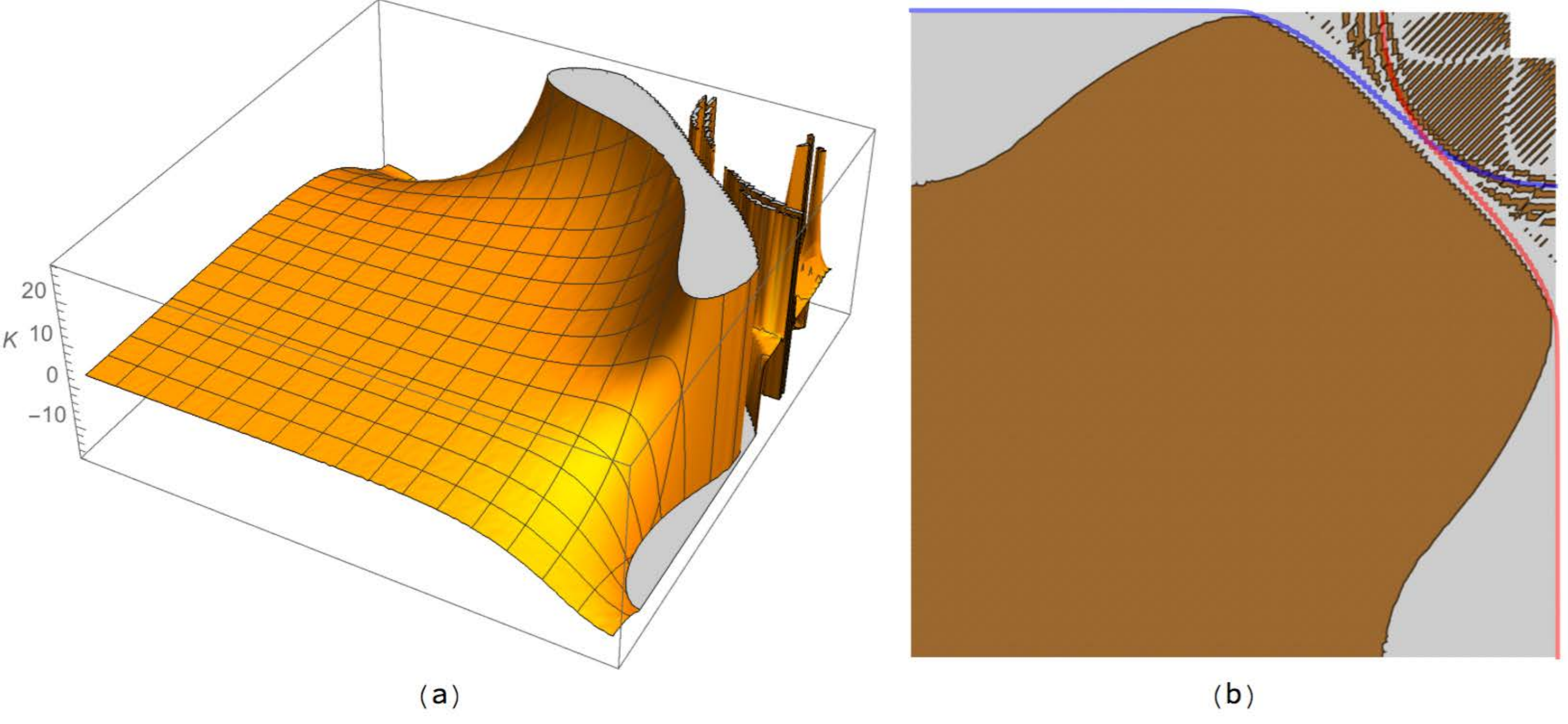}
\end{center}
\caption{\small 
(a) Gaussian curvature $K$ for the $h=1/16$ torus with boundary. (b) Regions with $K>0$ (brown) and $K<0$ (gray), with the last 1- and 2-geodesics shown. The complicated pattern in the regions foliated by a single geodesic (top right corner) is presumably due to numerical noise; the curvature of the exact solution must vanish there.
 }
\label{fig:PTcurv}
\end{figure}

We now return to our study of the geometry of the $h=1/16$ surface. The logarithm of line element $\rho$ is plotted in Figure \ref{fig:PTrhoplots}. It stays finite everywhere, and rises roughly linearly in the cylinder region. A crease is clearly visible, which coincides with the boundary of the region foliated by two geodesics. Since the Gaussian curvature is proportional to minus the Laplacian of $\ln\rho$, this crease implies a line curvature singularity with negative curvature, similar to what we saw in the Swiss cross described in subsection \ref{sec:metric}. Other than this crease, the metric appears to be smooth, so we expect the Gaussian curvature $K$ to be finite. $K$ is plotted in Figure \ref{fig:PTcurv}. The calculation of $K$ involves second derivatives of the metric; these derivatives, computed on the lattice, greatly amplify numerical noise; this is presumably the origin of the non-zero values for $K$ seen in the regions foliated by a single geodesics, which in the exact continuum solution must be flat. In the region foliated by both 1- and 2-geodesics, $K$ is nearly constant with a value around 1.8 in a region near the center, but then rises in the diagonal directions (toward the hole), and falls in the horizontal and vertical directions, becoming negative near the right and top edges of the fundamental domain. As discussed in subsection \ref{sec:spetorinthemodspa}, this bulk negative curvature is observed for all values of $h$ less than the critical value $h^{(1)}$ where the doubly-foliated region first touches itself (see \eqref{ells1}).

\begin{figure}[!ht]
\leavevmode
\begin{center}
\epsfysize=3.20cm
\epsfbox{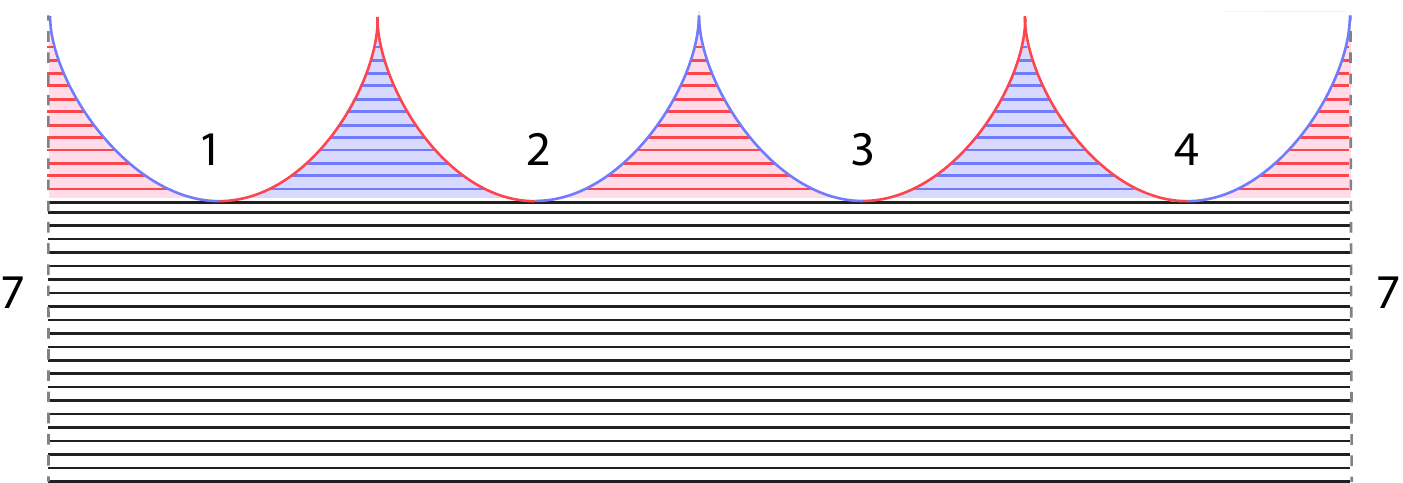}
\epsfysize=8.0cm
\epsfbox{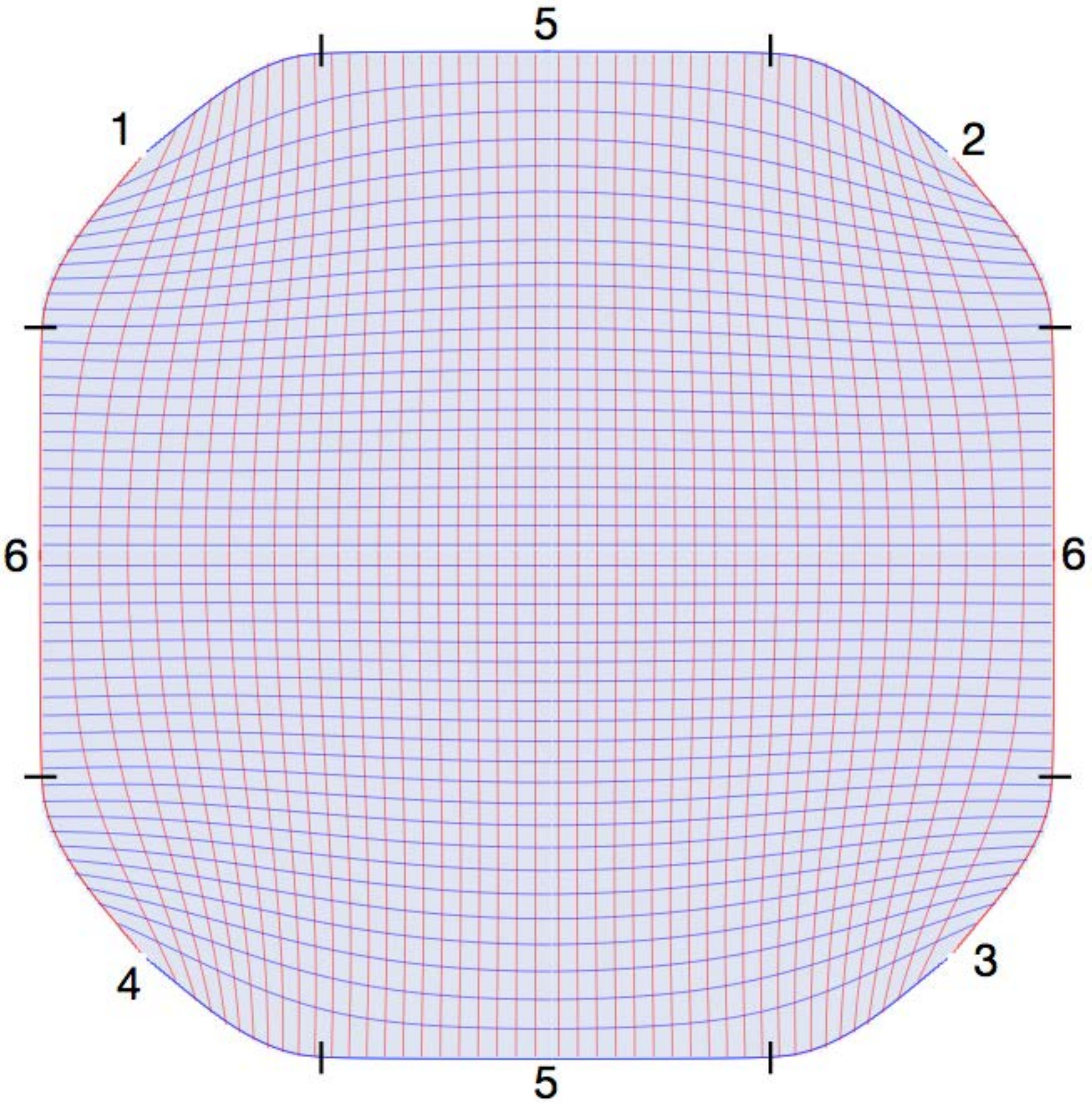}
\end{center}
\caption{\small 
Decomposition of the $h=1/16$ torus-with-hole into singly-foliated (left) and doubly-foliated (right) regions. The boundary segments with matching numbers are identified. The singly-foliated region is a cylinder (foliated by 3-geodesics) capped with four ``teeth'' (two foliated by 1-geodesics and two by 2-geodesics). The doubly-foliated region is a disk, whose boundary is divided into eight segments; four (numbered 5 and 6) are identified pairwise diametrically across the disk, and the other four are identified with the four parts of the boundary of the left figure.
 }
\label{fig:PTdiagrams}
\end{figure}

We can attempt to illustrate the geometry of this surface as follows. The cylinder is flat. The two pink and two light blue ``teeth'', foliated \emph{only} by the 1- and 2-geodesics respectively, are also flat, and join onto the cylinder without any line curvature. This part of the geometry is shown on the left panel of Figure \ref{fig:PTdiagrams}, where the left and right sides of the figure are identified. The doubly-foliated region is shown in the right panel. The boundary of this region is divided into eight segments, four of which are identified with each other pairwise diametrically across the disk and the other four with parts of the boundary of the figure on the left side. The boundary of the disk is a closed geodesic made up of pieces of 1- and 2-geodesics which meet tangently. Since it is topologically a disk, and its boundary has vanishing extrinsic curvature, by the Gauss-Bonnet theorem its total Gaussian curvature must be $2\pi$. It can be thought of roughly as a hemisphere with mostly uniform positive curvature but some ``lobes'' near its boundary that give it regions of large positive and negative curvature. It has $D_4$ dihedral symmetry. There is no line curvature where it is identified with itself, since it is identified along geodesics. There is line curvature, however, along the four lines where the left and right sides of Figure \ref{fig:PTdiagrams} are glued. Each such line is a geodesic from the point of view of the disk (right side) but has total extrinsic curvature $-\pi$ from the point of view of the ``teeth'' (left side), so carries total Gaussian curvature $-\pi$. The total curvature of the whole surface is thus $2\pi+4(-\pi)=-2\pi$, which of course agrees with that predicted by the Gauss-Bonnet theorem, given that the Euler character of the torus-with-hole is $-1$.

\section*{Acknowledgments}

Barton Zwiebach would like to acknowledge 
instructive discussions with Larry Guth and Yevgeny 
Liokumovich on systolic geometry.
The work of M.H is supported by the National Science Foundation through Career Award No. PHY-1053842, by the U.S.\ Department of Energy under grant 
DE-SC0009987, and by the Simons Foundation through a Simons Fellowship in Theoretical Physics. The work of B.Z.~is supported by the U.S.\ Department of Energy under grant Contract Number DE-SC0012567. M.H.\ would also like to thank MIT's Center for Theoretical Physics for hospitality and a stimulating research environment during his sabbatical year.

\begin{appendices}

\section{Numerical implementation}

In this appendix we give some details of the numerical solution of the primal and dual programs for the Swiss cross and punctured torus described in sections \ref{sec:programs} and subsection \ref{sec:puncturedtorusprograms} respectively. For both programs, the fundamental domain was discretized using a lattice, and the derivatives and integrals appearing in the constraints and objectives were replaced by lattice derivatives and integrals. Lowest-order derivatives and integrals were employed, both for simplicity and because---as can be seen in the results presented in subsections \ref{sec:metric} and \ref{sec:PTresults}---the functions being approximated are not necessarily smooth. The discretization scheme in shown in detail in subsection \ref{sec:discretization} for the illustrative case of the primal program for the Swiss cross in the L frame; the discretizations for the pentagon frame, the dual program, and the punctured torus are closely analogous.

The minimization and maximization in the two programs respectively was carried out in \emph{Mathematica} using the built-in functions FindMinimum and FindMaximum. The primal program is a constrained minimization problem, which \emph{Mathematica} solved with an interior point method. The dual program is an unconstrained maximization problem, which \emph{Mathematica} solved using a quasi-Newton method with Broyden-Fletcher-Goldfarb-Shanno updates.

In subsection \ref{sec:convergence}, we compare the results of the primal and dual at various resolutions, focusing on the case of the Swiss cross in the L frame, showing that the primal and dual agree to within numerical errors and show good convergence as the resolution is increased, both in the value of the extremal area and in the form of the optimal metric. Similar convergence behavior was observed for the punctured torus.

\subsection{Discretization}
\label{sec:discretization}

\begin{figure}[!ht]
\leavevmode
\begin{center}
\epsfysize=8.5cm
\epsfbox{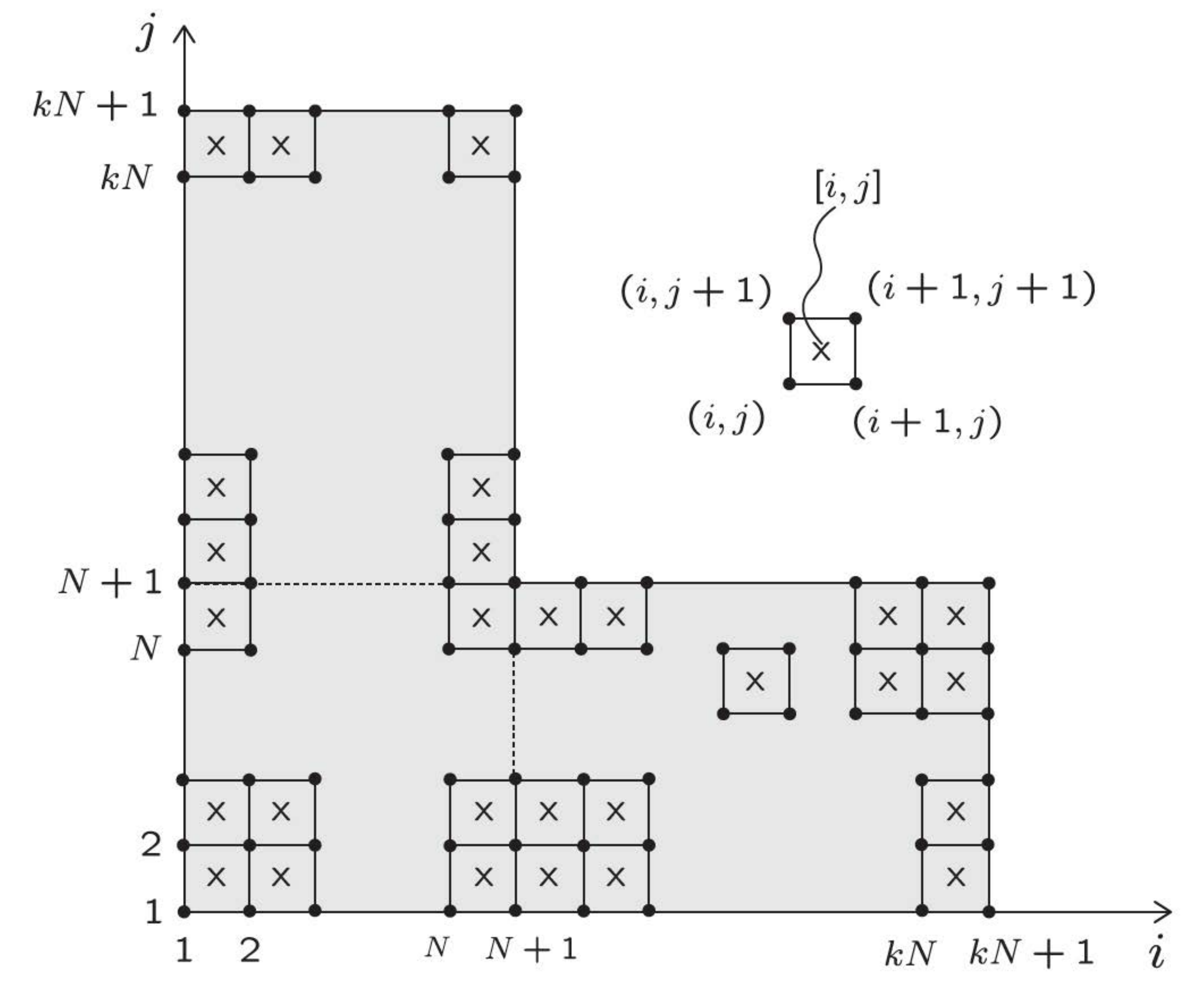}
\end{center}
\caption{\small  Discretization of the fundamental domain for the primal
program applied to the torus with a boundary.
The functions $\phi^1$ and $\phi^2$ entering the calibrations are defined
on the lattice points.  Derivatives and the metric are defined at the center
of the plaquettes.}
\label{ff8769}
\end{figure}

Here we describe in detail the discretization employed for the primal program on the Swiss cross in the L frame, a program we discussed in section \ref{sec:primalprogram}. The numerical solution requires
discretization of the region identified as a fundamental domain 
in Figure~\ref{ff876}.   We explain the discretization using Figure~\ref{ff8769}. 
The square region  $0 \leq x, y \leq 1/2$
has $N^2$ plaquettes, with $N$ some integer.  Each plaquette is a little square with edge $\Delta$ given by 
\begin{equation}
\Delta = {1\over 2N}\,. 
\end{equation}
We introduce a (rational) number $k$ such that $kN\in \mathbb{Z}$ is  the number of plaquettes that fit horizontally (or vertically).  
Since the distance from the
origin to the vertical edge is $\ell_s/2$ and
the edge of a plaquette is $\Delta$ we have
\begin{equation}
\tfrac{1}{2} \ell_s  \, = \, kN \Delta =  \tfrac{1}{2} k   \quad \to \quad  k = \ell_s
 \ \ (kN \in \mathbb{Z}) \,. 
\end{equation}
This allows us to do numerical 
work with rational values of $\ell_s$.  For $\ell_s= p/q$,
with $p$ and $q$ relatively prime integers, we can work 
with values of $N$ that are
arbitrary multiples of $q$.  The larger the value of $N$ the better
the accuracy. 

Each lattice point is labeled as $(i, j)$ where $i$ and $j$ are positive integers.
Notice that the origin is chosen to be
the point $(1,1)$.  The $(x,y)$ coordinates of 
the point $(i,j)$ are
\begin{equation}
(x[i] , \, y[i]) =  \Delta  \bigl( i-1\, , \, j-1  \bigr)  = {1\over 2N} \bigl(\, i-1\, , \, j-1 \bigr)  \,.  
\end{equation}
The function $\phi^1(x,y)$ is represented by its values $\phi[i,j]$ at each lattice point:
\begin{equation}
\phi[i,j]  \ = \ \phi^1 ( x(i), y(j) )  
\end{equation}
While the function $\phi^1$  lives on the lattice points, 
its derivatives and the metric are defined to live at the center of the
plaquettes.   We label the plaquette by the values $(i,j)$ of the lattice
point on the lower left of the plaquette.  Thus derivatives, for example,
are given by
\be
\begin{split}
(\partial_x \phi) [i,j]  =& \   {1\over 2\Delta} \bigl(  \phi[i+1,j] + \phi[i+1, j+1]
- \phi[i,j] - \phi[i, j+1]\bigr) \,, \\
(\partial_y \phi) [i,j]  =& \   {1\over 2\Delta} \bigl(  \phi[i,j+1] + \phi[i+1, j+1]
- \phi[i,j] - \phi[i+1, j]\bigr) \,. 
\end{split}
\ee
Note that on account of (\ref{exch-deriv-phi}) relating $\phi^1(=\phi)$ and $\phi^2$ derivatives we have
\be
(\partial_x \phi^2) [i,j] =  (\partial_y \phi) [j,i] \,, \quad \hbox{and}
\quad (\partial_y \phi^2) [i,j] =  (\partial_x \phi) [j,i] \,. 
\ee
The value of the scale factor $\Omega$ at the center of the $(i,j)$ plaquette is called $\Omega[i,j]$ and, as indicated in the program
(\ref{thirdprogram-vm99}), we have two inequalities to consider:  
\be
\begin{split}
\Omega[i,j] \ \geq \  & \ \bigl( 1 + (\partial_x \phi) [i,j]\bigr)^2  + \bigl( (\partial_y \phi) [i,j]\bigr)^2 \, , \\
\Omega[i,j] \ \geq \  & \ \bigl( (\partial_y \phi) [j,i]\bigr)^2 +
\bigl( 1 + (\partial_x \phi) [j,i]\bigr)^2 \,. 
\end{split}
\ee
Note that by defining 
\be
S[i,j] \ \equiv \  \bigl( 1 + (\partial_x \phi) [i,j]\bigr)^2  + \bigl( (\partial_y \phi) [i,j]\bigr)^2 \,, 
\ee
the inequalities for the metric take the simple form,
\be
\Omega[i,j] \geq S[i,j] \,,  \quad \hbox{and} \quad  
\Omega[i,j] \geq S[j,i]\,.
\ee
These constraints make it manifest that, as desired, the metric
has the expected reflection symmetry $\Omega[i,j] = \Omega[j,i]$. 
Moreover, $\Omega[i,i] = S[i,i]$.  The area functional to be minimized
is then
given by
\be
\hbox{Area}= \sum_{i,j\in F}  4 \Delta^2 \ \Omega[i,j]  \,. 
\ee
Here $F$ is the set of points labeling plaquettes in the fundamental domain and the factor of four is needed because the Swiss cross contains the fundamental domain four times.  More 
explicitly, the set of values in $F$ can be read from the figure and are:  
\be
F \ = \ \{i= 1, \ldots, N\,, j = 1\,, \ldots kN\} \cup \{i= N+1, \ldots, kN\,, j = 1\,, \ldots N\}\,. 
\ee
We need also the Gaussian curvature $K$ of our metric $ds^2 = \Omega (dx^2 + dy^2)$.  The value is 
\be
K \ = \ - {1\over 2 \Omega } \nabla^2  \ln \Omega \,. 
\ee
In a planar discretization the Laplacian of a function is computed using
5 points, one at the center where we evaluate the Laplacian.  There are two natural options, a cross configuration and a 
square configuration.  For a function $f(x,y)$ the Laplacian at the origin
using these configurations is 
\be
\label{laplacian-eqn}
\begin{split}
\nabla^2 f(0,0) \ = \ & \  {1\over h^2} \Bigl( f(h,0) + f(-h,0) + f(0,h) + f(0,-h) - 4 f(0,0) \Bigr)\,, \\
\nabla^2 f(0,0) \ = \ & \  {1\over 2h^2} \Bigl( f(h,h) + f(-h,-h) + f(h,-h) + f(-h,h) - 4 f(0,0) \Bigr)\,. \\
\end{split}
\ee 
Applied to our lattice $h= 1/(2N)$ and this gives us two options
\be
\label{curvature-eqn}
\begin{split}
K[i,j]  \ = \ & \ - {2N^2\over \Omega[i,j]}  \ln \biggl( \ {\Omega[i,j+1] \,\Omega[i, j-1] \, \Omega[i+1, j] \, \Omega[i-1, j] \over  \Omega^4[i,j] }\ \biggr) 
\,, \\[0.5ex]
K[i,j]  \ = \ & \ - {N^2\over \Omega[i,j]}  \ln \biggl( {\Omega[i+1,j+1] \,\Omega[i+1, j-1] \, \Omega[i-1, j+1] \, \Omega[i-1, j-1] \over  \Omega^4[i,j] }\biggr)\,.
\end{split}
\ee

\subsection{Convergence}
\label{sec:convergence}

In this appendix we present data showing that the discretized primal and dual programs agree with each other and converge as the lattice resolution is increased. For concreteness, we focus on the $\ell_s=2$ Swiss cross, in the L frame. Very similar convergence behavior was observed for the punctured torus.

\begin{figure}[!ht]
\leavevmode
\begin{center}
\epsfysize=4.25cm
\epsfbox{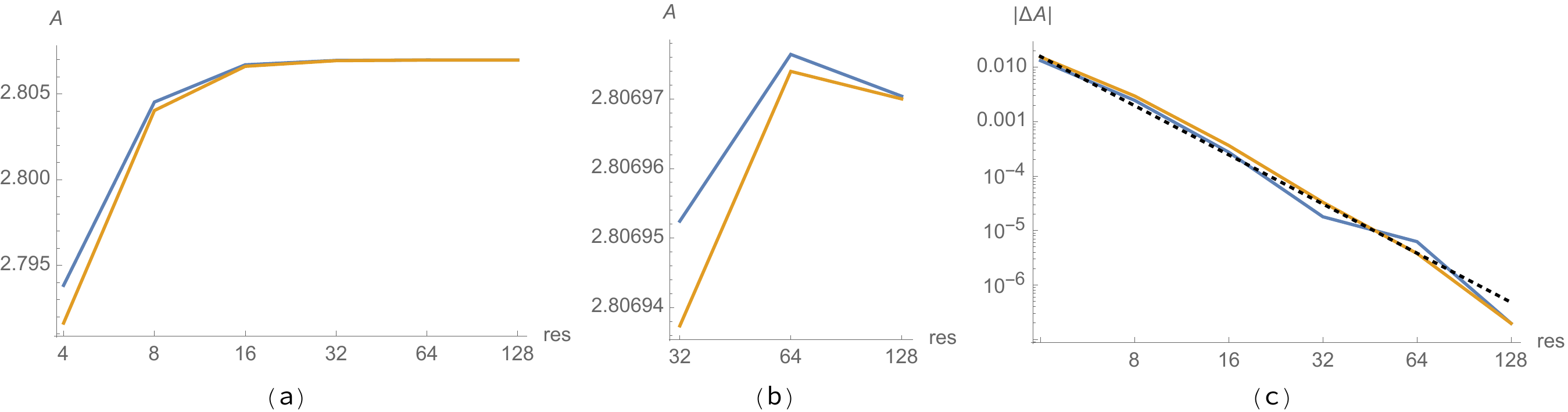}
\end{center}
\caption{\small 
(a) and (b) Extremal area $A$ on the $\ell_s=2$ Swiss cross obtained numerically from the primal (blue) and dual (orange) programs, versus the linear resolution 
``res'' of the lattice discretization. 
(c) Absolute value of difference between the optimal $A$ at a given
resolution and the optimum value 
$A_*$ in \eqref{extremalarea} (the average of the primal and dual results at the highest resolution). The dotted black line is (res)$^{-3}$, showing that convergence is in excellent agreement with a power law with an exponent around $-3$.
}
\label{fig:Aconvergence}
\end{figure}

We start by studying the convergence of the extremal area as the resolution is increased.  We define the {\em linear resolution} as the number of plaquettes on
the long edge of the fundamental domain.  In the notation of 
subsection \ref{sec:discretization} the linear resolution is $2N$.   
The total number of 
plaquettes in the fundamental domain is $3N^2$. 

Figure~\ref{fig:Aconvergence} shows the extremal area computed from the primal and dual programs at linear resolutions ranging from 4 to 128 plaquettes. The highest-resolution value  for the extremal area 
is obtained from either the primal or dual, which give the same result to seven significant figures:
\begin{equation}\label{extremalarea}
A_* = 2.806\,970\,. 
\end{equation}
The error will be estimated below as $\pm10^{-6}$, or $\pm1$ in the last digit.

There are two notable features in the numerical results.  
The first one is that the optimal values of the primal and the dual track each other quite closely as the resolution changes, with the primal value a tiny bit higher than the dual. At low resolution, both underestimate the extremal area. It might have been expected that the primal program, since it is a minimization program, would overestimate the area. However, while the primal objective evaluated exactly on any continuum configuration bounds the true minimum, in our numerical implementation the configuration is defined on a lattice and the objective involves discretized derivatives. Therefore there is no inconsistency in having a value of the objective which lies below the true continuum minimum. 
While we do not know why the primal and dual optimal values track each other so closely as the resolution is increased, one plausible explanation is that the \emph{discretized} programs, at the same resolution, are actually related to each other (perhaps only approximately) by strong duality.

The second notable result is that
the primal and dual optimal values 
both converge to a common value at a rate consistent with a power law with exponent $-3$.
Indeed, defining the error 
$\Delta A$ as a function of the 
linear resolution ``$\rm{res}$''
(equal to $2N$ in the language of appendix \ref{sec:discretization}) 
\be
\Delta A({\rm res}) :=   A({\rm{res}}) -A_*  \,,
\ee
with $A_*$ the highest resolution value in (\ref{extremalarea}), we find
\be
\label{error-estimate}
|\Delta A({\rm res})| \approx({\rm {res}})^{-3}  \,.  
\ee
It would be interesting to try to derive 
this convergence rate from a first-principles analysis of the discretized programs. 
We have not attempted such an analysis; instead we merely take the agreement between the primal and dual and their joint rapid convergence as indications that the errors introduced by the discretization are under control and the results are reliable.   In this spirit we estimate the error in (\ref{extremalarea}) as $\pm10^{-6}$, 
calculated by assuming that the error depends on the resolution as in (\ref{error-estimate}). A more conservative error estimate is to take the difference between the highest and second-highest resolutions (res $=128$ and res $=64$), about $5\times10^{-6}$.

\begin{figure}[!ht]
\leavevmode
\begin{center}
\epsfysize=4.25cm
\epsfbox{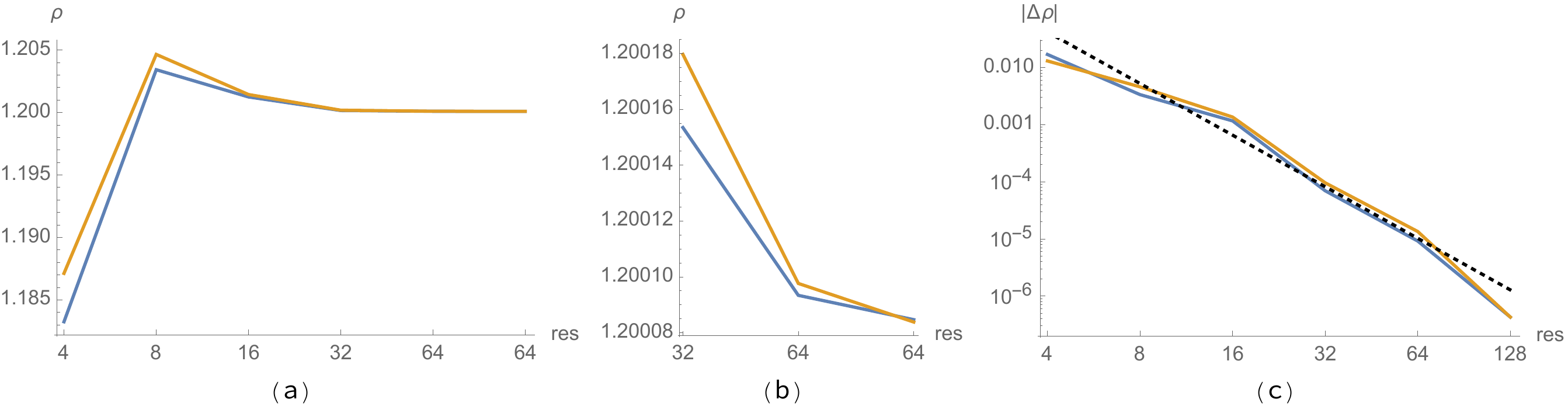}
\end{center}
\caption{\small 
(a) and (b) Value of the line element $\rho$ in the extremal metric at a representative point midway between the bottom-left corner and the inner corner of the fundamental domain, in the primal (blue) and dual (orange) programs, versus linear resolution. (c) Absolute value of difference between $\rho$ and the average of the highest-resolution values. The dotted black line is $\frac13$(res)$^{-3}$, showing that convergence is in good agreement with a power law with an exponent around $-3$.
}
\label{fig:rhoconvergence}
\end{figure}

\begin{figure}[!ht]
\leavevmode
\begin{center}
\epsfysize=9cm
\epsfbox{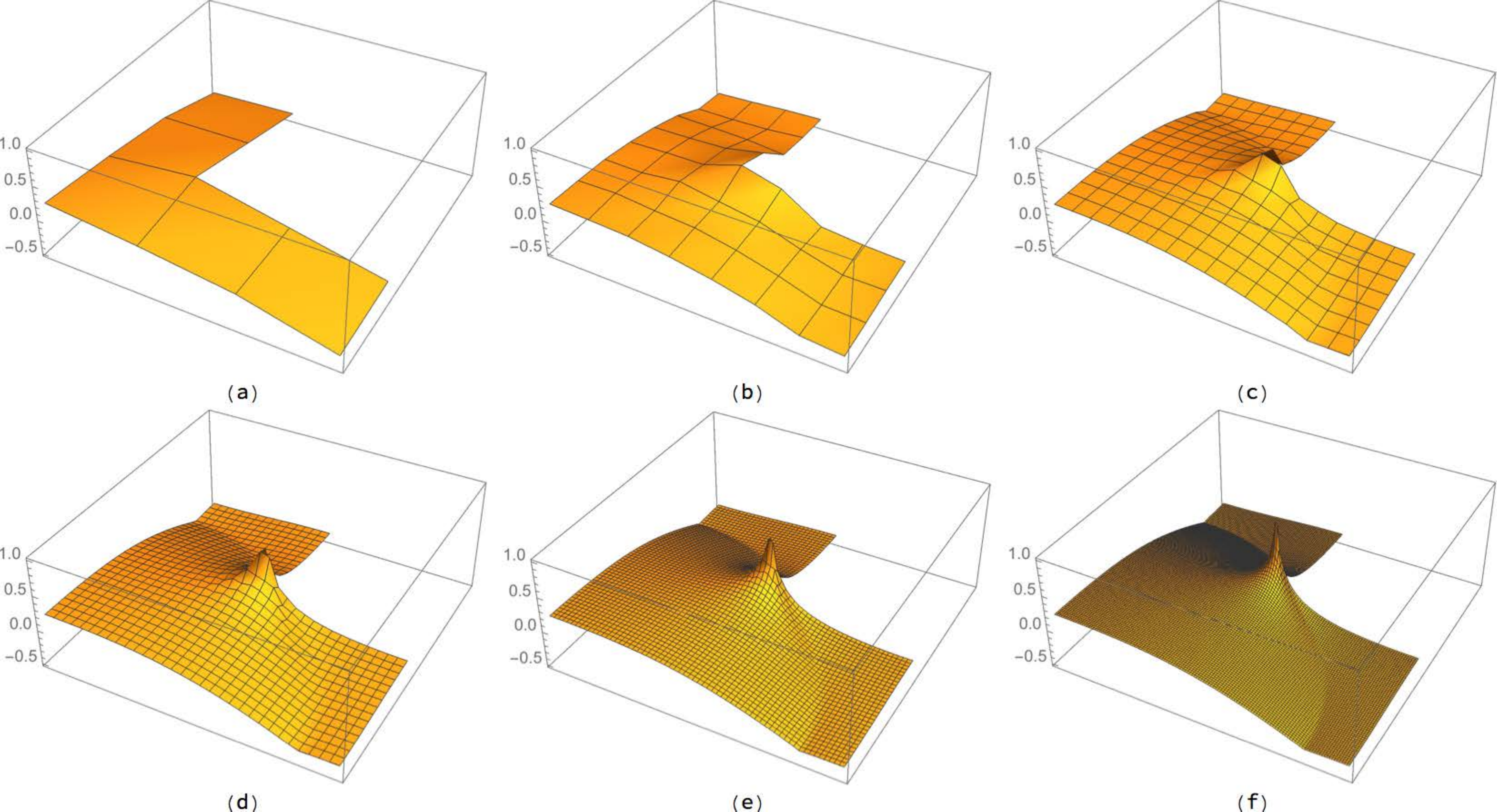}
\end{center}
\caption{\small 
$\ln\rho$ computed from the optimal configuration in the dual program at various linear resolutions: (a) 4; (b) 8; (c) 16; (d) 32; (e) 64; (f) 128.
}
\label{fig:rhoconvergence3D}
\end{figure}

Next, we study the extremal metric, or more precisely its line element $\rho=\sqrt\Omega$. In Figure\ \ref{fig:rhoconvergence}, we show the convergence of $\rho$ for both the primal and dual at a representative point, the midpoint between the bottom-left corner and the inner corner of the fundamental domain (corresponding to $x=y=\tfrac{1}{4}$ in Appendix \ref{sec:discretization}). 
Just as they did for the area, the metrics obtained from the primal and dual track each other closely and converge as a function of resolution like a power law with an exponent of roughly $-3$. At the highest resolution, the value of $\rho$ at the representative point differs between the primal and dual by less than 1 part in $10^6$. The difference between the primal and dual is not uniform over the fundamental domain, being largest near the inner corner, but is everywhere less than 1 part in $10^3$. Finally, in 
 Figure\ \ref{fig:rhoconvergence3D}, we show the convergence of $\ln\rho$ in the dual program over the whole fundamental domain as the resolution is increased.

\end{appendices}

\end{document}